# Comprehensive analysis of spherical bubble oscillations and shock wave emission in laser-induced cavitation

Xiao-Xuan Liang[1], Norbert Linz[1], Sebastian Freidank[1], Günther Paltauf[2] and Alfred Vogel[1]

[1]Institute of Biomedical Optics, University of Lübeck, Peter-Monnik-Weg 4, 23564 Lübeck, Germany

[2]Institute of Physics, Karl-Franzens-University Graz, Universitätsplatz 5, 8010 Graz, Austria

The dynamics of spherical laser-induced cavitation bubbles in water is investigated by plasma photography, time-resolved shadowgraphs, and single-shot probe beam scattering enabling to portray the transition from initial nonlinear to late linear oscillations. The frequency of late oscillations yields the bubble's gas content. Simulations with the Gilmore model using plasma size as input and oscillation times as fit parameter provide insights into experimentally not accessible bubble parameters and shock wave emission. The model is extended by a term covering the initial shock-driven acceleration of the bubble wall, an automated method determining shock front position and pressure decay, and an energy balance for the partitioning of absorbed laser energy into vaporization, bubble and shock wave energy, and dissipation through viscosity and condensation. These tools are used for analysing a scattering signal covering 102 oscillation cycles. The bubble was produced by a plasma with 1550 K average temperature and had 36 µm maximum radius. Predicted bubble wall velocities during expansion agree well with experimental data. Upon first collapse, most energy was stored in the compressed liquid around the bubble and then radiated away acoustically. The collapsed bubble contained more vapour than gas, and its pressure was 13.5 GPa. The pressure of the rebound shock wave initially decayed $\propto r^{-1.8}$, and energy dissipation at the shock front heated liquid near the bubble wall above the superheat limit. The shock-induced temperature rise reduces damping during late bubble oscillations. Bubble dynamics changes significantly for small bubbles with less than 10 µm radius.

_________________________________________________________________

## 1. Introduction

Laser-induced cavitation plays an important role in many fields, especially in laser materials processing in liquid environments (Barcikowski et al. 2019) and in biomedicine and biophotonics, where it enables to perform precise surgery on cells and within transparent tissues (Koenig et al. 1999; Vogel et al. 2005; Yanik et al. 2004; Tirlapur and Koenig 2002; Stevenson et al. 2010; Vogel et al. 1990; Juhasz et al. 1990; Chung and Mazur 2009; Palanker et al. 2010; Hoy et al. 2014). It involves localized energy deposition by laser pulses that results in a phase transition of the target material causing the bubble expansion, accompanied by the emission of an acoustic transient that often evolves into a shock wave. The bubble expansion displaces the breakdown medium, and the bubble collapse is again accompanied by acoustic transient emission. Shock waves and cavitation are inevitably linked to laser ablation in liquids (LAL) and to laser surgery. They may contribute to the desired effect but can also evoke undesired side effects, especially when very gentle cuts or perforations in biological environments are desired. Therefore, it is of great interest to systematically explore the dependence of laser-induced cavitation phenomena on laser parameters and bubble size. Moreover, the topic is of general physical interest as laser induced cavitation involves nonlinear optical, acoustical and hydrodynamic phenomena producing extreme states of matter.

Localized deposition of laser light energy into the bulk of water or transparent biological media relies on plasma formation by optical breakdown of the liquid and is associated with high volumetric energy densities, temperatures, and pressures (Vogel et al. 1996; Lauterborn and Vogel 2013). Therefore, bubbles often exhibit large-amplitude oscillations, and with tight focusing and an isotropic environment almost perfectly spherical bubbles can be produced that match the conditions assumed in theoretical models of spherical bubble dynamics. Sonochemistry via acoustically driven bubble oscillation also relies on spherical or approximately spherical bubble oscillations (Barcikowski et al. 2019; Suslick and Flannigan 2008), while strongly aspherical bubble dynamics is observed at surfaces in LAL and for bubbles in hydrodynamic cavitation that are involved in cavitation erosion (Lauterborn and Kurz 2010; Kanitz et al. 2019).

In the past, experimental studies of laser-induced cavitation in water have mostly been performed on millimetre-sized bubbles (Benjamin and Ellis 1966; Lauterborn and Bolle 1975; Tomita and Shima 1986; Vogel, Lauterborn and Timm 1989; Vogel, Busch and Parlitz 1996; Philipp and Lauterborn 1998; Baghdassarian, Tabbert and Williams 1999; Brujan and Vogel 2001; Lindau v Lauterborn 2003; Obreschkow et al. 2011; Obreschkow et al. 2013, Reuter et al. 2017, Sagar and Moctar 2020, Podbevsek et al. 2021). Here, surface tension and viscosity play only a minor role and the bubble dynamics exhibits self-similar features over a large range of bubble sizes (Plesset and Prosperetti 1977; Lauterborn and Kurz 2010). Energy deposition by laser pulses of femtosecond to nanosecond duration occurs in a time interval much shorter than the bubble oscillation time, which is 185 µs for a bubble with 1 mm maximum radius, and the bubble dynamics is the impulse response to an almost instantaneous energy deposition. Time scales are long enough to enable recording of the dynamics via high-speed photography. Laser surgery of biological tissues and, particularly, cells goes along with much smaller bubble sizes in the micro- and nanometre range and shorter oscillation times (120 ns for a bubble with 1 µm maximum radius), which requires faster experimental techniques than high-speed photography. The dynamics of such small bubbles is influenced by surface tension and viscosity because their contributions to the bubble wall pressure scale inversely with the bubble radius $R$ (Lauterborn and Kurz 2010). Moreover, the laser pulse duration can become a significant part of the bubble oscillation time, especially when ns laser pulses are used and the bubble expansion starts already during energy deposition. This establishes a need for a systematic investigation of the changes in bubble dynamics with decreasing bubble size, both in biological media and in water. Such investigations should include the temporal evolution of bubble radius, pressure, wall velocity, and of the pressure distribution around the bubble that may lead to shock wave formation. However, beyond that it is of utmost importance to establish an energy balance tracing the partitioning of absorbed laser energy into vaporization, shock wave emission, bubble formation, viscous damping, and condensation. Changes in the influence of viscosity and surface tension with deceasing bubble size will strongly affect the partitioning. An energy balance will thus help to understand the changing dynamics for the transition from micro- to nanocavitation, and it will elucidate the mechanisms governing cell and tissue surgery, and the accompanying side effects.

In this paper, we focus on bubble dynamics in water and present a set of experimental and simulation tools that enables a systematic study of the parameter dependence of bubble dynamics including the determination of peak pressures upon bubble generation and collapse, tracking of bubble oscillations and shock wave emission, and a complete energy balance. Simulations are based on the well-established Gilmore model (Gilmore 1952) that is extended in various aspects. Our simulations provide full information about the laser-induced bubble dynamics, shock wave emission and energy partitioning based on very few parameters that need to be determined experimentally. These are the size of the laser-produced plasma, which is translated into the start radius $R_0$ for the simulations, and the duration of the subsequent bubble oscillations, $T_{osci}$, where $i$ denotes the number of the oscillation. Numerous studies have confirmed that modelling predictions match experimental data for the initial bubble oscillations very well (Müller et al. 2009; Kröninger 2010; Obreschkow et al. 2013). Therefore, the bubble dynamics can be characterized by numerical simulations if the above mentioned benchmark data on plasma size and bubble oscillation times are available. The plasma size can be determined by time-integrated photography of the plasma luminescence or by time-resolved photography of the optical breakdown region, if the luminescence is too weak (Vogel, Busch and Parlitz 1996; Venugopalan et al. 2002; Schaffer et al. 2002). Bubble oscillation times can be determined with high temporal resolution through single-shot measurements detecting the forward scattering signal from a continuous wave (cw) probe laser beam (Vogel et al. 2008).

Probe beam detection schemes with different beam diameters at the bubble position and different angles of scattered light collection have been employed in the past (Barber et al. 1997; Matula 1999; Gompf and Pecha 2000; Weninger, Evans and Putterman 2000; Schaffer et al. 2002). In the present paper, we use the confocal scheme introduced by Vogel et al. (2008) in which the probe beam is collinear with the pump beam producing the bubble and the focus locations of both beams coincide. A fast photodetector provides high temporal resolution combined with a very good sensitivity that enables the detection of bubble oscillations within a large range of amplitudes from millimetres down to the nanometre range. When scattered light is detected in oblique forward direction as often done in studies of single-bubble sonoluminescence (SBSL), radius-time curves can be derived from the temporal evolution of the Mie-scattering signal. However, since the signal intensity depends on the cross section of the bubble and the scattering signal in oblique direction is several orders of magnitude weaker than in forward direction (Barber et al. 1997), the signal-to noise ratio is relatively small for micrometre-sized bubbles. By contrast, with confocal adjustment and collinear detection of the probe beam, the signal intensity is related to the bubble's cross section only when the bubble size is smaller than the probe beam waist, which for tight focusing is < 1 µm. For larger bubbles, most of the probe laser light is transmitted through the bubble but changes of Mie forward scattering during bubble oscillations will cause an intensity modulation of the light arriving at the detector. When the DC background is removed by AC coupling of the photodetector, very small oscillations in the late phase of the bubble lifetime can be detected. This way, we were able to trace the transition from the nonlinear large-amplitude oscillations immediately after laser-induced bubble generation to the linear small-amplitude oscillations of the long-lived residual gas bubble during more than 100 oscillation cycles.

The experimental generation of such highly spherical bubbles requires tight focusing of the pump laser pulses in order to guarantee the formation of compact plasma driving the bubble expansion (Venugopalan et al. 2002; Vogel et al. 2008; Obreschkow et al. 2013). Additionally, buoyancy effects that could lead to a movement out of the probe beam focus and to jet formation upon collapse (Benjamin and Ellis 1966, Zhang et al. 2015) must be avoided. Previous high-speed photographic studies on millimetre-sized bubbles eliminated buoyancy by investigating the bubble dynamics in a falling apparatus (Benjamin and Ellis 1966; Blake and Gibson 1987) or under zero gravity conditions during parabola flights (Obreschkow et al. 2011, 2013). We limit buoyancy effects by restricting investigations to small bubbles with maximum radius below 100 µm that are produced by focusing the pump laser pulse through a long-distance water-immersion microscope objective with large numerical aperture (NA). Tight focusing ensures the formation of spherical bubbles by the pump laser beam and provides a high sensitivity of the interferometric detection of bubble oscillations down to small bubble sizes. Time-resolved photography is employed to check the validity of the initial conditions used for the simulations of bubble dynamics.

Since the focus of the present paper is on spherical bubble dynamics, it is sufficient to use a spherical bubble model considering liquid compressibility. The model by Keller and Miksis (1980) does this assuming a constant sound velocity in the liquid, whereas the Gilmore model uses the local pressure conditions at the bubble surface. As a consequence, the Keller-Miksis equation suffers from the unphysical behaviour that acceleration and pressure difference have opposite sign when the bubble wall velocity exceeds the sound velocity in the liquid (Prosperetti and Hao 1999). Gilmore's approach avoids this and enables to follow shock wave propagation into the surrounding liquid even under extreme conditions of optical breakdown, where bubble wall velocities well above 1500 m/s were observed (Vogel, Busch and Parlitz 1996).

Advanced volume of fluid (VOF) models based on the solutions of the Navier-Stokes equations can be used both for aspherical and spherical bubbles (Mueller, Helluy and Ballmann 2010; Han et al. 2015; Koch et al. 2016; Lechner et al. 2017, Tian et al. 2020). Such models do not rely on the simplifications of the Kirkwood-Bethe hypothesis used in the Gilmore model for considering liquid compressibility but are computationally much more expensive. Koch et al. (2016) showed that this yields little gain of accuracy because the deviations between the results of both approaches for spherical bubble collapse are relatively small even in the most critical late collapse and early rebound phase. Moreover, present implementations of the VOF method are not suited to simulate the dynamics of very small bubbles that are of major interest for laser surgery. For bubble radii smaller than 27.8 µm, the pressure at the bubble wall becomes negative during the bubble oscillation owing to surface tension, and this cannot be handled by present VOF codes (Koch et al. 2016).

Models of spherical bubble dynamics including heat and mass transfer at the bubble wall (Fujikawa and Akamatsu 1980; Akhatov et al. 2001; Lauer et al. 2012, Zein, Hantke and Warnecke 2013, Peng et al. 2020, Zhong et al. 2020) are more complex and computationally expensive than the Gilmore model, which through its relative simplicity enables to effectively study parameter dependencies of bubble dynamics. Moreover, modelling of vapour condensation is hampered by the fact that the sticking coefficient for vapour molecules at the bubble wall depends on pressure and temperature, and experimental and theoretical data exhibit large variations (Marek and Straub 2001), making predictions uncertain. Akhatov et al. (2001) escaped the dilemma by using the sticking coefficient as fit parameter for achieving a good match between the predicted rebound behaviour and experimental observations. We follow a similar strategy by using the equilibrium bubble radii after optical breakdown, $R_{nbd}$, and during collapse, $R_{nc}$, as fit parameters, which are chosen such that model predictions match the observed oscillation times for the first and second oscillations, $T_{osc1}$ and $T_{osc2}$, respectively. This yields information on the evolution of the gas content during the bubble oscillations. The equilibrium radius during bubble expansion approximately represents the total amount of gas and vapour produced by water dissociation during breakdown and the phase change in the plasma volume, respectively. The vapour contained in the expanded bubble is obtained by assuming equilibrium vapour pressure at room temperature, and the amount of vapour plus non-condensable gas in the collapsed bubble is obtained by fitting $R_{nc}$ to $T_{osc2}$. The amount of non-condensable gas in the residual bubble is derived from the frequency of the late linear oscillations. Comparison of the latter two quantities then enables to distinguish the vapour and non-condensable gas fractions of the collapsed bubble.

Most previous models of bubble dynamics and shock wave emission started with the expanded bubble and focused on its collapse and rebound, whereas the present model includes also the initial expansion phase. It covers the continuous increase of the driving force for bubble expansion during the laser pulse (following Vogel et al. 1996) and the contribution of the particle velocity behind the detaching shock front to the bubble wall velocity. Gilmore (1952) already presented a first order approximation for the jump-start of bubble wall velocity but here we present a second order approximation that is shown to agree better with experimental results.

Previous simulations of acoustic transient emission after bubble collapse were often restricted to cases, where no shock front evolved (Hickling and Plesset 1964; Fujikawa and Akamatsu 1980; Akhatov et al. 2001; Koch et al. 2016), and in those simulations, where the pressure was high enough to result in shock formation, the position of the shock front and the peak pressure were determined manually (Akulichev et al. 1968; Ebeling 1978; Vogel, Busch and Parlitz 1996, Lai et al. 2021). Here we employ an automatized procedure that facilitates tracking of the shock front propagation and pressure decay during bubble expansion and rebound.

The potential of our hybrid approach based on the use of experimental data on $R_0$ and $T_{\text{osci}}$ as input for simulations with the extended Gilmore model is demonstrated through a detailed analysis of the scattering signal from a microbubble covering more than 100 bubble oscillation cycles. Tracking the partitioning of the laser energy absorbed via plasma formation revealed new insights about laser-induced bubble dynamics. During breakdown, all energy is stored in the laser plasma, and a large fraction can be converted into bubble energy, whereas during bubble collapse most of the energy of the expanded bubble goes into compression of the liquid surrounding the collapsed bubble and is emitted as shock wave. The energy of breakdown and rebound shock waves is rapidly dissipated as heat behind the shock front. This "convective" heat transport involves a larger energy fraction than conductive transport by heat diffusion from the plasma or the collapsed bubble. Heating of a liquid shell around the bubble induces a local reduction of surface tension and viscosity that explains the large number of oscillations observed for highly spherical bubbles.

## 2. Experimental methods

In this section, we first discuss the prerequisites that must be fulfilled for the investigation of spherical bubble dynamics. For this purpose, we explore the conditions under which the influence of buoyancy and adjacent boundaries is sufficiently small to avoid jetting or bubble disintegration during first collapse. Afterwards, we present the experimental setups used for laser-induced bubble generation, determination of plasma size, time resolved photography of the initial bubble and shock wave expansion, and introduce a sensitive single-shot technique enabling to detect bubble oscillation times within a very large range of oscillation amplitudes.

### 2.1. *Conditions for spherical bubble dynamics*

Factors that may cause deviations from a spherical bubble shape are asymmetric initial conditions (e.g. an elongated laser plasma) and asymmetric boundary conditions. During collapse, shape irregularities are amplified and the bubble gets deformed by a Rayleigh-Taylor instability of the bubble wall and may even disintegrate into fragments which often coalesce again during the rebound phase (Strube 1971, Prosperetti and Hao 1999; Yuan et al. 2001). The "stability crisis" occurring upon collapse prevents the bubble from reaching the extreme states of matter that can be achieved in a perfectly spherical oscillation.

The presence of free or solid boundaries at a distance $d$ from the bubble will cause deviations from spherical dynamics, if the dimensionless distance $\gamma = d/R_{max}$ becomes too small. Asymmetric boundary conditions will then induce dynamic pressure gradients resulting in Bjerkness forces that cause jet formation upon bubble collapse (Blake and Gibson 1987; Vogel, Lauterborn and Timm 1989; Lauterborn and Kurz 2010). In laser-induced cavitation, bubble dynamics may be affected by nearby focusing optics, the walls of the cell in which the bubbles are created, and the surface of the liquid in the cell. Obreschkow et al. photographically observed spherical collapse and rebound of millimetre-sized bubbles under micro-gravity conditions already for a relatively small standoff distance of $\gamma = 13.3$ (Obreschwkow et al. 2011, 2013). However, deviations from spherical shape are strongest at the collapse stage, which is hard to capture photographically. By contrast, observation of a large number of afterbounces is a simple practical criterion indicating that the spherical bubble shape has survived the stability crisis during the first collapse.

The bubble dynamics is also influenced by buoyancy constituting a static pressure gradient. Benjamin and Ellis observed pronounced jetting during the oscillation of a bubble with $R_{max} \approx 12$ mm although the bubble stayed far away from any boundary (Benjamin and Ellis 1966). The hydrostatic pressure difference between lower and upper bubble wall and the oscillation time were large enough to induce a significant upward bubble motion during one oscillation cycle. Upon collapse, the movement was then accelerated because of the conservation of Kelvin impulse. This induces a fast liquid jet that penetrates the bubble and becomes visible when it rebounds after the collapse. The influence of buoyancy can be assessed by the parameter

$$\delta = \sqrt{\frac{\rho_0 \, g \, R_{max}}{p_\infty - p_v}}, \qquad (2.1)$$

which expresses the ratio of bubble collapse time to the time it takes an inviscid bubble of radius $R_{max}$ to rise one radius from rest driven by buoyancy forces (Blake, Taib and Doherty 1986; Best and Kucera 1992). Here $p_\infty$ denotes the ambient pressure, $p_v$ the vapour pressure at ambient conditions, $\rho_0$ is the mass density of the liquid, and $g$ the gravitational constant. Calculations balancing the Bjerkness force exerted by a solid boundary below the bubble and the buoyancy force showed that, at $\gamma \geq 2$, both are equally strong for $\delta \cdot \gamma = 0.392$ (Brujan et al. 2005). However, the net balancing of both forces does not lead to spherical collapse but goes along with the formation of jets in opposite directions that maintains zero Kelvin impulse (Brujan et al. 2005; Zhang et al. 2015). Outward jet propagation with bubble splitting is observed for $\delta > 0.22$, while for $\delta < 0.22$, the jets propagate towards each other, and the bubble maintains an approximately (but not perfectly) spherical shape. Perfectly spherical collapse requires $\gamma \rightarrow \infty$, and $\delta \rightarrow 0$, which can experimentally only be approximated.

To assess, under which experimental conditions the bubble may survive the "stability crisis" during the first collapse, we use Benjamin and Ellis' experiment as a starting point, where for $R_{max} \approx 12$ mm and $\delta = 0.034$ pronounced jetting was observed (Benjamin and Ellis 1966). The buoyancy parameter decreases with decreasing bubble size but indications for deviations of the rebounding bubble from spherical shape have still been found in high-speed photographic series of bubbles with 0.75 mm maximum radius (Kroninger et al. 2010; Lauterborn and Vogel 2013). Also at $R_{max} = 0.65$ mm we still observed slight bubble movement and jetting during collapse and rebound (unpublished results). For $R_{max} = 0.65$ mm, $\delta = 0.008$ and, according to Brujan et al. (2005), the strength of the buoyancy effect resembles the influence of a solid boundary at dimensionless distance $\gamma = 55$. Thus, the bubble's dimensionless standoff distance from solid or free boundaries must be larger than this value to provide approximately spherical dynamics throughout collapse and rebound.

In sonoluminescence, the bubble shape must remain approximately spherical upon collapse to maintain the bubble's integrity during many oscillations. Disintegration of SBSL bubbles arises from surface instabilities that slowly grow in size during a sequence of many oscillations (Lauterborn and Kurz 2010; Holzfuss 2008; Brenner, Hilgenfeldt and Lohse 2002). With laser-induced bubbles undergoing just one major oscillation before the stability crisis at first bubble collapse, the upper $R_{max}$ limit for spherical bubble collapse could be somewhat higher. Stable sonoluminescence has been observed for bubbles up to $\approx 90$ µm maximum radius (Gompf and Pecha 2000, Weninger, Evans and Putterman 2000). We assume that laser-induced bubbles remain spherical up to $R_{max} = 100$ µm. Here, the buoyancy parameter has a value of $\delta = 0.003$, and we take this as upper limit for spherical bubble dynamics in water under the influence of gravity.

The elliptical or conical shape of the laser plasma initiating the bubble expansion is another factor impeding spherical bubble formation. Close to the breakdown threshold, the plasma is elliptical, corresponding to the irradiance distribution in the laser focus. Well above threshold, it acquires a more conical shape because the breakdown front moves upstream during the pulse into the cone angle of the incoming laser beam (Docchio et al. 1988, Vogel et al. 1996). An elongated shape of the plasma results in different and asymmetric $R(t)$ curves in axial and radial direction (Tinne et al. 2013), and affects both shape and vigour of the bubble collapse. Even if the expanded bubble looks fairly spherical, the shape distortions become relevant again in the collapse phase (Lauterborn and Kurz 2010; Sinibaldi et al. 2019), dampen the vigour of the collapse and may lead to bubble breakup. If the plasma elongation is too strong, the bubble acquires a dumbbell shape upon collapse and breaks up in two or more parts (Lindau Dissertation 2001). The ratio between length and diameter of the laser focus is inversely proportional to the numerical aperture (*NA*) of the optical system used for bubble generation (Vogel et al 2005; Vogel et al 2008). Therefore, the *NA* must be as large as possible to achieve a spherical bubble shape.

For focusing at large NAs, it is challenging to guarantee a sufficiently large standoff distance $\gamma$ and to avoid spherical aberrations that produce an enlarged focus with several hot spots (Vogel et al. 1999a). Obreschkow et al. (2011, 2013) and Sinibaldi et al. (2019) used a parabolic mirror with $NA = 0.6$ to achieve this task for millimetre-sized bubbles, and Vogel et al. (2008) employed long-distance water-immersion microscope objective for aberration-free micro-bubble generation at $NA = 0.8$. In the present paper, we use water-immersion objectives with *NA*s up to 0.9.

Reproducible bubble generation over a large energy range requires the creation of homogeneous plasmas with little pulse-to pulse variations. This requires the use of lasers with good beam quality. Reproducible plasma formation is easier to achieve with ultrashort laser pulse durations, where seed electrons are abundant while with ns pulses, especially at IR wavelengths, the generation of seed electrons is the critical hurdle for plasma formation (Vogel et al. 1996; Noack and Vogel 1999; Linz et al. 2015; Linz et al. 2016). As a consequence, multiple plasmas are frequently formed, especially at moderate or small NA (Vogel et al. 1996; Tagawa et al. 2016). Spherical bubble generation with ns pulses is possible at large NA under two conditions: close to the breakdown threshold, where multiple plasma cannot yet form, or at large energies well above threshold, where they merge (Sinibaldi et al. 2019).

2.2. *Setup for the generation of highly spherical bubbles, plasma photography and oscillation tracking*

The experimental arrangements for the investigation of the behaviour of spherical laser-induced cavitation bubbles in water are depicted in figure 1. We used different laser systems and detection schemes, depending on the investigation task. Small, highly spherical bubbles were generated using the setup of figure 1 (a). The laser source was a Ti:sapphire femtosecond laser (Spectra Physics Spitfire) pumping a traveling-wave optical parametric amplifier of superfluorescence (TOPAS; Light Conversion, TOPAS 4/800) as described by Linz et al. (2016). At a wavelength of $\lambda = 775$ nm and at 1 kHz repetition rate, this laser system delivers pulses of 265 fs duration and up to 20 µJ pulse energy. Single laser pulses are selected from the pulse train using a mechanical shutter. Pulse energy is adjusted by a combination of rotatable Fresnel rhomb retarder and thin film polarizers and measured by diverting a part of the pump beam onto an energy meter (Ophir PD10-pJ).

The core of the setup for plasma photography and probe beam measurements of the subsequent bubble oscillations is a water-filled cuvette with three confocally adjusted water immersion microscope objectives (Leica HCX APO L U-V-I) built into the cuvette wall. The pump laser beam is first expanded by a telescope consisting of a biconcave lens (f = – 40 mm) and a laser achromat (f = 200 mm), and then focused into deionized and filtered (0.2 µm) water. For this purpose, we used either a 40x, *NA* = 0.8 objective with 3.3 mm working distance, or a 63x, *NA* = 0.9 objective with a working distance of 2.2 mm. The rear entrance pupil of the objective was overfilled to create a uniform irradiance distribution corresponding to an Airy pattern in the focal plane. The cw probe laser beam (CrystaLaser, 658 nm, 40 mW) is aligned collinear and confocal with the fs-pump beam. The transmitted probe laser light is collected by a 10x, *NA* = 0.3 objective built into the opposite cell wall and imaged onto a photoreceiver (Femto HCA-S-200M-SI) connected to a digital oscilloscope (Tektronix DPO 70604). The photoreceiver is protected from the fs laser irradiation by a blocking filter.

Plasma luminescence was photographed through a third microscope objective (20x, NA = 0.5, 3.5 mm working distance) that was oriented perpendicular to the optical axis of the pump and probe beams, and recorded by a digital SLR camera (Canon EOS 5D). To be able to image very small features near the optical resolution limit, the image formed by the 20x objective and tube lens was further magnified using a Nikkor objective (63mm/1:2,8), which is corrected for 8x magnification. This way, we achieved a total magnification factor of 162 and a diffraction-limited spatial resolution of 0.5 µm. A confocal arrangement of all three water immersion objectives could only be achieved, when the 40x objective was used to focus the fs-pulses. For tighter focusing with the 63x objective, the 20× imaging objective had to be removed and we could only perform probe beam scattering measurements.

The absorbed laser energy was obtained from measurements of the plasma transmittance $T_{tra}$ using the relation $E_{abs} = E_L (1-T_{tra})$. For transmission measurements, the photoreceiver was exchanged by a calibrated energy meter, and the 10x objective was replaced by a 63x water immersion objective ($NA = 0.9$) that collected all transmitted light. Calibration accounted for light losses by reflections at optical surfaces and by absorption in the microscope objective and in water.

### 2.3. *Single-shot recording of bubble oscillations via probe beam scattering*

In the present study, we are not interested in determining the bubble's radius time curve from the intensity of the scattering signal but rather want to detect bubble oscillations with maximum sensitivity. Figure 1(b) depicts the path of the probe laser beam through a bubble produced by a confocal pump laser pulse. The confocal adjustment guarantees that even very tiny bubbles at the optical breakdown threshold can be detected with high sensitivity (Vogel et al. 2008; Linz et al. 2016). For larger spherical bubbles, the probe beam passes perpendicularly through the bubble wall, and most of the probe laser light is transmitted through the focal region. Only a small portion is scattered, or reflected at the bubble walls and interferes with the directly transmitted beam. For bubbles larger than the Rayleigh range, up to 96% of the incident light is transmitted and 0.04% interferes with the transmitted beam after being reflected at the rear and then the anterior bubble wall. The bias is removed by AC coupling of the photoreceiver but the coherent mixing of scattered and reflected light with the transmitted beam creates a homodyne gain, which largely improves the signal-to-noise ratio of the interference modulation (Paschotta 2008). During bubble oscillations, this causes a small interference modulation of the probe beam signal at the detector. For bubbles larger than the beam waist diameter but smaller than the Rayleigh range, changes in the angular distribution of Mie forward scattering lead to pronounced fluctuations of the light amount transmitted through the collection aperture. During the oscillation of large bubbles, this leads to transient strong signal modulations after the start and towards the end of each oscillation that enable to determine the oscillation period, $T_{osc}$ (Vogel et al. 2008). The collection $NA$ should be chosen such that these modulations are maximized while the direct light transmission is attenuated as much as possible. In our experiments, a value of $NA \approx 0.1$ provided the best results. It was realised by placing an aperture with 4 mm diameter behind the 10x objective. The amplified AC-coupled photoreceiver had a signal bandwidth reaching from 25 kHz to 200 MHz. An additional monitor output of the DC signal was used for the basic alignment of the entire setup.

Due to buoyancy, small bubbles move a little upward during their oscillations although they retain a spherical shape. Nevertheless, late bubble oscillations can still be detected if the buoyancy is small enough such that the bubble stays within the probe beam focus. According to Eq. (2.1), a bubble with $R_{max} = 100$ µm moves about 0.63 µm during the first oscillation but much less during later oscillations, when the radius is significantly smaller. The diffraction-limited focus diameter $d = \lambda/NA$ in our setup is 0.73 µm. Thus, with horizontal probe beam orientation, late oscillations can be investigated for $R_{max} < 100$ µm. With a vertical probe beam, the buoyant bubble stays longer within the focal region, and the measurement range is extended to larger bubble sizes. In our experiments, we used a horizontally oriented beam. Here, the light passage through the bubble becomes slightly asymmetric after the large initial oscillation during which buoyancy is strongest, and the forward Mie scattering lobe will then be obliquely oriented. Therefore, both width and orientation of the central lobe change during bubble oscillations, what enhances the intensity fluctuations of the light transmitted through the collecting aperture of $NA \approx 0.1$. For 30 µm < $R_{max}$ < 50 µm, more than 100 oscillations could be traced in "lucky shots", and radius variations well below 1 nm during late oscillations were detected.

### 2.4. *Time-resolved photography of shock wave emission and bubble dynamics*

The influence of plasma shape and boundary conditions on the geometrical shape of the bubble and shock wave was investigated by time-resolved shadowgraph photography, as shown in figure 1(c). A collimated and optically delayed frequency-doubled portion of the pump laser beam was used for illumination at short delay times up to 120 ns. For longer delay times, we employed a plasma discharge flash lamp (High-Speed Photosystems, Nanolite KL-L) with 18 ns pulse duration. For each shot, the bubble oscillation time was monitored by recording hydrophone signals of breakdown and collapse shock waves using a PVDF hydrophone (Ceram) with 12 ns rise time.

With the collimated illumination beam, we could not use the 20x microscope objective for imaging as done in Fig, 1(a) because its back focal plane lies inside the objective and it can easily be damaged, when the collimated laser beam is focused on an interior lens. Instead, we used an external 7x macro objective (Leitz Photar) for photography that provided a spatial resolution of 2 µm. Use of an external imaging lens made it possible to utilize the 63x, $NA = 0.9$ microscope objective for plasma generation, which enabled to create highly spherical bubbles. Elongated plasmas were produced by focusing at smaller numerical apertures between $NA = 0.16$ and $NA = 0.25$. In this experimental series, bubbles were generated by a Nd:YAG laser (Continuum YG 671-10), which delivers pulses of 1064 nm wavelength with either 6 ns or 30 ps duration at pulse energies of up to 250 mJ and 10 mJ, respectively.

Pulse energies of up to 250 µJ were employed to investigate the influence of plasma shape and boundary conditions on the geometrical shape of shock wave and bubble during initial expansion and after rebound. More energetic 10-mJ and 20-mJ nanosecond laser pulses focused at $NA = 0.25$ were used to create large high-density plasmas that enable to visualize the bubble wall formation in the early phase of plasma expansion as well as shock wave-induced phase transitions, which may enlarge the vaporized liquid volume and shift the bubble wall location. The energetic laser pulses were focused through a combination of a laser achromat and an ophthalmic contact lens built into the cuvette wall, as described by Vogel & Busch & Parlitz (1996). The pulse energy was measured using a pyroelectric energy meter (Laser Precision Rj 7100).

## 3. Theoretical analysis and numerical methods

After introducing the Gilmore model of cavitation bubble dynamics, we present an extension of the model considering the rapid increase of bubble wall velocity during laser-induced energy deposition in a compressible liquid. Acoustic and shock wave emission are then described based on the extended equation of motion. Water vapour generation by vaporization of the liquid in the plasma and its progressive condensation during the bubble oscillations is considered by fitting the equilibrium bubble radii during expansion and collapse/rebounds such that the model predictions agree with measured values of the first and later oscillation times. Finally, we present a theoretical framework for tracking the partitioning of deposited laser energy into vaporization energy and various components of mechanical energy as well as their dissipation pathways.

### 3.1. *Equations governing the cavitation bubble dynamics*

We used the Gilmore model of cavitation bubble dynamics (Gilmore 1952; Lauterborn and Kurz 2010) to calculate the temporal development of the bubble radius and the pressure inside the bubble, as well as the pressure distribution in the surrounding liquid. The model considers the compressibility of the liquid surrounding the bubble, viscosity and surface tension. Sound radiation into the liquid from the oscillating bubble is incorporated based on the Kirkwood-Bethe hypothesis (Cole 1948). The Gilmore model assumes a constant gas content of the bubble, neglecting evaporation, condensation, gas diffusion through the bubble wall, and heat conduction. Heat and mass transfer strongly influence the pressure reached at maximum bubble expansion and during bubble collapse (Fujikawa and Akamatsu 1985) but are of little importance for the dynamic behaviour during the initial stages of bubble expansion. For strong oscillations, i.e. strong compression of the contents inside the bubble, the model is augmented by a van der Waals hard core law to account for a non-compressible volume of the inert gas inside the collapsing bubble (Löfstedt, Barber and Putterman 1993; Lauterborn and Kurz 2010).

The bubble dynamics is described by the equation

$$\left(1-\frac{U}{C}\right)R\dot{U} + \frac{3}{2}\left(1-\frac{U}{3C}\right)U^2 = \left(1+\frac{U}{C}\right)H + \frac{U}{C}\left(1-\frac{U}{C}\right)R\frac{dH}{dR}. \quad (3.1)$$

Here, $R$ is the bubble radius, $U = dR/dt$ is the bubble wall velocity, an overdot means differentiation with respect to time, $C$ is the speed of sound in the liquid at the bubble wall, and $H$ is the enthalpy difference between the liquid at pressure $p(R)$ at the bubble wall and at hydrostatic pressure

$$H = \int_{p|_{r\to\infty}}^{p|_{r=R}} \frac{dp(\rho)}{\rho}, \quad (3.2)$$

whereby $\rho$ and $p$ are the density and pressure within the liquid, and $r$ is the distance from the bubble centre. The driving force for the bubble motion is expressed through the difference between the pressure within the liquid at the bubble wall and at a large distance from the wall (static pressure). Assuming an ideal gas inside the bubble, the pressure $P$ at the bubble wall is given by

$$P = p\big|_{r=R} = \left(p_\infty + \frac{2\sigma}{R_n}\right)\left(\frac{R_n^3 - R_{vdW}^3}{R^3 - R_{vdW}^3}\right)^\kappa - \frac{2\sigma}{R} - \frac{4\mu}{R}U, \quad (3.3)$$

where $\sigma$ denotes the surface tension, $\mu$ the dynamic shear viscosity, and $\kappa$ the ratio of the specific heat at constant pressure and volume. The term $R_{vdW}^3 = (b\,R_n)^3$ describes the size of the van der Waals hard core, with van der Waals coefficient $b$. The pressure is assumed to be uniform throughout the volume of the bubble. The pressure far away from the bubble is $p|_{r\to\infty} = p_\infty$. The symbol $R_n$ denotes the equilibrium radius of the bubble at which the bubble pressure balances the hydrostatic pressure. The equation of state (EOS) of water is approximated by the Tait equation, with $B = 314$ MPa, and $n = 7$ (Ridah 1988)

$$\frac{p+B}{p_\infty+B} = \left(\frac{\rho}{\rho_\infty}\right)^n, \tag{3.4}$$

which leads to the following relationships for the sound velocity $C$ and enthalpy $H$ at the bubble wall

$$C = \sqrt{c_\infty^2 + (n-1)H}, \tag{3.5}$$

$$H = \frac{n(p_\infty+B)}{(n-1)\rho_\infty}\left[\left(\frac{P+B}{p_\infty+B}\right)^{(n-1)/n} - 1\right], \tag{3.6}$$

with $c_\infty$ and $\rho_\infty$ denoting the sound velocity and mass density in the liquid at normal conditions. The term $dH/dR$ in Eq. (3.1) can be derived from Eqs. (3.3) and (3.6) by calculating $(dH/dP) \times (dP/dR)$. It reads

$$\frac{dH}{dR} = \frac{1}{\rho_0}\left(\frac{p_\infty+B}{P+B}\right)^{\frac{1}{n}} \times \left(-3\kappa R^2 \left(p_\infty + \frac{2\sigma}{R_n}\right)\frac{\left(R_n^3 - R_{vdW}^3\right)^\kappa}{\left(R^3 - R_{vdW}^3\right)^{\kappa+1}} + \frac{2\sigma}{R^2} + \frac{4\mu U}{R^2}\right). \tag{3.7}$$

### 3.2. *Description of laser-induced bubble initiation*

Direct modelling of the temporal evolution and spatial distribution of energy deposition during bubble generation via optical breakdown is complicated, and the details depend strongly on the laser pulse duration. Following Vogel, Busch and Parlitz (1996), we neglect the details of the breakdown process and refer only to the plasma size at the end of the laser pulse, and to the maximum radius reached by the cavitation bubble as a consequence of plasma expansion. The extent of the plasma marks the volume into which laser energy is deposited, and the size of the expanded cavitation bubble is an indicator for the conversion efficiency of light energy into mechanical energy. Calculations start with a (virtual) bubble nucleus at equilibrium with radius $R_0$, whereby the volume of this nucleus is identified with the photographically determined plasma size in the liquid. The energy input during the laser pulse is simulated by raising the value of the equilibrium radius $R_n$ from its small initial value $R_0$ at the beginning of the pulse to a much larger final value $R_{nbd}$. The underlying assumption is that the absorbed laser energy is proportional to the amount of liquid vaporized by the laser pulse, which in turn is proportional to the equilibrium volume of the laser-induced bubble given by $4/3\,\pi\,R_{nbd}^3$. This assumption holds when the energy deposited into the plasma is significantly larger than the sum of the energy needed for heating the vaporized liquid volume to the boiling temperature plus the latent heat of vaporization.

Since $R_n$ is a measure of the gas content of the cavitation bubble, an increase of its value beyond $R_0$ implies that the pressure inside the bubble rises and that the bubble starts to expand. Iteratively, we determine the $R_{nbd}$ value for which the calculation yields the same oscillation time $T_{osc1}$ or maximum radius $R_{max1}$ as determined experimentally. In practice, we match model predictions to $T_{osc}$ rather than to $R_{max}$ because this is the primary information provided by the light scattering signal.

While energy deposition by ultrashort laser pulses can be regarded as quasi-instantaneous, the finite duration of the laser pulse must be considered for nanosecond breakdown. The temporal evolution of the laser power $P_L$ during the pulse is modelled by a $\sin^2$ function with duration $\tau_L$ (full-width at half-maximum) and total duration $2\tau_L$

$$P_L = P_{L0} \sin^2\left(\frac{\pi}{2\tau_L}t\right), \qquad 0 \leq t \leq 2\tau_L. \tag{3.8}$$

Assuming that the cumulated volume increase of the equilibrium bubble at each time $t$ during the laser pulse is proportional to the laser pulse energy $E_L$ absorbed up to this time, Vogel & Busch & Parlitz (1996) derived an equation for the temporal development of the equilibrium radius $R_n$ during the laser pulse

$$R_n(t) = \left\{R_0^3 + \frac{R_{nbd}^3 - R_0^3}{2\tau_L}\left[t - \frac{\tau_L}{\pi}\sin\left(\frac{\pi}{\tau_L}t\right)\right]\right\}^{1/3}. \tag{3.9}$$

For laser-induced bubble formation in an incompressible liquid, the pressure discontinuity at the plasma border would be felt throughout the entire volume of the liquid once it is allowed to take effect in the simulation. The cavity wall then starts to be accelerated outward from rest, i.e. $U = 0$ at $t = 0$. By contrast, for a compressible liquid, the shock front represents an "event horizon" up to which the breakdown effects are "felt" by the liquid. As the plasma border is sharp, a shock front will form immediately, and the initial bubble wall velocity $U_0$ equals the initial particle velocity $u_p$ behind the shock front, whereby the shock pressure is identical with the initial plasma and bubble pressure, $p_s = P$.

Gilmore considered the case, where the internal bubble pressure, $P_i$, is suddenly changed to a new constant value, which produces a finite velocity jump in an infinitesimal time. Considering only large terms in the equation of motion for the bubble wall and using the Tait equation that links $H$ and $C$ to $P$, he derived the first order approximation

$$U_0 = \int_0^H \frac{dh}{c} \approx \frac{H}{C} \approx \frac{P - p_\infty}{\rho_\infty c_\infty}. \tag{3.10}$$

The approximate expression is accurate when $|H| \ll C^2$, which for water corresponds to $|P_i - p_\infty| \ll 2000$ MPa (Gilmore 1952). This is not sufficient for modelling laser-induced breakdown, where much larger plasma pressures may be involved. Therefore, we will present a derivation of $U_0$ based on the Hugoniot curve data from Rice and Walsh (1957). It yields Eq. (3.10) as first order approximation and enables to formulate a second order approximation, which is accurate up to much higher pressure values.

Rice and Walsh fitted their Hugoniot curve data by the analytical expression

$$u_p = c_1\left(10^{(u_s - c_\infty)/c_2} - 1\right), \tag{3.11}$$

where $u_s$ is the shock wave velocity and the constants are $c_1 = 5190$ m/s, $c_2 = 25306$ m/s and $c_0$ is the sound velocity, $c_\infty = 1483$ m/s. By rearranging Eq. (3.11), $u_s$ can be expressed as a function of $u_p$

$$u_s = c_\infty + c_2 \log_{10}\left(u_p/c_1 + 1\right). \tag{3.12}$$

Using Eq. (3.11) and the conservation of momentum at a shock front, $p_s - p_\infty = u_s u_p \rho_\infty$ (Duvall and Fowles 1963), one can link $p_s$ to $u_s$

$$p_s = c_1 \rho_\infty u_s \left(10^{(u_s - c_\infty)/c_2} - 1\right) + p_\infty, \tag{3.13}$$

where $\rho_\infty = 998$ kg/m$^3$ is the mass density of water and $p_\infty = 10^5$ Pa is the hydrostatic pressure. Inserting Eq. (3.12) into Eq. (3.13), one finally obtains a relation between $p_s$ and $u_p$

$$p_s = \rho_\infty u_p \left[c_\infty + c_2 \log_{10}\left(u_p/c_1 + 1\right)\right] + p_\infty. \tag{3.14}$$

If $u_p \ll c_1$, the second term in the bracket can be dropped, which leads to

$$p_s = \rho_\infty u_p c_\infty + p_\infty. \tag{3.15}$$

Immediately after breakdown, $u_p = U_0$, and $p_s = P$. After resolving Eq. (3.15) for $U_0$, we get

$$u_p = U_0 = (P - p_\infty)/\rho_\infty c_\infty, \tag{3.16}$$

which equals Gilmore's first order approximation in Eq. (3.10).

Before deriving a higher order approximation of Eq. (3.14), let us first see how we can integrate the rapid start of the bubble wall velocity during the laser pulse into the equation of motion (3.1). For this purpose, we rewrite the equation such that it describes the evolution of $\dot{U}$ and add a term $\dot{u}_p$ that expresses the evolution of the particle velocity at the bubble wall going along with the increase of $R_n$

$$\dot{U} = -\frac{3}{2}\frac{U^2}{R}\frac{C - U/3}{C - U} + \frac{H}{R}\frac{C + U}{C - U} + \frac{U}{C}\frac{dH}{dR} + \dot{u}_p \tag{3.17}$$

The term $\dot{u}_p$ is derived from Eq. (3.16) as

$$\dot{u}_p = \begin{cases} \dfrac{\dot{P}}{\rho_\infty c_\infty} & \text{for } 0 \leq t \leq 2\tau_L, \\ 0 & \text{otherwise} \end{cases} \tag{3.18}$$

The time interval $0 \leq t \leq 2\tau_L$ corresponds to the duration of the laser pulse as defined by Eq. (3.8). We shall now look at the pressure evolution. For very short pulse durations, energy deposition is inertially confined and we can neglect the bubble wall movement during the pulse and use the approximation $R = R_0$. Since the fluid does not yet move, we can neglect also viscosity. Assuming $\kappa = 4/3$, we obtain from Eq. (3.3) for the time evolution of the bubble pressure during the laser pulse

$$P = \frac{p_\infty}{R_0^4} R_n^4(t) + \frac{2\sigma}{R_0^4} R_n^3(t), \tag{3.19}$$

with $R_n(t)$ given by Eq. (3.9). The time derivative of Eq. (3.19) reads

$$\dot{P} = \frac{4 p_\infty R_n(t) + 6\sigma}{R_0^4} R_n^2(t) \dot{R}_n(t), \tag{3.20}$$

and the time derivative of Eq. (3.9) is

$$\dot{R}_n(t) = \frac{1}{3} R_n^{-2}(t) \frac{R_{nbd}^3 - R_0^3}{2\tau_L} \left[1 - \cos\left(\frac{\pi}{\tau_L} t\right)\right]. \tag{3.21}$$

By inserting Eq. (3.21) into (3.20) one gets

$$\dot{P} = \frac{2p_\infty R_n(t) + 3\sigma}{3\tau_L R_0^4} \left(R_{nbd}^3 - R_0^3\right) \left[1 - \cos\left(\frac{\pi}{\tau_L}t\right)\right], \tag{3.22}$$

and by inserting Eq. (3.22) into Eq. (3.18) one finally obtains

$$\dot{u}_p = \frac{2p_\infty R_n(t) + 3\sigma}{3R_0^4 \tau_L \rho_\infty c_\infty} \left(R_{nbd}^3 - R_0^3\right) \left[1 - \cos\left(\frac{\pi}{\tau_L}t\right)\right], \tag{3.23}$$

which enables to numerically integrate Eq. (3.17).

A simulation of the bubble wall movement based on the first order approximation is shown as dash-dotted curve in figure 2, together with simulation results not considering the "jump start" of the bubble wall and the results of the second order approximation that will be presented below. The first order approximation yields a start velocity $U_0 = 886$ m/s, which is much higher than the particle velocity behind a shock front having a pressure equal to the bubble pressure at the end of the laser pulse. For the starting conditions of figure 2, this pressure is $P_{max} = 1.31$ GPa, and the corresponding particle velocity is about 500 m/s (Rice and Walsh 1957).

Besides overestimating $U_0$, the first order approximation predicts a continuous drop of the bubble wall velocity after the jump-start, which contradicts the physical picture of the sequence of events. After optical breakdown, the shock front immediately detaches from the plasma because its velocity is much larger than the particle velocity behind the front (Cole 1948; Rice and Walsh 1957). When the shock wave moves away from the plasma border, ever more fluid is incorporated into the flow and the pressure at the shock front as well as the particle velocity behind the shock front drop rapidly. Nevertheless, the bubble wall may, for a short while, continue to be accelerated by the internal bubble pressure, which should result in a peak of the $U(t)$ curve a short while after the jump start. With increasing bubble radius, the kinetic energy imparted to the liquid is distributed among an ever-larger liquid mass and the bubble wall velocity will then start to decrease although the bubble pressure is still higher than the hydrostatic pressure.

In order to improve the accuracy of the model predictions, we go back to the relationship between $P$ and $u_p$ in Eq. (3.14) and formulate a second order approximation considering the second term in the bracket through its Taylor expansion

$$\log_{10}\left(\frac{u_p}{c_1} + 1\right) = \frac{u_p}{\log(10)c_1} - \frac{u_p^2}{\log(100)c_1^2} + \frac{u_p^3}{\log(1000)c_1^3} - \cdots. \tag{3.24}$$

If $u_p$ is well below $c_1 = 5190$ m/s, higher order terms of the Taylor expansion can be dropped. Keeping the first term and inserting Eq. (3.24) into Eq. (3.14), we obtain

$$P = \rho_\infty c_\infty u_p + \frac{\rho_\infty c_2}{\log(10)c_1} u_p^2 + p_\infty. \tag{3.25}$$

For $P \gg p_\infty$, we can ignore $p_\infty$, which enables to formulate a quadratic equations of type ($Ax^2 + Bx - P = 0$)

$$\underbrace{\frac{\rho_\infty c_2}{\log(10)c_1}}_{A} u_p^2 + \underbrace{\rho_\infty c_\infty}_{B} u_p - P = 0. \tag{3.26}$$

For A, B, P > 0, such equations have a positive real and a negative imaginary root

$$u_p = \frac{-B \pm \sqrt{B^2 + 4AP}}{2A}. \tag{3.27}$$

Inserting A and B in Eq. (3.26) into the positive root of Eq. (3.26), we obtain

$$u_p = \frac{\sqrt{\rho_\infty^2 c_\infty^2 + 4\dfrac{\rho_\infty c_2}{\log(10)c_1}P} - \rho_\infty c_\infty}{\dfrac{2\rho_\infty c_2}{\log(10)c_1}}. \tag{3.28}$$

The time derivative of this equation is

$$\dot{u}_p = \frac{\dot{P}}{\sqrt{\rho_\infty^2 c_\infty^2 + \dfrac{4\rho_\infty c_2}{\log(10)c_1}P}}. \tag{3.29}$$

Eq. (3.29) equals the first order approximation result in Eq. (3.18) when the second term in the denominator is neglected. Inserting Eq. (3.22) into Eq. (3.29), we finally obtain

$$\dot{u}_p = \frac{1}{\sqrt{\rho_\infty^2 c_\infty^2 + \dfrac{4\rho_\infty c_2}{\log(10)c_1}P}} \frac{2p_\infty R_n(t) + 3\sigma}{3R_0^4 \tau_L}\left(R_{nbd}^3 - R_0^3\right)\left[1 - \cos\left(\frac{\pi}{\tau_L}t\right)\right], \tag{3.30}$$

with $P$ given by Eq. (3.19).

Numerical integration of Eq. (3.17) with $\dot{u}_p$ from Eq. (3.30) yields the solid curve in figure 2, with start velocity $U_0 = 513$ m/s in good agreement with Hugoniot data, and a time evolution $U(t)$ that corresponds well to the expected physical scenario described above. In the following, the second order approximation of the jump-start of bubble wall velocity will be used in all numerical simulations, if not otherwise mentioned.

### 3.3. Acoustic and shock wave emission

The solution of Eq. (3.17) with $\dot{u}_p$ from Eq. (3.30) and $R_n(t)$ from Eq. (3.9) was used to calculate the pressure distribution in the liquid surrounding the cavitation bubble (Gilmore 1952, Knapp et al. 1970). The calculation is based on the Kirkwood–Bethe hypothesis which expresses that the quantity $y = r(h + u^2/2)$ propagates outward along a "characteristic," traced by a point moving with velocity $c + u$. Here $c$ is the local velocity of sound in the liquid, $u$ is the local liquid velocity, and $h$ is the enthalpy difference between liquid at pressures $p$ and ambient pressure $p_\infty$ (Cole 1948). The Kirkwood–Bethe hypothesis leads to the differential equations

$$\dot{u} = -\frac{1}{c-u}\left[(c+u)\frac{y}{r^2} - \frac{2c^2 u}{r}\right], \qquad \dot{r} = u + c, \tag{3.31}$$

with

$$c = c_\infty\left(\frac{p+B}{p_\infty + B}\right)^{(n-1)/2n}. \tag{3.32}$$

The pressure $p$ at $r = r(t)$ is given by

$$p = (p_\infty + B)\left[\left(\frac{y}{r} - \frac{u^2}{2}\right)\cdot\frac{(n-1)\rho_\infty}{n(p_\infty + B)} + 1\right]^{n/(n-1)} - B. \tag{3.33}$$

Numerical solution of Eqs. (3.17) and (3.31) with the bubble radius $R$, the bubble wall velocity $U$, and the quantity $y = R(H + U^2/2)$ at the bubble wall as initial conditions yields the velocity and pressure distribution in the liquid along one characteristic. Solution of the equation for many initial conditions, i.e., along many characteristics, allows computation of $u$ and $p$ for a network of points $(r, t)$. To determine $u(r)$ and $p(r)$ at a certain time, one has to collect a set of points with $t = $ constant from this network.

When the bubble pressure is high, the pressure profiles in the liquid become steeper with time until a shock front is formed. Afterward, the calculations yield ambiguous pressure values, because they do not consider the energy dissipation at the shock front. The ambiguities have no physical meaning but simply indicate the presence of a discontinuity. The position of the shock front and the peak pressure at the front can be determined using the conservation laws for mass-, impulse-, and energy-flux through the discontinuity. As illustrated in figure 3, it is defined by a vertical line in the $u(r)$ plots cutting off the same area from the ambiguous part of the curve as that added below the curve (Rudenko and Soluyan 1977; Landau and Lifschitz 1987). The location of the front was determined in the $u(r)$ plots and transferred to the $p(r)$ plots. The method of determining the shock front position by subtracting/adding equal areas under the $u(r)$ curve is in accordance with conservation of momentum. The progressive reduction of peak pressure values going along with this procedure represents dissipation effects at the shock front, which are associated with an abrupt temperature rise (Brinkley and Kirkwood 1947; Cole 1948; Rice and Walsh 1957, Duvall and Fowles 1964; Müller 2007).

We employed a commercial Matlab software package for the numerical integration of Eqs. (3.17) and (3.31). The constants used for water at a temperature of 20°C are: density of water $\rho_\infty = 998$ kg/m$^3$, surface tension $\sigma = 0.073$ Nm$^{-1}$, adiabatic exponent for water vapour $\kappa = 4/3$, coefficient of the dynamic shear viscosity $\mu = 0.001$ Nsm$^{-2}$, velocity of sound $c_\infty = 1483$ m/s, static ambient pressure $p_\infty = 100$ kPa, vapour pressure $p_v = 2.33$ kPa, and van der Waals coefficient $b = 1/9$. A van der Waals hardcore is used in the calculations of bubble collapse but it is not needed for modelling the bubble expansion. Therefore, the van der Waals radius reads $R_{vdW} = 1/9\, R_{nc}$, where $R_{nc}$ is the equilibrium radius of the bubble relevant for the collapse phase. It is considerably smaller than $R_{nbd}$ immediately after optical breakdown because most of the water vapour produced during bubble generation condenses during the oscillation (Ebeling 1978). The $R_{nc}$ value is chosen such that the calculation yields the same oscillation time of the rebounding bubble, $T_{osc2}$, as determined experimentally.

### 3.4. *Choice of the adiabatic exponent*

We use the room temperature value of the adiabatic exponent for water vapour, $\kappa = 4/3$, for the modelling of laser-induced bubble oscillations although the adiabatic exponent of water drops with increasing temperature (due to the increasing influence of molecular vibrations) and the temperature during breakdown and bubble collapse reaches much higher values than room temperature. This simplification is justified by the fact that part of the water molecules will dissociate for $T > 3000$ K (Mattson and Desjarlais 2007; Mattson and Desjarlais 2006), resulting in diatomic molecules such as $H_2$ and $O_2$, which have a larger adiabatic exponent of 1.4 at room temperature (Fujikawa and Akamatsu 1980). Since both changes will, at least partly, compensate each other, the choice $\kappa = 4/3$ appears reasonable even for a large temperature range. Bubble oscillation is isothermal ($\kappa = 1$) most of the time and adiabatic only during early expansion and late collapse/early rebound (Prosperetti and Hao 1999; Brenner, Hilgenfeldt and Lohse 2002). Therefore, some researchers use a continuously changing time-dependent value of $\kappa$ (Brenner, Hilgefeldt and Lohse 2002), and others switch from the isothermal to the adiabatic value, when the bubble radius passes the equilibrium radius (Barber et al. 1997; Yuan et al. 2001). Following Lauterborn and Kurz (2010), we use a constant value because a variation of $\kappa$ will largely complicate the tracking of energy partitioning during the bubble oscillations while it has little influence on $R(t)$ and $P(t)$. The strongest influence is seen in the temperature evolution, which is of major interest for understanding the mechanism of sonoluminescene but less important for plasma-mediated laser surgery that relies mainly on the mechanical effects of the bubble oscillations.

### 3.5. *Indirect consideration of condensation in the transition from nonlinear to linear bubble oscillations*

Laser-induced cavitation bubbles are generated via vaporization and dissociation of the water inside the plasma volume. Therefore, they initially contain a large amount of vapour together with some non-condensable gas arising from the dissociation of water molecules in the hot plasma (Roberts et al. 1996; Mattsson and Desjarlais 2006; Elles et al. 2007; Müller et al. 2009; Sato et al. 2013; Barmina, Simakin and Shafeev 2016). The vapour largely condenses during the first nonlinear bubble oscillations but non-condensable gas remains, and finally the bubble exhibits small-amplitude linear oscillations around the equilibrium radius of the residual gas bubble, $R_{res}$. The use of different $R_n$ values for the calculation of laser-induced bubble expansion and for its dynamics during the first collapse and later collapse events parametrizes vapour condensation during the oscillation cycles (Ebeling 1978). The $R_{nbd}$, $R_{nc1}$ and $R_{nc2}$ values are chosen by fitting the predicted bubble dynamics to measured values of $T_{osc1}$, $T_{osc2}$, and $T_{osc3}$, respectively. After the second collapse, the $R_n$ value is kept constant because we assume that condensation is now approximately complete and that the residual bubble is mainly filled with non-condensable gas.

A reduction of $R_n$ between breakdown and collapse seems to contradict the assumption of adiabatic expansion and collapse implied in Eq. (3.3). In fact, expansion and collapse can be approximated as adiabatic processes only during the initial expansion and the final collapse phase. In the expanded stage, heat and mass transfer at the bubble wall resemble an isothermal scenario. It is usually assumed that at $R = R_{max}$ the vapor inside the bubble is at equilibrium with the liquid outside the bubble such that the bubble pressure corresponds to the equilibrium vapour pressure at room temperature (Akhatov et al. 2001; Fujikawa and Akamatsu 1980, Prosperetti and Hao 1999). However, the change between adiabatic and isothermal conditions hardly affects the bubble motion during the expanded stage because after the initial expansion phase, the ongoing bubble expansion is driven by inertia. Thus, condensation and heat exchange may take place without any major influence on $R(t)$. Vice versa, the driving force for bubble collapse is hardly influenced by phase transitions at the bubble wall as long as the internal bubble pressure is much smaller than the external pressure, which is the case for most of the collapse time. However, the amount of gas and vapour in the bubble becomes crucial in the final collapse phase, where it determines the peak pressure upon collapse and the amount of energy carried away by shock wave emission. Therefore, the use of Eq. (3.3) together with a reduction of $R_n$ at the stages of maximum bubble expansion provides a realistic description of the bubble oscillation with implicit consideration of the net amount of condensation taking place during the first oscillations. The actual decrease of $R_n$ is gradual and not stepwise as assumed in our simulations, where $R_n$ is reduced at $R_{max1}$ and $R_{max2}$, but the stepwise reduction does not influence the predicted dynamics

Fujikawa and Akamatsu (1980) and Yasui (1995) demonstrated that the bubble collapse is more vigorous when mass transfer by condensation and heat conduction are considered in the simulations because the reduction of the bubble's gas content by condensation reduces the buffering effect of the gas. Heat conduction reduces the temperature at collapse, which facilitates condensation and leads to higher collapse velocity and peak pressure. However, their model including evaporation, condensation and heat conduction is highly complex and, therefore, not well suited for extensive parametric studies. In our approach, the influence of condensation and heat conduction is indirectly accounted for by fitting $R_{nc}$ to match measured oscillation times. Direct modelling of condensation is hampered by the fact that the precise determination of evaporation and condensation coefficients of water is still an active field of research (Eames, Marr and Sabir 1997; Storey and Szeri 2000; Toegel et al. 2000; Marek and Straub 2001; Akhatov et al. 2001; Lauer et al. 2012, Zein, Hantke and Warnecke 2013). Akhatov et al. (2001) looked closely into the condensation process during the collapse of laser-induced bubbles but – in lack of precise theoretical derivations - had to use the sticking factor of vapour molecules to the bubble wall as a fitting factor. This resembles the approach of using $R_{nc}$ as fitting factor.

The mass reduction of the bubble content during collapse must be considered for obtaining realistic values of the collapse temperature. Determination of $R_{nc}$ by fitting $R(t)$ to the measured value of $T_{osc2}$ makes it possible to circumvent the explicit modelling of mass transfer. This is done by assuming that the collapse proceeds as adiabatic process, which starts at room temperature for $R = R_{max}$ with a (virtual) bubble pressure $p_{R\max,\text{virt}}$ corresponding to the amount of gas represented by $R_{nc}$. The virtual starting pressure is lower than the real pressure given by the equilibrium vapour pressure at $R = R_{max}$. The equilibrium radius $R_{nc}$ refers to a bubble with internal pressure $p = p_\infty$ at room temperature. To obtain $p_{R\max,\text{virt}}$, which also refers to room temperature, we must, hence, relate the pressure in bubbles of different size at equal temperature, as described by Boyle's law. That leads to

$$p_{R\max,\text{virt}} = p_\infty \left( \frac{R_{nc}}{R_{max}} \right)^3 . \tag{3.34}$$

For an adiabatic process, pressure and temperature are linked by

$$p_1^{(1-\kappa)} T_1^\kappa = p_2^{(1-\kappa)} T_2^\kappa , \tag{3.35}$$

which provides 
$$T_{coll} = 293\text{K} \left( \frac{p_{R\max,\text{virt}}}{p_{coll}} \right)^{(1-\kappa)/\kappa} \tag{3.36}$$

for the collapse temperature. With $\kappa = 4/3$ and by inserting Eq. (3.34) into Eq. (3.36) we get

$$T_{coll} = 293\text{K} \left( \frac{p_{coll}}{p_\infty} \right)^{1/4} \left( \frac{R_{max}}{R_{nc}} \right)^{3/4} . \tag{3.37}$$

This approach provides an upper estimate of the collapse temperature, as heat conduction is neglected.

At a later stage, when it exhibits small-amplitude linear oscillations, the bubble is filled mostly with non-condensable gas. It originates largely from water dissociation in the laser plasma; Akhatov et al. (2001) showed that rectified diffusion of dissolved air into the laser induced cavitation bubble is negligibly small. Besides the non-condensable gas, a small fraction of vapour will also be present in the residual bubble. Its amount is given by the equilibrium pressure corresponding to the temperature of the liquid at the bubble wall. The linear resonance frequency of the residual bubble reads as (Lauterborn and Kurz, 2010)

$$\nu_0 = \frac{1}{2\pi R_{\text{nres}} \sqrt{\rho_\infty}} \sqrt{3\kappa \left( p_\infty + \frac{2\sigma}{R_{\text{nres}}} - p_v \right) - \frac{2\sigma}{R_{\text{nres}}} - \frac{4\mu^2}{\rho_\infty R_{\text{nres}}^2}} \tag{3.38}$$

Measurement of the bubble oscillation time at late stages yields $\nu_0$, and by inserting this value into Eq. (3.38), $R_{\text{nres}}$ can be determined with high precision. Comparison of the radius $R_{\text{nres}}$ of the residual gas bubble and $R_{nc}$ then enables to discriminate between the gas and vapour content at the first bubble collapse. This is possible, as the gas content will remain approximately constant between the first collapse and later oscillations because the permanent gas dissolves only very slowly into the liquid surrounding the residual bubble.

### 3.6. *Energy balance for laser-induced bubble formation and oscillations*

Already decades ago theoretical studies have shown that during the expansion of large bubbles driven by underwater explosions or induced by optical breakdown the largest part of the initial energy is radiated away as a shock wave and degraded into heat by dissipative processes as the wave propagates outward (Cole 1948; Ebeling 1978). A smaller fraction of the initial energy remains as bubble energy, and upon collapse and rebound of the bubble the largest part of the remaining energy is again radiated away acoustically. The same applies for subsequent afterbounces until the bubble dynamics finally changes into small-amplitude linear oscillations of the remaining gas bubble for which viscous damping dominates.

Research in the 1980's and 1990's focused on an experimental investigation of energy partitioning for millimetre-sized bubbles (for which surface tension and viscosity may be neglected) by measuring the bubble's potential energy

$$E_{\text{pot}} = (4/3)\pi R_{\text{max}}^3 (p_\infty - p_\text{v})$$
(3.39)

and shock wave energy

$$E_{\text{SW}} = \frac{4\pi R_{\text{m}}^2}{\rho_\infty c_\infty} \int p_\text{s}^2 \, dt, \qquad (3.40)$$

whereby $R_{\text{m}}$ denotes the distance of the measurement location from the emission center (Vogel and Lauterborn 1988; Vogel, Busch and Parlitz 1996; Vogel et al. 1999b). While the determination of $E_{\text{pot}}$ is quite easy, measurement of the temporal shock wave profile needed for the determination of $E_{\text{SW}}$ is challenging already in the far-field (Vogel and Lauterborn 1988; Tinguely et al. 2012; Lauterborn and Vogel 2013) but even more in the near field close to the source. However, both near- and far field data are needed to assess the total emitted shock wave energy and the rate of energy dissipation upon wave propagation (Cole 1948; Vogel, Busch and Parlitz 1996; Vogel et al. 1999b). One way out was to determine near-field pressure profiles by numerical calculations using the Gilmore model and far-field profiles by hydrophone measurements (Vogel, Busch and Parlitz 1996). Another alternative was to determine the energy dissipation from the decay of shock wave pressure with propagation distance that was obtained by measuring $u_\text{s}(r)$ (Vogel et al. 1999b). These investigations confirmed the theoretical prediction that shock wave emission is the dominant mechanism of energy dissipation in spherical laser-induced bubble oscillations and supported earlier statements that the dissipation is largest near the source where pressure and velocity gradients at the shock front are largest (Cole 1948). Up to a distance of 10 times the plasma radius, 80 – 90% of the initial shock wave energy is dissipated (Vogel et al. 1999b).

A simpler approach for establishing an energy balance of bubble collapse and rebound was taken by Tinguely et al. (2012). They identified the emitted shock wave energy with the difference of the bubble energies before collapse and after rebound (i.e. at $R_{\text{max1}}$ and $R_{\text{max2}}$), considering the change of internal energy $\Delta U_{\text{int}}$ arising from the work done by the liquid on the gas in the bubble between the two stages. Hereby the gas in the bubble was assumed to be non-condensable, with constant amount, and adiabatic changes during collapse and rebound. The shock wave energy is then given by $E_{\text{SW}} = E_{\text{pot}}^{\text{max1}} - E_{\text{pot}}^{\text{max2}} - \Delta U_{\text{int}}$. Static pressure and gas pressure were varied in the range $p_{\text{stat}} \in [1,100]\,\text{kPa}$, and $p_{R\text{max1}} \in [1,100]\,\text{Pa}$, respectively. It turned out that $\Delta U_{\text{int}}$ was negligible (< 1%) for the range of parameters investigated, which justified the approximation $E_{\text{SW}} \approx E_{\text{pot}}^{\text{max1}} - E_{\text{pot}}^{\text{max2}}$.

The above approach is adequate for large bubbles but too simple for $R_{\max} \to 0$, where viscous damping and surface tension must be considered. Moreover, it neglects the energy flow by water vaporization and condensation, and provides no information on the energy partitioning between shock wave emission and bubble formation after breakdown, Finally, it would be interesting to track the energy flow through the collapse phase itself, distinguishing between the energy stored in the compressed bubble content and in the liquid surrounding the bubble. In the following, we present a complete treatment of the energy flow and partitioning for laser-induced bubbles based on the Gilmore model described in section 4.2.1.

3.6.1. *Overview over the energy partitioning*

Figure 4 presents a flow diagram for the partitioning of the absorbed laser energy, $E_{\text{abs}}$. During optical breakdown, the energy absorbed in the plasma volume $V_{\text{P}} = (4/3)\pi R_0^3$ partitions into vaporization energy

$$E_{\text{v}} = \rho_{\infty} (4/3)\pi R_0^3 \left[ C_{\text{p}}(T_2 - T_1) + L_{\text{V}} \right] \qquad (3.41)$$

and an internal energy gain $\Delta U_{\text{int}}$ of the heated, pressurized gas volume. The isobaric heat capacity of water is $C_{\text{p}} = 4187$ J/(K·kg) at 20°C and the latent heat of vaporization is $L_{\text{V}} = 2256$ kJ/kg at 100°C. An equation for $\Delta U_{\text{int}}$ will be given further below.

The expanding bubble content does work, $W_{\text{gas}}$, on the surrounding liquid, and the internal energy decreases accordingly. The index "gas" refers here to both water vapour and the non-condensable gas produced by plasma-mediated water dissociation. The total energy involved in the bubble oscillation is

$$E_{\text{abs}} = E_{\text{v}} + \Delta U_{\text{int}}(t) + W_{\text{gas}}(t). \qquad (3.42)$$

For isochoric energy deposition with ultrashort laser pulses, $E_{\text{abs}} = E_{\text{v}} + \Delta U_{\text{int}}$, and the work on the liquid starts only after the end of the laser pulse, when the energy of the free electrons in the laser plasma has been thermalized. In the general case, however, conversion of $\Delta U_{\text{int}}$ into $W_{\text{gas}}$ starts already during the laser pulse. During bubble expansion, the gas does work on the liquid, whereas during collapse the inrushing liquid does work by compressing the bubble content.

During bubble expansion, the gas compresses the surrounding liquid and overcomes the liquid viscosity, the hydrostatic pressure $p_{\infty}$, and the pressure $p_{\text{surf}}$ arising from the surface tension at the bubble wall. In doing this, it creates kinetic energy $E_{\text{kin}}$ of the accelerated liquid, potential energy $E_{\text{pot}}$ of the expanding bubble, drives the emission of a shock wave with energy $E_{\text{SW}}$, and does the work $W_{\text{visc}}$ by overcoming viscous damping. Altogether, the work done on the liquid involves the components

$$W_{\text{gas}} = E_{\text{SW}}^{\text{bd}} + E_{\text{kin}} + W_{\text{visc}} + E_{\text{pot}}, \text{ with } E_{\text{pot}} = W_{\text{stat}} + W_{\text{surf}}, \qquad (3.43)$$

where $W_{\text{stat}}$ and $W_{\text{surf}}$ denote the work done against hydrostatic pressure and surface tension. At $R = R_{\max}$, the kinetic energy is zero and the potential energy reaches its maximum value $E_{\text{pot}}^{\text{max1}}$. At this stage, the energy of the breakdown shock wave, $E_{\text{SW}}^{\text{bd}}$, can be obtained by evaluating all other terms in Eqs. (3.42) and (3.43) and subtracting them from $E_{\text{abs}}$.

The vaporization energy $E_v$ is needed for the phase transition itself and cannot be converted into mechanical energy of shock wave emission and bubble oscillation. It is released during the bubble oscillations by condensation of vapour at the bubble wall, and heat conduction into the surrounding liquid. The total amount of energy dissipated by condensation is composed of latent heat and internal energy stored in the vapour. We denote the part lost during bubble expansion $E_{\text{cond}}^{\text{exp}}$. The release of latent heat during the bubble expansion can be assessed by comparing the amount of vapour corresponding to the liquid in the plasma volume to the amount contained in the expanded bubble at $R_{\text{max1}}$, assuming equilibrium vapour pressure at ambient temperature for $R = R_{\text{max}}$.

During *collapse*, the potential energy of the expanded bubble partitions into energy needed to overcome liquid viscosity, a part $E_{\text{compr}}^{\text{coll1}}$ consumed for compression of the liquid surrounding the bubble, and another part increasing the internal energy during the compression of the bubble content. That increase of internal energy is, however counteracted by losses from vapour condensation. Together with the latent heat released during condensation, it constitutes the energy fraction $E_{\text{cond}}^{\text{coll1}}$. The release of latent heat is assessed by comparing the amount of vapour in the bubble at $R_{\text{max1}}$ with the amount contained in a bubble with equilibrium radius $R_{\text{nc1}}$ that is obtained by fitting the predicted bubble dynamics to the measured value of $T_{\text{osc2}}$. We can distinguish between water vapour and non-condensable gas content of the bubble at first collapse if we assume that the residual bubble contains mostly non-condensable gas and that all vapour is condensed when the bubble reaches its minimum size during the second collapse. The gas content of the residual bubble is obtained by evaluating the late bubble oscillations with the help of Eq. (3.38), and the vapour content at first collapse is given by the volume difference of the equilibrium bubbles with radii $R_{\text{nc1}}$ and $R_{\text{nc2}} = R_{\text{nres}}$.

The energy partitioning during *rebound* resembles the events after optical breakdown, with one crucial difference: shock wave emission is now driven not only by the re-expanding bubble content but also by the re-expansion of the compressed liquid surrounding the bubble. Therefore, a much larger energy fraction is radiated away acoustically during rebound than after optical breakdown. The parts of the rebound shock wave energy, $E_{\text{SW}}^{\text{reb}}$, which originate from the re-expansion of the bubble and liquid are denoted $E_{\text{SWB}}$ and $E_{\text{SWL}}$, respectively.

During later bubble oscillations ("afterbounces"), the partitioning proceeds in a similar way as described for the first bubble collapse and rebound. However, with decreasing amplitude of the bubble oscillations, shock wave emission turns into linear acoustic emission. We treat the energy of the acoustic waves emitted after second collapse and during later oscillations as one entity, denoted as $E_{\text{acoust}}$. Finally, a small bubble remains that contains non-condensable gas and vapour at equilibrium conditions. Since the potential energy of this bubble is zero, its energy is solely given by the residual internal energy $U_{\text{int}}^{\text{res}}$. As already mentioned, the residual bubble with radius $R_{\text{nres}}$ contains mostly non-condensable gas and little vapour. The vapour content is determined by the equilibrium vapour pressure at the temperature of the liquid at the bubble wall. We will see in chapter 4.2.3 that this temperature is significantly larger than room temperature due to heat dissipation from the bubble content and at the front of the outgoing shock waves. However, for simplicity, we assume room temperature of the residual bubble's content when we establish the energy balance.

In the following, we will step-by-step present the equations describing energy partitioning during the laser-induced bubble oscillations. We start by tracking the internal energy.

### 3.6.2. *Changes of internal energy, and latent heat released into the liquid*

For a spherical oscillating bubble containing an ideal gas under adiabatic conditions, the change of internal energy from state 1 ($p_{\text{gas1}}$, $V_1$) to state 2 ($p_{\text{gas2}}$, $V_2$) is

$$\Delta U_{\text{int}} = \frac{4\pi}{3(\kappa-1)}\left(p_{\text{gas}2}R_2^3 - p_{\text{gas}1}R_1^3\right). \tag{3.44}$$

The energy absorbed during plasma formation is deposited into a small volume with equivalent spherical radius $R_0$ that is regarded as initial size of the laser-induced cavitation bubble. We assume a pressure

$$p_{\text{gas}|R=R_0} = p_\infty + 2\sigma/R_0 \tag{3.45}$$

for the bubble nucleus at time $t_0 = 0$ before the laser pulse, and express the time-dependent gas pressure during the pulse through the time evolution of equilibrium radius $R_n$ as

$$p_{\text{gas}}(t) = \left(p_\infty + \frac{2\sigma}{R_n(t)}\right)\left(\frac{R_n^3(t)}{R^3(t)}\right)^\kappa. \tag{3.46}$$

The internal energy of the bubble at time $t$ during bubble expansion is then given by

$$\Delta U_{\text{int}}^{\text{exp}}(t) = \frac{4\pi}{3(\kappa-1)}\left\{\left(p_\infty + \frac{2\sigma}{R_n(t)}\right)\left(\frac{R_n^{3\kappa}(t)}{R^{3\kappa}(t)}\right)R^3(t) - \left(p_\infty + \frac{2\sigma}{R_0}\right)R_0^3\right\}. \tag{3.47}$$

Since the second term in Eq. (3.47) is very small, we will neglect it in the following.

For the bubble collapse and subsequent bubble oscillations, the van der Waals hardcore $R_{\text{vdW}}$ must be considered. In the numerical integration of Eq. (3.17) and in the calculation of $U_{\text{int}}(t)$, it is introduced at the time corresponding to $R = R_{\text{max}1}$. After deleting the second term in Eq. (3.47) and considering the van der Waals hardcore, it reads

$$\Delta U_{\text{int}}^{\text{coll}1}(t) = \frac{4\pi}{3(\kappa-1)}\left\{\left(p_\infty + \frac{2\sigma}{R_n(t)}\right)\left(\frac{R_n^3(t) - R_{\text{vdW}}^3}{R^3(t) - R_{\text{vdW}}^3}\right)^\kappa \times (R^3(t) - R_{\text{vdW}}^3)\right\}. \tag{3.48}$$

In order to quantify the loss of internal energy of the bubble by condensation of vapour at the bubble wall during bubble expansion and collapse we first define the internal energy at the time of maximum bubble expansion for which equilibrium conditions at ambient temperature, i.e. isothermal conditions are assumed:

$$U_{\text{int}}^{\text{max}1}\bigg|_{p=p_v} = \frac{4\pi}{3(\kappa-1)}\,p_v\,R_{\text{max}1}^3 = 4\pi\,p_v\,R_{\text{max}1}^3. \tag{3.49}$$

The internal energy loss during bubble expansion is given by the difference between the energy of an adiabatically expanding bubble at $R_{\text{max}}$ (which is obtained by evaluating Eq. (3.47) at the time of maximum bubble expansion for $R_n = R_{\text{nbd}}$) and the energy corresponding to isothermal conditions at $R_{\text{max}}$ from Eq. (3.49):

$$\Delta U_{\text{int,cond}}^{\text{exp}} = U_{\text{int}}\bigg|_{R_n=R_{\text{nbd}}} - U_{\text{int}}^{\text{max}1}\bigg|_{p=p_v} \tag{3.50}$$

In a similar fashion, the internal energy lost during bubble collapse is calculated by subtracting the energy of an adiabatically collapsing bubble with gas content corresponding to the equilibrium bubble radius at collapse (which is obtained by evaluating Eq. (3.48) for $R_n = R_{\text{nc}1}$) from the energy corresponding to isothermal conditions at $R_{\text{max}1}$

$$\Delta U_{\text{int,cond}}^{\text{coll}1} = U_{\text{int}}^{\text{max}1}\bigg|_{p=p_v} - U_{\text{int}}\bigg|_{R_n=R_{\text{nc}1}} \tag{3.51}$$

The changes of internal energy during rebound and second collapse are

$$\Delta U_{\text{int,cond}}^{\text{reb}} = U_{\text{int}}\Big|_{R_n = R_{\text{nc1}}} - U_{\text{int}}^{\max 2}\Big|_{p=p_v}, \tag{3.52}$$

and
$$\Delta U_{\text{int,cond}}^{\text{coll2}} = \Delta U_{\text{int,cond}}^{\text{res}} = U_{\text{int}}^{\max 2}\Big|_{p=p_v} - U_{\text{int}}\Big|_{R_n = R_{\text{nc2}}}, \tag{3.53}$$

respectively. We assume that condensation is complete after second collapse such that a constant amount of residual internal energy remains.

The internal energy lost by condensation is dissipated as heat into the surrounding liquid. In addition, we need to look at the latent heat released into the liquid. For this purpose, we express the amount of vapour contained in the bubble at different instants in time (directly after the laser pulse, at $R_{\max 1}$, $R_{\min 1}$, and $R_{\max 2}$) through the radius of a vapour bubble at room temperature with pressure 0.1 MPa. The vapour bubble radius after breakdown is calculated considering conservation of mass during vaporization of the liquid in the plasma volume. With

$$\rho_\infty \frac{4}{3}\pi R_0^3 = \rho_v \frac{4}{3}\pi (R_v^{\text{bd}})^3, \text{ we obtain } R_v^{\text{bd}} = R_0 (\rho_\infty / \rho_v)^{1/3}. \tag{3.54}$$

The mass density of vapour at $p_v = 0.1$ MPa and $T = 20°C$ is $\rho_v = 0.761$ kg/m$^3$. The amount of vapour in the expanded bubble can be assessed by assuming that the vapour pressure at $R_{\max 1}$ and $R_{\max 2}$ equals the equilibrium vapour pressure at room temperature, $p_v = 2.33$ kPa. The corresponding bubble radii for vapour at ambient pressure are then

$$R_v^{\max i} = R_{\max i}(p_v / p_\infty)^{1/3}, \text{ with i = 1 and 2}. \tag{3.55}$$

The loss of latent heat by condensation during bubble expansion is given by

$$\Delta E_v^{\exp} = E_v - E_v^{\max 1}, \text{ with } E_v^{\max 1} = \left(\frac{R_v^{\max 1}}{R_v^{\text{bd}}}\right)^3 E_v \tag{3.56}$$

and $E_v$ from Eq. (3.41). Note, that this does not include the loss of internal energy by condensation, which will be presented in the next section. The energy transfer during the first collapse is

$$\Delta E_v^{\text{coll1}} = E_v^{\max 1} - E_v^{\text{coll1}}, \text{ with } E_v^{\text{coll1}} = \left(\frac{R_v^{\text{coll1}}}{R_v^{\text{bd}}}\right)^3 E_v, \text{ and } R_v^{\text{coll1}} = (R_{\text{nc1}}^3 - R_{\text{nc2}}^3)^{1/3}. \tag{3.57}$$

The calculation of $R_v^{\text{coll1}}$ is based on the assumption that at second collapse the condensation process is completed and the bubble contains only non-condensable gas. The latent heat released during rebound and second collapse is given by

$$\Delta E_v^{\text{reb}} = E_v^{\text{coll1}} - E_v^{\max 2}, \text{ with } E_v^{\max 2} = \left(\frac{R_v^{\max 2}}{R_v^{\text{bd}}}\right)^3 E_v \tag{3.58}$$

and
$$\Delta E_v^{\text{coll2}} = E_v^{\max 2}, \tag{3.59}$$

respectively. The latent heat remaining at $R_{\max 2}$ is dissipated during the transition into linear bubble oscillations. For each stage, the total condensation loss $E_{\text{cond}}$ is the sum of internal energy loss and release of latent heat

$$E_{\text{cond}} = \Delta U_{\text{int,cond}} + \Delta E_v. \tag{3.60}$$

### 3.6.3. *Work done by the gas, and shock wave energies*

The work done by the gas during the bubble oscillations is given by

$$W_{gas}(t) = \int p_{gas}(t)\,dV = \int 4\pi R^2 p_{gas}(t)\,dR. \tag{3.61}$$

For numerical integration, this relation must be rewritten as time integral. By transforming the differential variable from $dR$ to $dt'$ using $dR = (dR/dt') \times dt' = U(t')dt'$, we obtain

$$W_{gas}(t) = \int_0^t 4\pi R(t')^2 U(t') p_{gas}(t')\,dt', \tag{3.62}$$

where $p_{gas}$ is given by Eq. (3.46).

The work $W_{gas}$ done on the liquid during bubble expansion (or by the liquid on the collapsing bubble) is composed of $W_{stat}$, $W_{surf}$, and $W_{visc}$, as expressed by Eq. (3.43). The work done at a given time to overcome the hydrostatic pressure is

$$W_{stat}(t) = \int p_\infty\,dV = \int 4\pi R^2 p_\infty\,dR = \int_0^t 4\pi R(t')^2 U(t') p_\infty\,dt'. \tag{3.63}$$

The work done to overcome the surface tension is

$$W_{surf}(t) = \int p_{surf}\,dV = \int 4\pi R^2 p_{surf}\,dR = 8\pi \sigma \int_0^t R(t')U(t')\,dt', \tag{3.64}$$

and the potential bubble energy at any given time is therefore

$$E_{pot}(t) = W_{stat}(t) + W_{surf}(t). \tag{3.65}$$

Finally, the work required to overcome viscosity is

$$W_{visc}(t) = \int p_{visc}\,dV = \int 4\pi R^2 p_{visc}\,dR = 16\pi \mu \int_0^t R(t')U(t')^2\,dt'. \tag{3.66}$$

The energy balance for the conversion of the absorbed energy $E_{abs}$ during the expansion process is obtained by numerical integration of the above equations up to the time $t_{max1}$:

$$E_{abs} = \underline{E_{SW}^{bd}} + \underline{E_{cond}^{exp}} + \underline{W_{visc}} + E_{pot}^{max1} + U_{int}^{max1}\Big|_{p=p_v} + E_v^{max1} \tag{3.67}$$

The underscored terms are energy parts that are dissipated during expansion, the other terms remaining at $t = t_{max1}$ are parts of the total bubble energy $E_B^{max1}$, which is the sum of the bubble's internal and potential energy. The parts of $E_{pot}^{max1}$ related to hydrostatic pressure and surface tension are given by Eqs. (3.59) and (3.60). Note that the energy change by condensation in Eq. (3.63) refers only to the part related to the change of internal energy; the loss of latent heat by condensation is already covered by Eq. (3.54). The breakdown shock wave energy, $E_{SW}^{bd}$, cannot be calculated directly but it can be determined from Eq. (3.63) since all other terms are known.

The energy balance for the collapse process starts with the energy remaining at $R_{min1}$ and tracks their dissipation and conversion during collapse. It is obtained by numerical integration of Eqs. (3.58) – (3.62) up to $t = t_{min1}$:

$$E_B^{max1} = E_{pot}^{max1} + U_{int}^{max1}\Big|_{p=p_v} + E_v^{max1} = \underline{E_{cond}^{coll1}} + \underline{W_{visc}} + E_{compr}^{coll1} + U_{int}^{min1} + E_v^{coll1}. \tag{3.68}$$

The underscored terms represent energy parts that are dissipated during collapse, the other terms remaining at $t = t_{min1}$ describe parts of the total amount of energy contained in the compressed liquid and gas, $E_{\text{compr}}^{\text{total}}$. The energy stored in the compressed liquid, $E_{\text{compr}}^{\text{coll1}}$, cannot be calculated directly but it can be obtained from Eq. (3.64) because the other terms are known.

The energy balance for the rebound starts with the energy contained in the compressed bubble and liquid and tracks its dissipation and conversion during re-expansion. The balance at $R_{\text{max2}}$ is determined by numerical integration up to $t = t_{\text{max2}}$:

$$E_{\text{compr}}^{\text{total}} = E_{\text{compr}}^{\text{coll1}} + U_{\text{int}}^{\text{min1}} + E_{\text{v}}^{\text{coll1}} = \underline{E_{\text{SW}}^{\text{reb}}} + \underline{E_{\text{cond}}^{\text{reb}}} + \underline{W_{\text{visc}}} + E_{\text{pot}}^{\text{max2}} + U_{\text{int}}^{\text{max2}}\Big|_{p=p_{\text{v}}} + E_{\text{v}}^{\text{max2}}. \quad (3.69)$$

The total rebound shock wave energy, $E_{\text{SW}}^{\text{reb}}$, is determinable from Eq. (3.69), since all other terms are known. We can even distinguish between the parts of the shock wave energy originating from the re-expansion of the compressed bubble with energy $U_{\text{int}}^{\text{min1}}$ and the compressed liquid with energy $E_{\text{compr}}^{\text{coll1}}$, which are denoted $E_{\text{SWB}}$ and $E_{\text{SWL}}$, respectively:

$$E_{\text{SWL}} = E_{\text{compr}}^{\text{coll1}}, \qquad \text{and} \qquad E_{\text{SWB}} = E_{\text{SW}}^{\text{reb}} - E_{\text{SWL}}. \quad (3.70)$$

In order to obtain the energy balance for the afterbounces, the numerical integration is conducted up to a time at which the bubble oscillations have ceased and only a residual gas bubble remains. For this time period we have

$$E_{\text{B}}^{\text{max2}} = E_{\text{pot}}^{\text{max2}} + U_{\text{int}}^{\text{max2}}\Big|_{p=p_{\text{v}}} + E_{\text{v}}^{\text{max2}} = \underline{E_{\text{acoust}}} + \underline{E_{\text{cond}}^{\text{coll2}}} + \underline{W_{\text{visc}}} + U_{\text{int}}^{\text{res}}\Big|_{R=R_{\text{nc2}}}. \quad (3.71)$$

The energy $E_{\text{acoust}}$ of the acoustic radiation after second collapse and during later oscillations is calculated from the other known values in Eq. (3.67). Assuming that condensation is completed at second collapse, we have $R_{\text{nres}} = R_{\text{nc2}}$. Note that we need to know $R_{\text{max3}}$ to determine $R_{\text{nc2}}$. For nanobubbles, experimental values of $R_{\text{max3}}$ may not be available. In that case, we identify the equilibrium radius during afterbounces and the residual bubble radius with $R_{\text{nc1}}$.

The above approach for establishing an energy balance of laser-induced bubble oscillations is valid as long as heat conduction out of the energy deposition volume during the laser pulse can be neglected and when the laser pulse duration $\tau_{\text{L}}$ is much shorter than the bubble oscillation time. The characteristic thermal diffusion time for a spherical absorber is $\tau_{\text{D}} = d^2/8\kappa$, where $d$ is the focal diameter and $\kappa$ is the thermal diffusivity. For pulse durations $\tau_{\text{L}} \geq \tau_{\text{D}}$, the dynamics changes from approximately adiabatic towards isothermal conditions, and the bubble expands significantly already during laser pulse. As a consequence, less work is done by the expanding gas and both acoustic radiation and the overshoot over the equilibrium radius are largely reduced. Models of such dynamics must explicitly consider heat and mass transfer. The approach presented here is valid only as long as energy deposition is thermally confined and $\tau_{\text{L}} \ll T_{\text{osc}}$.

## 4. Results

We first show the results of plasma photography providing $R_0$ data, and of time-resolved photography of the initial bubble expansion and shock wave emission. The images show, which focusing angles and standoff distances from boundaries are needed to ensure highly spherical bubble dynamics, and visualise shock wave-induced phase transitions outside the plasma-heated region. Then we present one selected probe beam scattering signal that traces the dynamics of a highly spherical bubble over more than 100 oscillations. This signal is analysed in detail by numerical simulations using the tools described in section 3.2. The analysis includes the evolution of bubble radius, wall velocity and internal pressure, the collapse temperature upon first collapse, and the shock wave emission at breakdown and after the first bubble collapse. Analysis of the transition from nonlinear to linear bubble oscillations and of late bubble oscillations provides insights about an elevated liquid temperature near the bubble wall during the late oscillations and on the relative content of vapour and non-condensable gas during the first bubble collapse. Finally, we establish a complete energy balance for the absorbed laser energy by tracking its partitioning and dissipation throughout the entire bubble lifetime. In this section, results will be discussed as far as necessary for their interpretation, and in section 5, they will then be put in a larger context.

### 4.1. *Experiments*

#### 4.1.1. *Plasma size and initial bubble radius*

Figure 5 (a) shows luminescent plasmas in water produced by 1040-nm fs pulses of energies up to 600 nJ that were focused at $NA = 0.8$. The bubble threshold was at $E_{\text{th,bubble}} = 25$ nJ, corresponding to a threshold irradiance of $8.0 \times 10^{12}$ W/cm$^2$. For pulse energies $\geq 3 \times E_{\text{th,bubble}}$, luminescence could be detected photographically by integrating over many breakdown events. We identified the plasma boundary with the outer bound of the region in which luminescence could be clearly distinguished from the uniform background. The volume of the luminescent region determined from the photographs is plotted in figure 5(b) as a function of laser pulse energy, and figure 5 (c) shows the corresponding radius values of spheres with same volume that define the initial bubble radius $R_0$ for the numerical simulations.

#### 4.1.2. *Influence of plasma shape and nearby boundaries on radial symmetry*

The plasma shape influences the initial phase of shock wave emission and cavitation bubble expansion after laser-induced breakdown. Time-resolved shadowgraphs in figure 6 show the transition from non-spherical near-field dynamics to approximately spherical far-field dynamics for ps and ns breakdown after the generation of elongated plasmas at $NA \leq 0.25$. The transition from near-field to spherical far-field shock wave emission takes about 100 ns. The stronger elongation of the ps plasma in figure 6(a) has little influence on this transition time because the hydrodynamic events are dominated by a relatively small region with high volumetric energy density located upstream from the plasma tip near the beam waist. The aspherical dynamics lasts less than 1/100 part of the expansion phase ($R_{\text{max}}$ is reached after $\approx 50$ µs) but influences a significant part of the $R(t)$ curve. Deviations from spherical shape show up again upon bubble collapse and rebound because shape instabilities are enhanced during the collapse phase (Strube 1971).

Although the bubbles produced by elongated plasmas become approximately spherical during expansion, they retain a slightly oblate shape even at $R = R_{\text{max}}$, with larger diameter perpendicular to the optical axis than along the axis. If the plasma eccentricity is sufficiently large, an equatorial jet can evolve during the final collapse stage that transforms into two axial jets in opposite directions that lead to bubble splitting (Blake et al 1997; Brujan et al. 2001a).

At *NA* = 0.9, the plasma exhibits little elongation and the shock wave is almost perfectly spherical already after 30 ns, as seen in figure 7. Here the transition from near-field to spherical far-field shock wave emission is completed in less than 1/300 part of the expansion phase, and the expanded bubble with 473 µm radius is perfectly spherical at $t = 40$ µs. Nevertheless, pronounced jetting towards the objective is observed during rebound after the first collapse because the dimensionless stand-off distance from the front lens of the objective was only $\gamma = 4.65$. The jet is horizontal, which indicates that the solid boundary in the bubble's vicinity has a much stronger influence than buoyancy. In section 2.1, we argued that a standoff distance of $\gamma > 55$ would be needed to guarantee spherical dynamics, and this assessment is corroborated by the strong jetting seen in figure 7. For a working distance of 2.2 mm, the requirement $\gamma > 55$ is fulfilled if the bubble size stays below $R_{max} = 40$ µm.

4.1.3. *Pressure evolution within the breakdown region, and bubble wall formation*

In the theoretical description of bubble dynamics presented in section 3, a homogeneous pressure distribution inside the laser-produced bubble is assumed throughout the entire bubble lifetime. The initial bubble wall position is identified with the outer boundary of the luminescent plasma region, and it is assumed that the location of the bubble wall is affected only by the pressure difference between inner and outer pressure but not shifted by phase transitions arising from the shock wave passage. Both assumptions will be checked in the following by evaluating the time-resolved photographs of the initial hydrodynamic processes induced by 10-mJ and 20-mJ laser pulses shown in figure 8. The energetic high-density plasmas are sufficiently large to resolve details of the laser-induced sequence of events including plasma luminescence, shock wave emission, pressure equilibration in the breakdown region behind the shock front, and the appearance of a black region between plasma and shock wave that delineates the bubble wall location.

To support the analysis of these image features, we determined the plasma energy density, $\varepsilon$, by relating the luminescent plasma volume determined from the photographs to the amount of energy absorbed in this volume that is obtained from transmission measurements (Nahen and Vogel 1996). The average energy density is $\varepsilon \approx 40$ kJ/cm$^3$ for the 10-mJ pulse in figure 8 (a), and $\varepsilon \approx 35$ kJ/cm$^3$ for the 20-mJ pulse in figure 8 (b). Under the assumption of isochoric energy deposition, we can derive the plasma pressure from the energy density and obtain values of 11.3 GPa and 10.1 GPa, respectively using the IAPWS-95 formulation of the water EOS (Wagner and Pruss 2002). These data are an upper estimate; the actual pressure is somewhat lower because the bubble starts to expand already during the ns laser pulse. The large pressure jump at a shock front results in rapid energy dissipation and in a temperature rise after shock wave passage, which for $\Delta p = 10$ GPa amounts to 576°C (Rice and Walsh 1957). Therefore, the dissipated shock wave energy will create a phase transition in a thin zone extending beyond the rim of the expanding plasma, up to which vaporization is induced directly by the absorbed laser energy. Based on this background information, we will now step-by-step analyse the image series.

Plasma luminescence is visible on all images although it rapidly ceases after the end of the pulse because the photographs were taken with open shutter in a darkened room. The time given on the images refers to the time delay between the pump pulse producing the plasma and the illumination pulse for shadowgraph photography. The photographic exposures were adjusted to provide similar background brightness in all pictures. Due to the divergence of the illuminating laser beam, the illuminated spot is larger for longer delay times. This lowers its irradiance and makes the self-luminescent plasma appear brighter relative to the background although its actual luminescence remains the same.

The shock wave emission from the expanding plasma is visible due to the light deflection by the steep refractive index gradients at the shock front and in the shock waves' trailing edge.

Plasma formation starts at the laser beam waist (on the left) and moves upstream towards the incoming laser beam while the laser power increases during the pulse (Docchio 1988, Vogel et al. 1996). The movement of the breakdown wave results in a delayed shock wave emission from the upstream part of the plasma. The geometrical form of the shock wave reflects both the overall plasma shape and inhomogeneities of the energy density distribution within the plasma. Such inhomogeneities can result in the release of pressure transients propagating in the zone between plasma and outer shock front that are visible as dark structures on the images. Because of the pressure dependence of sound velocity, the transients propagate at high speed and finally catch up with the shock front.

In figure 8(a), the plasma exhibits an inhomogeneous energy distribution with a high-density region close to the beam waist and a larger and more pronounced hot spot located further upstream. The inhomogeneity was reproducible from shot to shot, which is due to the temporal pulse shape of the 6-ns pulse exhibiting two peaks (Vogel et al. 1996). Therefore, the breakdown wave moving upstream against the incoming laser beam produces two spatially separated high-density regions corresponding to the two peaks of the laser pulse. The hot spot in the upstream region is generated during the second half on the laser pulse, and the fast transient emerging from it propagates into the high-pressure region behind the shock wave emitted earlier during the pulse near the beam waist. The velocity of this transient is indicative for the speed of pressure equilibration within the breakdown region and provides information about the average pressure in this region. The pressure transient traversing the breakdown region in axial direction has passed the supercritical fluid in the plasma region after about 40 ns, and after about 120 ns its front will reach the trailing edge of the shock wave. The average speed of the transient in axial direction during the first 60 ns is about 300 µm/60 ns, i.e. ≈5000 m/s. For the water Hugoniot centred at ambient conditions (20°C and 0.1 MPa), this value corresponds to a pressure of ≈10 GPa (Rice and Walsh 1957), which is consistent with the estimated pressure value derived from the plasma energy density. It is also consistent with pressure values obtained for equal pulse energy by analysis of the initial shock wave speed close to the plasma rim through time-resolved photography (Vogel, Busch and Parlitz 1996) and streak photography (Noack and Vogel 1998). The above analysis shows that the pressure within the breakdown region rapidly equilibrates (within 1/2800 part of the bubble expansion time of 168 µs that was obtained from the hydrophone signal). This justifies the assumption of a homogenous bubble pressure made in the Gilmore model. By contrast, deviations of the high pressure region from spherical geometry still persist after pressure homogenisation and affect the bubble expansion and shock wave emission for a longer time.

The picture series in figure 8(a) shows a black region between the luminescent plasma region and the shock wave already at $t = 7$ ns, shortly after the peak of the 6-ns (FWHM) laser pulse. This outer border of this region is usually identified with the wall of the expanding cavitation bubble but it appears already before a phase boundary is formed in the fluid. A phase change (formation of the bubble wall) occurs when and where pressure and temperature both drop below the critical point (374°C, 22 MPa). For the 10-mJ, 6-ns laser pulse, this happens only after 70-80 ns as shown by numerical simulations of $p(t)$ in (Vogel, Busch and Parlitz 1996), much later than the time when a black region becomes visible on shadowgraphs. Before the formation of a phase boundary, we already see the border of the expanding supercritical fluid. Within that region, the refractive index is lowered because of the reduced mass density. The sharp grey level transition on the shadowgraphs thus matches the border of the expanding "plasma region" that is taken as "bubble wall" in the simulations.

Due to the energy dissipation at the shock front, a phase transition will occur also in a thin layer around high-density plasmas, as discussed above. Like the phase change within the plasma volume, its onset is delayed but the heated region shows up already shortly after the shock wave passage. It is well visible in figure 8 (b), where the plasma was produced with a 20-mJ laser pulse. Here, the outer border of the heated region appears rugged, especially in the picture taken at $t = 12$ ns. In locations, where the temperature rises above the superheat limit (kinetic spinodal), which lies at ≈ 300 °C (Vogel and Venugopalan 2003; Vogel et al. 2005; Debenedetti 1996), the instable liquid starts to expand immediately (Zhigilei et al. 2003). Therefore, the refractive index drops, which become visible as a dark region on the photographs, similar to the supercritical fluid in the plasma region. The rugged appearance of the borderline of the shock-induced phase transition is likely due to inhomogeneities in the plasma producing pressure fluctuations in the near-field and local variations of the temperature rise.

We conclude that around high-density plasmas, the shock wave can enlarge the volume of vaporized liquid driving the bubble expansion. This "convective" heat transport by shock wave propagation and energy dissipation at the shock front is much faster than heat diffusion. It produces a thermal gradient that is less steep than the initial gradient arising from the spatial distribution of free electron density in the laser plasma. As a consequence, the "bubble wall" appears fuzzy in figure 8 during the first 50-60 ns. It smoothens out later during the ongoing plasma expansion, when a phase boundary is formed, the bubble content cools down adiabatically and surface tension can smoothen out local irregularities.

The additional vapour mass produced by energy dissipation at the shock front is hard to quantify and, therefore, not included in the energy balance presented in this paper. However, the phase transition occurs only after the pressure in the breakdown region has dropped below the critical point, i.e. when most of the deposited laser energy is already converted into the mechanical energy of the outgoing shock wave and the outward fluid flow of the liquid. Therefore, it will have little influence on the bubble expansion. However, by raising the temperature in the liquid around the bubble, it will influence the damping behaviour during the late oscillations (see section 4.2.3 further below).

### 4.1.4. *Single-shot probe beam signal*

Figure 9 shows a probe beam signal representing the evolution of oscillation times $T_{osc,i}$ during the life time of an almost perfectly spherical bubble with 35.8 µm maximum radius and a dimensionless stand-off distance $\gamma = 70$ from the microscope objective's front lens. The signal seen in figure 9(a) covers 102 oscillations and portrays the transition from initial nonlinear cavitation bubble oscillations to linear oscillations around the equilibrium radius of the residual gas bubble. The bubble was produced by a 755-nm, 155-nJ fs pulse focused at $NA = 0.9$. The equivalent spherical plasma radius for this pulse energy read from figure 5(*c*) is $R_0 = 1.33$ µm, and the absorbed energy is 86 nJ. With the plasma volume $V_p = 9.85$ µm$^3$ from figure 5(*b*), this yields an average volumetric energy density of 8.73 kJ/cm$^3$. The internal energy density $U_{int}/V_p = (E_{abs} - E_v)/V_p$ with $E_v = 25.49$ nJ from Eq. (3.41) is 6.14 kJ/cm$^3$, the difference of both values corresponds to the vaporization enthalpy. Since energy deposition is isochoric, the average plasma temperature can be determined from $U_{int}/V_p$ and the water EOS. Using the IAPWS-95 formulation (Wagner and Pruss 2002), we obtain $T_{avg} = 1550$ K. The peak temperature will be somewhat larger, as suggested by the inhomogeneous brightness distribution in the photographs of figure 5(*a*).

The *NA* used for the acquisition of the signal in figure 9 is slightly larger than that used for taking the photographs in figure 5 but the plasma shape will presumably still be somewhat elongated. The elongated shape causes a stability crisis in the first collapse that is reflected by the asymmetry of the probe beam signal during second and third oscillation seen in figure 9(b). However, the signal symmetry is regained in the fourth oscillation, which indicates that surface tension has restored the spherical shape.

The signal undulations during the first cavitation bubble oscillation are interference fringes reflecting the radius-time evolution. Their modulation is largest around the maximally expanded stage of the bubble. However, the region with detectable fringe separation is too small to gain significant information about the *R*(*t*) curve. Later, each signal undulation represents one period of the small-amplitude oscillations of the residual bubble around its equilibrium radius. The undulations arise from the interference between bubble wall reflections with the directly transmitted beam combined with changes in the angular distribution and orientation of the central Mie scattering lobe, as described in section 2.2. The slow undulation of the average signal level within the first 40 µs is a consequence of the capacitive AC coupling having a lower cut-off frequency at 25 kHz and has no physical meaning. It does not affect the determination of bubble oscillation times.

### 4.2. Numerical calculations

#### 4.2.1. Evolution of bubble radius, wall velocity and internal pressure; collapse temperature

TABLE 1 summarizes characteristic breakdown and bubble parameters corresponding to the probe beam signal of figure 9, and figure 10 shows simulation results for the time evolution of cavitation bubble radius, internal pressure, and bubble wall velocity obtained with these parameters. Enlarged views of $R(t)$, $P(t)$, and $U(t)$ for time intervals of 20 ns after optical breakdown and 1 ns around the first collapse are presented in figure 11.

Because of the relatively small plasma temperature of 1550 K, the breakdown pressure is merely 1.25 GPa, much lower than in previous experiments with bubbles generated by 10-mJ IR ns laser pulses (Vogel, Busch and Parlitz 1996). Water dissociation relies on free-electron-mediated pathways (see section 5.2) as thermal dissociation sets in only at 3000 K (Mattsson and Desjarlais 2006). Compared with the amount of vaporized liquid, only a small amount of non-condensable gas is produced by dissociation. As a consequence, the bubble collapse is only weakly damped by its permanent gas content, and the collapse pressure reaches 13.5 GPa, which is about 11 times larger than the plasma pressure upon breakdown. The minimum bubble radius at collapse is $R_{min}$ = 419 nm, less than one third of the plasma size that defines the initial bubble radius. The duration of the collapse pressure peak is much shorter (FWHM = 76.5 ps) than the pressure peak after breakdown (FWHM = 920 ns). It is interesting to note that models of bubble collapse in compressible liquids that are based on full solutions of the Navier-Stokes equations rather than on the Kirkwood Bethe hypothesis yield somewhat higher collapse pressures than the Gilmore model (Fuster, Dopazo and Hauke 2011; Koch et al. 2016). Thus, the actual collapse pressure may be even larger than 13.5 GPa.

During breakdown, the bubble wall velocity performs a jump-start and reaches a peak value of 540 m/s. At this time, the bubble content is still a supercritical fluid, and a phase boundary between vapour and liquid water does not yet exists. It forms after 6.85 ns, when pressure and temperature inside the bubble drop below the critical point. After the bubble wall has reached its velocity peak, the internal pressure of the plasma/bubble region still imparts kinetic energy to the surrounding liquid but the velocity of the boundary between both regions already decreases because an ever-larger mass is involved in the radial outward flow. Due to conservation of momentum, this goes along with a decrease of fluid velocity. During the final collapse stage, the phase boundary disappears again because both the bubble content and a thin liquid shell around the bubble become supercritical. The liquid/gas boundary reappears during rebound. The growing bubble pressure in the final collapse phase rapidly reverses the direction of the bubble wall velocity. It changes within ≈ 40 ps from a peak collapse value of -1788 m/s to a peak rebound value of 369 m/s, which goes along with an acceleration of $5.4 \times 10^{13}$ m/s$^2$.

During collapse, the bubble wall velocity becomes supersonic with respect to the sound velocity in the gaseous bubble content at ambient conditions, $c_0$. This happens at a bubble pressure of 0.45 MPa, when the mass density of the bubble content is still relatively low. Later during collapse, when the bubble content becomes a supercritical fluid, the sound velocity inside the bubble rapidly increases to value larger than the bubble wall velocity. The thermodynamic data reported by Rice and Walsh (1957) for the water Hugoniot centred at 20°C and 0.1 MPa cannot be directly applied to the bubble collapse because the temperature in the collapsed bubble is much higher than for the 20°C Hugoniot at a pressure of 13.5 GPa. Images in figure 8(a) indicate that the pressure transient emitted from the hot spot propagates even faster through the hot plasma region than through the cooler liquid around. Thus, $c$ will be in the order of 5000 m/s or faster also in the collapsed bubble. The high sound velocity promotes a rapid pressure equilibration within the collapsed bubble, which again justifies the assumption of a homogeneous bubble pressure made in the Gilmore model.

The collapse temperature calculated using Eq. (3.37) for an adiabatic collapse with an amount of vapour corresponding to $R_{nc1}$ is $T_{coll} = 31400$ K. This is more than ten times larger than what would be found under the assumption that the collapse starts with the equilibrium vapour pressure at room temperature and proceeds with constant vapour content, neglecting condensation. Under those conditions, the buffering by the larger vapour content results in a collapse pressure $p_{coll} = 0.061$ GPa, and a temperature of 2995 K.

The above data are obtained under the assumption of a van der Waals hardcore in the collapsing bubble. Simulations without van der Waals hardcore yield a much smaller minimum bubble radius of $R_{min} = 139$ nm, a much larger collapse pressure of 40.8 GPa, and peak bubble wall velocities of -3010 m/s and 2017 m/s during collapse and rebound, respectively. The collapse temperature obtained with Eq. (3.37) would be $T_{coll} = 45970$ K.

4.2.2. *Shock wave emission at breakdown and upon bubble collapse*

Figures 12 to 14 present the evolution of the velocity distributions $u(r)$ and pressure distributions $p(r)$ after the optical breakdown, and during the final collapse and early rebound phase for parameters corresponding to the signal in figure 9. The location of the shock front has been determined using the procedure described in section 3.3.

As seen in figure 12, a shock front forms within a few picoseconds after breakdown, owing to the ultrashort laser pulse duration. Shock wave detachment can be followed by looking at the $u(r)$ distributions in figure 12(*a*). After 4.7 ns, a velocity minimum has evolved between shock front and bubble wall that separates the particle velocity behind the shock front from the outward flow around the expanding bubble. The minimum becomes more pronounced with increasing propagation distance.

The situation is more complex in the case of bubble collapse and rebound, as seen in figures 13 and 14. The maximum velocity of the inrushing flow is reached ≈ 75 ps before collapse. The high bubble pressure evolving during collapse stops the inward flow at the bubble wall and drives an outward flow that reaches its peak velocity 90 ps after the collapse at a location slightly ahead of the bubble wall. The outward flow collides with the still incoming flow from outer liquid regions at the location of the shock front (figure 13). In the compressible liquid, the change of motion of the bubble wall is communicated to the liquid by the passage of the shock wave, and the flow reversal occurs, therefore, at the shock front.

For the transients emitted after breakdown and upon the bubble's rebound, the pressure decay is faster than for acoustic waves for which the attenuation would be proportional to $r^{-1}$. However, once the rebound shock front has formed, its peak pressure decays initially much faster (proportional to $r^{-1.75}$ in the steepest region) than for the shock wave emitted after optical breakdown ($\propto r^{-1.27}$). This is because the collapse pressure is one order of magnitude larger than the breakdown pressure. In both cases, the shock front persists also in the far field, where the pressure has dropped to a few MPa. In the pressure range around 20 MPa, the peak amplitude decays proportional to about $r^{-1.15}$. This value is typical for the regime of weak shock wave propagation (Arons 1954; Rogers 1977).

For the breakdown shock wave, the slope of the $p_{peak}(r)$ curve is steepest after about 4 µm propagation distance, where it has a value of -1.27. However, most of the shock wave energy is dissipated even closer to the plasma because the dissipation rate is proportional to the pressure jump at the shock front (Vogel et al. 1999b). Upon rebound, a shock front develops within about 50 ps and 750 nm propagation distance. At this time, it exhibits a pressure jump of ≈ 8 GPa, corresponding to a temperature jump to 436°C (Rice and Walsh 1957). The pressure decay is fastest in the range between 1 µm and 3 µm. The strong energy dissipation at the shock front heats a few micrometre thick liquid layer around the expanding bubble. This will vaporize a thin liquid shell and reduce surface tension and local viscosity in the liquid near the bubble wall, which may influence subsequent bubble oscillations as discussed in section 4.2.3.

The difference between the acoustic emission after breakdown and bubble collapse seen in figures 10 to 14 diverges from previous reports on similar pressure amplitudes after breakdown and collapse (Vogel et al 1988; Lauterborn and Vogel 2013). This is because the plasma energy densities in the earlier studies were much higher than in the present investigations. Furthermore, the initial bubble radius $R_0$ (given by the plasma size) is usually larger than the minimum collapse radius $R_{min}$, which is often comparable to or even smaller than the diffraction-limited optical resolution. Therefore, the breakdown events can be optically resolved but the amplitude of the rebound shock wave drops significantly already while it propagates within the optical resolution limit. As a consequence, the peak pressure produced during collapse tends to be underestimated. We conclude that numerical simulations of the bubble dynamics based on measured oscillation times can provide unique information not available by direct measurements of bubble wall movement and shock wave emission.

4.2.3. *Transition from nonlinear to linear oscillations, and late bubble oscillations*

The initial bubble oscillations are strongly damped by shock wave emission and decay into linear small-amplitude oscillations around the equilibrium radius of the residual gas bubble. During the initial nonlinear oscillations, most of the vapour produced during plasma formation condenses (Table 1, right columns) and at second rebound, the vapour content has already dropped to 1/200 of the initial value. This finding justifies our assumption, that condensation is complete at second collapse and the residual bubble contains only non-condensable gas.

It is interesting to note that $R_v$ after plasma formation is larger than $R_{nbd}$. The entity $R_{nbd}$ does not express the exact amount of gas and vapour in the bubble but rather measures the bubble's internal energy that does work on the surrounding liquid [Eq. (3.48)]. The difference between $R_v$ and $R_{nbd}$ is small for high-density plasmas but it becomes ever larger towards the bubble threshold where the energy density decreases. Upon first collapse, the vapour content of the bubble is already much smaller than after breakdown, and the amount of non-condensable gas becomes relevant. Nevertheless, we see that $R_{nc1}$, which reflects the total gas content of the bubble, is not much larger than $R_{vc1}$, which indicates that water vapour significantly contributes to buffering the collapse.

The radius of the residual bubble, $R_{res}$, which is assumed to contain only non-condensable gas, can be derived from the frequency of the bubble oscillations at late times using Eq. (3.38). Figure 9c shows that the oscillation frequency remains approximately constant after 5-10 oscillations, which indicates that the transition from nonlinear dynamics to linear gas bubble oscillations has been completed at this time. The mean oscillation time from $50^{th}$ to $100^{th}$ oscillation is 813.3 ± 6.7 ns. Based on this value, Eq. (3.38) with room temperature values of surface tension and viscosity yields a value $R_{nres} = 2.60$ µm for the equilibrium radius of the residual gas bubble under isothermal conditions ($\kappa = 1$).

Figure 15(*a*) shows that the $R(t)$ curve predicted for bubble oscillations in water at room temperature exhibits strong damping both for adiabatic conditions and for isothermal conditions that prevail during the late oscillations. Damping is slightly less pronounced for the isothermal curve but even here the peak-to-peak oscillation amplitude drops below 1 nm after little more than 30 µs (30 oscillations). This finding disagrees with the experimental observation that a modulated probe beam signal could be detected even after 100 oscillations. We attribute the disagreement to the neglect of a temperature rise around the bubble in the simulations. In reality,

energy is dissipated near the bubble wall through condensation and heat conduction, by viscous damping, and by dissipation at the shock front. The resulting temperature rise reduces surface tension and, particularly, viscosity as shown in figure 15(*b*).

In order to investigate the influence of temperature on the damping behaviour, we performed calculations with surface tension and viscosity values at elevated temperatures. To assess the possible temperature rise in the liquid around the bubble, we must leave the simplifying assumption made for the energy balance that the residual bubble contains exclusively non-condensable gas. We see from the approximately constant oscillation frequency observed experimentally in figure 9(*b*) during the late bubble oscillations that the sum of gas pressure and equilibrium vapour pressure must remain approximately constant. This works only if the partial vapour pressure in the residual bubble is significantly smaller than the gas pressure because the vapour pressure will drop with deceasing temperature during the bubble lifetime of 90 µs, which would change the total amount of gas if vapour pressure would dominate. Vapour pressure stays below the (ambient) gas pressure if the average temperature during the investigated oscillation time lies well below 100°C. An upper bound for the bubble wall temperature can also be assessed by looking at the condition that the equilibrium vapour pressure must be balanced by the sum of hydrostatic and Laplace pressure from surface tension; otherwise, the bubble would grow in size. At $R_{res}$ = 2.5 µm, this is the case for $T$ = 113°C.

The possible temperature rise can also be estimated from the plasma energy. In the following section, we will see that the absorbed laser energy producing the bubble of figure 9 amounts to 64.3 nJ. It can produce an average temperature rise of 48.7 K in a 2 µm thick liquid shell around a bubble with 2.5 µm radius, corresponding to a final temperature of 68.7°C. The thermal relaxation time for a spherical source of 9 µm diameter is 71 µs (in our rough estimate we neglect the missing central volume because the volume of the surrounding 2-µm shell is 5.8 times larger), which is similar to the observed oscillation time. Since the peak temperature at the bubble wall is much larger than the average temperature of 68.7 °C, the temperature at the bubble wall will remain elevated during its linear oscillations.

Based on the above findings, we simulated the bubble dynamics under isothermal conditions for temperature values at the bubble wall of $T_W$ = 60°C and $T_W$ = 110°C prevailing during the entire bubble lifetime. The results are presented in figure 16. Bubble oscillations cease after 70 µs for $T_W$ = 60°C but last for 90 µs for $T_W$ = 110°C. However, the predicted peak-to-peak oscillation amplitude after 90 µs is only 0.06 nm even for $T_W$ = 110°C. The calculations provide only a rough assessment of the influence of the elevated temperature in the bubble's surroundings as they assume a constant value of $T_W$ neglecting heat diffusion. Nevertheless, we may conclude that a good fit of calculated $R(t)$ data to experimental observations can be achieved only with a fairly large temperature elevation, which strongly reduces viscous damping. Even then the predicted oscillation amplitude at $T_{osc,100}$ is below 0.1 nm. This demonstrates the high sensitivity of our interferometric probe beam technique.

The residual bubble radius determined from the period of the late bubble oscillations, $R_{nres}$ = 2.60 µm, is slightly larger than the value of the equilibrium radius at the second bubble collapse, $R_{nc2}$ = 2.44 µm, that was obtained by fitting the $R(t)$ curve to the measured value $T_{osc,3}$ assuming adiabatic conditions for nonlinear oscillations (table 1). We assume that during the first and second vigorous collapse phases almost all vapour has condensed at the bubble wall, whereas later some re-evaporation from the heated liquid into the residual bubble occurred. The assumption that the bubble contains hardly any vapour at the second collapse provides an upper estimate for the amount of non-condensable gas produced by water dissociation during optical breakdown. For the first collapse, with $R_{nc2}$ = 3.6 µm, it implicates a vapour/gas ratio of 68.9% vapour and 31.1% non-condensable gas. The result that most of the bubble content at first collapse is still water vapour is consistent with simulation results obtained in the context of single bubble sonoluminescence (Storey and Szeri 2000).

In the above considerations, we have assumed that all non-condensable gas was produced by water dissociation in the breakdown plasma and neglected the production of additional gas during the first and second collapse, where temperatures are even higher than in the breakdown plasma and a large number of chemical reactions take place (Yasui 1995; Toegel and Lohse 2003). This neglect is justified by the fact that the water mass contained in the collapsed bubble is only 0.57% of the mass within the laser plasma (table 1).

4.2.4. *Energy partitioning*

The partitioning of the absorbed laser energy and the energy flow during the entire bubble lifetime that is schematically depicted in figure 4 has been quantitatively analysed for the signal of figure 9, using the tools presented in section 3.6. Numerical values for energy fractions integrated over the entire bubble lifetime, and over the expansion, collapse, rebound, and afterbounce phases are listed in table 2, as well as the energy balance at three specific instants ($R_{max1}$, $R_{min1}$, and $R_{max2}$). At these instants, a complete balance can be established, because the fluid movement has stopped, $E_{kin}= 0$, and the amount of gas and vapour can be assessed based on simple assumptions without explicit modelling the kinetics of heat and mass transfer during the bubble oscillations. Figure 17 focuses on the partitioning of internal bubble energy, i.e. that part of the deposited energy, which can do work on the surrounding liquid. It presents the time evolution of those energy fractions that can be tracked continuously.

The incident energy of the laser pulse producing the bubble with $R_{max}$ = 35.8 µm was 155 nJ. The total amount of deposited energy derived from the plasma radius $R_0$ taken from figure 5 and the model fit of $R(t)$ to measured oscillation times is $E_{abs} = E_v + \Delta U_{int}$ = 25.5 nJ + 37.0 nJ = 62.5 nJ. Because of the moderate plasma energy density, 40.8% of the absorbed energy is needed to vaporize the liquid in the plasma volume, and only 59.2% are available for doing work on the surrounding liquid. A fraction of 55.4% of $\Delta U_{int}$ (32.8% of $E_{abs}$) appears as potential energy $E_{pot}^{max1}$ of the expanded cavitation bubble at $R_{max1}$, and 38.9% of $\Delta U_{int}$ (23.0% of $E_{abs}$) are emitted as shock wave. Over the entire bubble lifetime, 91.5% of the initial internal energy is carried away by shock waves emitted after optical breakdown and bubble collapse, and only 4.7% are lost by viscous damping. A small fraction of $\Delta U_{int}$ (3.8%) is dissipated by condensation of the vapour inside the bubble. This adds to the energy required for vaporization of the liquid in the plasma volume that is later released by condensation. We have seen in section 4.2.2 that most of the shock wave energy is also dissipated as heat in close vicinity to the optical breakdown site.

Thermal dissipation raises the temperature of the liquid around the oscillating residual bubble to a value significantly higher than room temperature, which explains why more than 100 after-bounces could be detected. The heated liquid shell around the bubble is relatively thick during the late oscillations when the bubble is small, and the temperature dependence of surface tension and viscosity thus plays a significant role. By contrast, during the first oscillations, when the bubble is large and the heated boundary layer is very thin, the bubble dynamics can be well described under the assumption that the bubble oscillates in water at room temperature.

The energy balance enables to track the energy stored upon collapse in the compressed bubble content and the surrounding liquid, and to discriminate between the energy components of the rebound shock wave originating from both sources. It turns out that the energy content of the compressed liquid, $E_{compr}^{coll}$, is 12.6 times higher than the inner energy of the bubble at $R_{min}$ (table 2). Therefore, most energy is radiated away acoustically upon rebound and only a small fraction originating from the internal energy of the compressed bubble content can contribute to bubble formation. As a consequence, the fraction of the rebound shock wave energy $E_{SW}^{reb}$ originating from the re-expansion of the compressed liquid ($E_\mathrm{SWL}$) is 18.45 times larger than the part provided by the rebounding bubble ($E_\mathrm{SWB}$).

During spherical laser-induced bubble oscillations in a homogeneous medium, most deposited energy is soon transformed into heat even though transiently a large percentage appears as mechanical energy of bubble and shock wave. However, in laser surgery, the mechanical energy of the bubble and shock waves will partly do mechanical work on the cellular and tissue structures surrounding the laser focus instead of being thermally dissipated. The possible mechanical destruction range is relatively small, especially for the shock wave effects that are related to the large particle velocity behind the shock front and to pressure gradients arising when the shock waves passes through regions with inhomogeneous acoustical properties (Vogel, Busch and Parlitz 1996; Lokhandwalla and Sturtevant 2001). Yet, also bubble-mediated effects are confined to a region smaller than $R_\mathrm{max1}$ if the material properties in the breakdown region are isotropic. The picture changes when asymmetric boundary conditions induce aspherical bubble dynamics and jetting, which can concentrate the mechanical energy of the bubble onto a small area apart from the laser focus (Vogel et al. 1989; Vogel et al. 1990; Brujan et al. 2001; Lauterborn and Kurz 2010, Barcikowski et al. 2019, Lechner et al. 2020, Reuter and Ohl 2021).

## 5. Discussion

The detailed characterisation of spherical cavitation bubble dynamics in this paper relies on a 'hybrid' approach in which numerical simulations are fitted to readily accessible experimental data on plasma size and bubble oscillation times. Valuable information on nonlinear energy deposition, phase transitions during bubble generation, and on heat and mass transfer during the bubble oscillations is obtained without an explicit, complex description of these processes. In section 5.1, we first compare our hybrid approach to previously published explicit models of bubble generation, and in 5.2 we discuss the origin of non-condensable gas in the bubble by plasma-mediated water dissociation. In 5.3, we then describe how the amount of non-condensable gas produced during breakdown influences the vigour of the bubble collapse and compare the extreme conditions produced by the collapse of laser-induced bubbles with the conditions created during acoustically driven single bubble sonoluminescence (SBSL). When the laser-induced bubble contains little non-condensable gas, its collapse is very violent and results in the emission of shock waves exhibiting a more rapid pressure decay than acoustic waves. In 5.4, this result is compared to the findings in previous studies on acoustic emission by collapsing and rebounding bubbles. Our approach can distinguish between energy fractions stored in the compressed content of the collapsed bubble and the compressed liquid surrounding the bubble, and track their conversion into bubble and shock wave energy during rebound. In section 5.5 we use this feature to elucidate the differences between energy partitioning after optical breakdown and bubble collapse, and compare our findings with the results of explicit approaches based on the solution of the Navier Stokes equations.

### 5.1. *Bubble generation*

In "ab initio" models of laser-induced cavitation, the modelling starts by describing the process of nonlinear energy deposition and subsequent phase transitions leading to bubble formation and continues with the bubble dynamics driven by the phase transition (Glinsky et al. 2001; Byun and Kwak 2004; Dagallier et al. 2017). Starting conditions are incident laser pulse energy, focus shape, and parameters describing the nonlinear absorption properties. Simulation results on energy deposition must then be appropriately linked to the input parameters needed for the simulation of bubble dynamics such as initial bubble radius, and the driving force for bubble expansion (bubble pressure or, in the framework of the Gilmore model, equilibrium radius). This approach provides insights into the mechanisms of energy deposition but imperfections in the complicated modelling of nonlinear energy deposition and in the transformation of the laser-induced temperature elevation into parameters of the bubble model may affect the accuracy of the final predictions on bubble dynamics.

In the present paper, the process of bubble generation is not explicitly modelled. Instead, the expansion of a virtual seed bubble is linked to two experimental parameters: the breakdown volume derived from photographs of plasma luminescence (or from the size of the breakdown region in shadowgraph photos), and the duration of the first bubble oscillation determined from probe beam scattering. The seed bubble radius $R_0$ is obtained by identifying its volume with the breakdown volume, and model predictions for $T_{osc1}$ are fitted to experimental values by varying $R_{nbd}$. This way, complete information about the amount of absorbed laser energy and the laser-induced bubble dynamics is available without a need for modelling nonlinear energy deposition and phase transitions. The absorbed laser energy is determined 'retrospectively' by analysing the $R(t)$ curve fitted to $T_{osc1}$. Information on the first bubble collapse and later phases of the bubble lifetime is available, if not only $T_{osc1}$ but also later oscillations are measured. The accuracy of the hybrid approach mainly depends on the precision with which the breakdown volume and bubble oscillation times can be determined. While $T_{osc}$ can be measured with high precision, determination of $R_0$ is more critical because it may be difficult to precisely delineate the borders of plasma or the breakdown region on photographs, the plasma may exhibit spatial inhomogeneities and hot spots, and an aspherical plasma shape that needs to be transformed into an equivalent spherical radius. As a consequence, the model predictions on laser induced bubble expansion and breakdown pressure, which rely on $R_0$ and $T_{osc1}$ data, bear more uncertainties than simulation results on bubble collapse and rebound that are derived from $T_{osc1}$ and $T_{osc2}$ data.

We extended the equation of motion of the Gilmore model by a term describing the initial jump of the bubble wall velocity to a large finite value equal to the particle velocity behind the plasma-induced shock front, which is due to liquid compressibility. The jump-start of the bubble wall was observed experimentally by Vogel et al. (1996) but at that time not yet included in the modelling of bubble generation. Figure 18 compares predicted $U(t)$ curves with and without jump start to experimental data from Vogel, Busch and Parlitz (1996). For all investigated laser parameters, consideration of the jump start provides much better agreement between experimental data and simulated $U(t)$ curves. It is interesting to note that for the bubble in figure 18 (d), where a 10-mJ, 6-ns pulse produced an average plasma energy density of 40 kJ/cm$^3$, both measured and predicted bubble wall velocity exceed the sound velocity in the liquid. In this case, the Keller-Miksis model that considers compressibility assuming a constant sound velocity in the liquid would not be able to describe the laser-induced bubble expansion and shock wave emission correctly.

In figure 18 (a), where a 30-ps, 50-µJ laser pulse produced a very small bubble, optical resolution did not suffice to resolve the velocity peak in the very early expansion phase. However, in (b) to (d) the velocity peak could be resolved and compared with model predictions. Measured peak velocities are larger than predicted values, even when the jump-start is considered. This is because laser pulses were only moderately focused at $NA = 0.25$, which resulted in elongated plasmas, especially at higher pulse energies. Under these conditions, the bubble wall velocity in the direction perpendicular to the long plasma axis is initially faster than for spherical symmetry. In a later stage of the bubble expansion, when the bubble acquires an approximately spherical shape, the bubble wall velocity is equal in all directions. In the transition phase, the measured velocity must be slower than for spherical expansion to compensate for the initial overshoot. This behaviour is indeed observed in all cases presented in Fig. 18: Experimental $U(t)$ data initially exceed the predicted values for spherical expansion, then drop below the simulated $U(t)$ curve, and finally both curves converge.

Figure 19 compares the evolution of bubble pressure and shock wave profiles with and without bubble wall jump-start for the 10-mJ, 6-ns pulse of figure 18 (*d*). The shock front forms within 8 ns in both cases but the jump start of the bubble wall results in a considerably lower maximum bubble pressure (4.75 GPa compared to 8.8 GPa without consideration of the particle velocity behind the shock front). The experimental value for the pressure at the bubble rim determined by measuring the shock front speed was 7.15 GPa (Vogel, Busch and Parlitz 1996). Simulations would yield this value with a start radius $R_0 = 30$ µm instead of $R_0 = 37$ µm, which provides the value of 4.75 GPa. As mentioned above, determination of $R_0$ from plasma photographs is critical, and hot regions within the luminescent plasma may produce higher peak pressures than the average values inferred from the total volume of the luminescent region. Independent measurements of other breakdown parameters such as $p_s(r)$ can support the choice of an $R_0$ value that provides the most realistic model predictions on the dynamics of shock wave emission and bubble expansion.

In our model, energy deposition during the laser pulse is implemented by taking the time evolution of the volume of the equilibrium bubble with radius $R_n$ as a measure for the deposited energy. This approach avoids explicit modelling of the interplay between pressure, temperature, and density in the growing bubble that would be required if bubble pressure instead of $R_n$ was employed as starting condition. It is valid as long as energy deposition is thermally confined and the initial bubble expansion can be treated as adiabatic process.

The assumption of isotropic bubble pressure is inevitable for a spherical bubble model in which one single parameter, $R_{nbd}$, is fitted such that predicted $T_{osc1}$ values match experimental values. Aberration-free focusing of ultrashort laser pulses at large *NA* produce compact plasmas with little elongation that approximate this assumption (see section 2.1 and figure 7). However, with decreasing *NA* and increasing laser pulse duration, the energy deposition becomes ever more non-isotropic, especially at energies well above breakdown threshold. Here, plasma formation starts at the beam waist, the breakdown front moves upstream towards the incoming beam while the laser power increases, and after the peak of the pulse incoming energy is deposited into plasma produced earlier (Docchio et al. 1988, Vogel et al. 1996). For ns pulses, this does not only result in a non-isotropic energy and pressure distribution but plasma generation still goes on in the upstream region while the hydrodynamic response has already started at the beam waist, as seen in figure 8 (a) and described by Vogel et al. (1996). Only ≈ 40 ns after the end of the laser pulse, a homogeneous pressure distribution has evolved in the breakdown region (section 4.1.3). Under those conditions, spherical bubble models yield only an approximate picture of the very early stage of laser-induced bubble formation. Ab-initio modelling of plasma formation and phase transitions would be needed to obtain realistic spatio-temporal pressure distribution driving bubble expansion. Future modelling approaches could combine spatiotemporal simulations of plasma formation (Arnold et al. 2007; Bulgakova et al. 2015) considering the specific band structure of water (Linz et al. 2015, Linz et al. 2016, Liang et al. 2019) with VOF models including liquid compressibility (Koch et al. 2016; Lechner et al. 2017). Experimental validation of such model predictions has recently become possible through pump-probe holographic imaging with fs free-electron laser pulses that provides the evolution of energy and pressure distribution during and immediately after laser-induced breakdown (Vassholz et al. 2021).

### 5.2. *Evolution of vapour and non-condensable gas content*

During laser-induced plasma formation, liquid water within the plasma volume is vaporized and partially dissociated into gaseous products. Atomic hydrogen and oxygen will largely recombine to form water but some molecular hydrogen and oxygen remain as long-lived gaseous products (Nikogosyan, Oraevsky and Rupasov 1983; Barmina, Simakin and Shafeev 2016, 2017). For luminescent plasmas, the energy density within the luminescent region by far exceeds the value needed for complete vaporization, and the vapour mass after the phase transition equals the liquid mass contained in the plasma volume of radius $R_0$. The situation becomes more complex close to the bubble threshold. Here, the conditions for bubble nucleation must be included into the analysis (Kiselev 1999; Vogel et al. 2005, Vanraes and Bogaerts 2018) and it must be considered that only part of the liquid volume in the plasma region will be vaporized. The model presented in this paper covers only the first scenario involving complete vaporization of the liquid within high-density plasma; scenarios involving partial vaporization require explicit modelling of the energy deposition process.

In high-density plasma, water dissociation can proceed through thermally driven hydrolysis (Mattsson and Desjarlais 2006, 2007) and by free-electron-mediated bond breaking. Liang & Zhang & Vogel (2019) showed that the average kinetic energy of free electrons (i.e. conduction band electrons) in luminescent plasmas in bulk water is 6.8 eV, and the high-energy tail of their energy spectrum reaches up to 14 eV. This energy is sufficient to break bonds by dissociative electron attachment (Cobut et al. 1996; Fedor et al. 2006; Ram et al. 2009). When conduction band electrons solvate, they go through a process of "hydration" until they are trapped in a relatively stable configuration of surrounding water molecules at an energy level of 6.5 eV above the valence band (Linz et al. 2015). The hydrated electrons may also contribute to dissociation and subsequent gas formation (Draganic & Draganic 1971; Nikogosyan, Oraevsky and Rupasov 1983; Elles et al. 2007) because their energy (≥ 6.5 eV) is larger than the O-H bonding energy of 5.1 eV (Toegel, Hilgenfeldt and Lohse 2002; Maksyutenko, Rizzo and Boyarkin 2006). During adiabatic expansion, most dissociation products recombine and form water again, while some form molecular hydrogen and oxygen through a chain of chemical reactions that remains in the bubble as non-condensable gas. Unlike for SBSL, where a sequence of many acoustically driven oscillations allows for rectified diffusion of dissolved air into the bubble (Brenner, Hilgenfeldt and Lohse 2002), diffusion of dissolved gas into a laser-induced cavitation bubble is negligibly small (Akhatov et al. 2001).

Free-electron-mediated gas generation dominates at lower temperatures while for $T \geq 3000$ K thermal dissociation becomes the dominant mechanism (Lédé, Lapique and Villermaux 1983; Mattsson and Desjarlais 2006). For $T > 4500$ K, most of the chemical bonds are broken, and only radicals of H and O are found (Lédé, Lapique and Villermaux 1983; Jung, Jang and You 2013). Correspondingly, the amount of non-condensable gas in laser-produced cavitation bubble increases with growing plasma temperatures (Sato et al. 2013).

In the case of plasma-mediated bubble generation that is portrayed in detail in the present paper, the plasma temperature is relatively low ($T_{avg}$ = 1550 K). Although it suffices to induce complete vaporization of the water within the plasma volume, it is so low that the bubble's non-condensable gas content must have been produced exclusively by free-electron-mediated water dissociation. For the same reason, the plasma luminescence seen in figure 5 cannot be blackbody radiation but must be Bremsstrahlung emitted from conduction band electrons before they recombine and thermalize. The emission of Bremsstrahlung resulting from the interaction of ultrashort laser pulses with dielectrics has been analysed by Petrov, Palastro and Penano (2017) for $SiO_2$, which has a bandgap of 9.5 eV, similar to that of water.

Non-condensable gas can leave the bubble only by dissolution, which takes much longer than the lifetime of the transient cavitation bubble and its afterbounces (Baffou et al. 2014). Measurements of the gas content at late stages of the bubble lifetime thus provides information on the dependence of water dissociation on plasma energy density.

The water vapour content changes rapidly during the bubble oscillation. Most of the vapour produced during optical breakdown condenses during the initial adiabatic expansion. However, during the subsequent isothermal expansion up to $R_{max}$, vapour again invades the bubble, and the vapour content of the maximally expanded bubble by far exceeds the amount of non-condensable gas (Storey and Szeri 2000). At collapse, vapour condenses out of the bubble again but during the final collapse phase, the time scale of the bubble wall movement becomes much faster than that of the water vapour transport out of the bubble, and some vapour is trapped inside the bubble. This process was modelled explicitly in several studies (Fujikawa and Akamatsu 1980; Yasui 1995; Storey and Szeri 2000; Toegel et al. 2000; Akhatov et al. 2001; Lauer et al. 2012, Zein, Hantke and Warnecke 2013). Unfortunately, the values of evaporation and condensation coefficients depend on pressure and temperature, and their values are still uncertain (Eames, Marr and Sabir 1997; Marek and Straub 2001). Therefore, Akhatov et al. (2001) used it as a free parameter for fitting simulation results to the experimentally observed bubble rebound, which is actually a way to determine its value. Our approach of using $R_n$ as fit parameter in conjunction with Eqs. (3.52) to (3.55) enables to determine the amount of vapour condensing during the bubble's expansion, collapse and rebound. Results are in good agreement with the findings of Akhatov et al. (2001).

### 5.3. *Extreme conditions during generation and collapse of laser-induced bubbles*

Optical breakdown by tightly focused laser pulses produces high-density plasmas with average volumetric energy densities reaching from a few $kJ/cm^3$ to about 40 $kJ/cm^3$, depending on laser pulse duration, wavelength, focusing angle, and pulse energy (Vogel et al. 1996). In this paper, we investigated the cavitation events produced by a weakly luminescent plasma with relatively small temperature and small non-condensable gas content that exhibits a particularly vigorous bubble collapse. The bubble collapse concentrates the potential energy of the expanded bubble at $R_{max1}$ into the much smaller volume of the bubble at $R_{min1}$ and the compressed liquid in the bubble's immediate vicinity. For the vigorously collapsing bubble investigated in this paper, the volume ratio $(R_{max1}/R_{min1})^3$ is $6.28 \times 10^5$. This energy concentration produces a peak pressure of 13.5 GPa and a temperature of 31400 K. Under these conditions, plasma forms in the collapsed bubble (Brenner, Hilgenfeldt and Lohse 2002; Mattsson and Desjarlais 2006), which is consistent with the observation of a brightly luminescent flash such as reported by Ohl et al. (1999) and Baghdassarian et al. (1999, 2001).

The vigour of the first bubble collapse depends on the ratio $R_{max1}/R_{nc1}$ (which is 9.76 for the bubble of figure 9) and is reflected in the ratio of the maximum radii of first and second oscillation, $R_{max1}/R_{max2}$. When the collapse pressure is very high, much energy is carried away by shock wave emission and little energy remains for the rebounding bubble, which results in a large $R_{max1}/R_{max2}$ ratio. Figure 20 compares ratios from previous publications with values obtained in the present study. In previous studies with larger bubbles, buoyancy effects distorted the spherical bubble shape, and the $R_{max1}/R_{max2}$ ratios were significantly smaller than in the present paper. For large spherical laser-induced bubbles, where surface tension and viscosity can be ignored, the vigour of collapse is mainly determined by the amount of permanent gas produced during breakdown and the amount of water vapour trapped in the collapsing bubble that both dampen the collapse (Toegel et al. 2000). The relative importance of water vapour increases for larger bubbles because the amount of water contained in the expanded bubble scales with $R_{max}^3$, while the surface area through which condensing vapour can escape during collapse scales proportional to $R_{max}^2$. The parameter combination explored in the present study (small bubble and relatively low plasma energy density) results in a comparatively strong collapse.

How does the collapse of laser-induced bubbles compare to SBSL bubbles? The dynamics of SBSL bubbles strongly depends on the frequency, $f$, and pressure, $p_a$, of the driving sound field that are restricted by stability requirements for the bubble oscillation (Brenner, Hilgenfeldt and Lohse 2002). With excitation in the kHz range, the most vigorous dynamics occurs around $f \approx 16$ kHz (Toegel et al. 2000) and $p_a \approx 0.145$ MPa (Matula 1999; Toegel and Lohse 2003). Matula (1999) predicted a ratio $R_{max}/R_{nc} = 9.94$ for a SBSL bubble driven at $f \approx 25$ kHz and $p_a \approx 0.142$ MPa. Since a similar ratio was observed in the present study for a laser-induced bubble, similar collapse pressures are expected in both cases. Streak-photographic measurements of SBSL collapse pressure support this conclusion. Pecha and Gompf (2000) found a shock wave velocity of $u_s = 4000$ m/s at $f \approx 20$ kHz and $p_a \approx 0.139$ MPa, and Weninger, Evans and Putterman (2000) reported $u_s = 5930$ m/s (Mach 4) for $f \approx 16.5$ kHz and $p_a \approx 0.145$ MPa. An evaluation of these $u_s$ data using the Hugoniot data of Rice and Walsh (1957) yields $p_s = 5.3$ GPa and $p_s = 15.4$ GPa, respectively, close to the collapse pressure of 13.5 GPa obtained in the present paper. A more vigorous SBSL collapse dynamics can be achieved only through larger driving pressures, which can be realized with smaller bubbles driven at very high frequencies (Camara, Putterman and Kirilov 2004). Also for laser-induced bubbles, the collapse pressure will increase with decreasing $R_{max}$, when the pressure exerted by surface tension becomes significant. The data in figure 20 show a strong increase of $R_{max1}/R_{max2}$ for $R_{max} \leq 6$ μm, indicative for high collapse pressure leading to strong energy dissipation by shock wave emission.

In the context of SBSL, researchers pointed out that the wall of the collapsing bubble can launch an internal shock wave when its velocity exceeds the sound velocity inside the bubble (Roberts and Wu 1996, Lin and Szeri 2001, Brenne, Hilgenfeldt and Lohse 2002). It was postulated that the geometrical focusing of the internal shock wave could produce a tiny spot in the bubble centre with strongly elevated pressure and temperatures up to $10^6$ K (Wu and Roberts 1993). However, Geers, Lagumbay and Vasilyev (2012) showed that internal wave effects play no role for relatively weak bubble collapse with $R_{max}/R_n \leq 4$, where a sharp boundary between liquid and gas exists during collapse. Storey and Szeri (2000) demonstrated for SBSL bubbles with larger $R_{max}/R_n$ ratio that endothermic reactions of water vapour trapped in the collapsing bubble would significantly reduce the peak temperature and that the hot spot – if present – would involve only 0.1% of the volume of the collapsed bubble. An equilibrating factor for the pressure distribution within the bubble is the rapid increase of sound velocity upon compression of the bubble content. Its density exceeds the liquid density both for collapsing gas bubbles (Yuan et al. 2001) and vapour bubbles, and the sound velocity thus assumes much higher values than under ambient conditions. This effect is further enhanced by the rise of sound speed with increasing temperature (Vuong, Szeri and Young 1999). We showed that laser-induced bubbles contain mainly water vapour during first collapse and that less than one third of their content is non-condensable gas (section 4.2.3). Therefore, it seems appropriate to use EOS data for water to analyse the sound speed at collapse. Based on the EOS data by Rice and Walsh (1957) and the collapse pressure value of 13.6 GPa from figure 9, we estimate that the sound velocity in the collapsing bubble reaches a value of 5400 m/s at $R = R_{min1}$. This value is much larger than the peak bubble wall velocity of 1793 m/s, which impedes the formation of an inner shock wave. Lack of an inner shock wave justifies the assumption of a homogeneous bubble pressure made in the Gilmore model. Lin, Storey and Szeri (2002) showed, furthermore, that even when spatial non-uniformities of the pressure within the bubble are present, they have little influence on the accuracy of $R(t)$ solutions of uniform-pressure assuming Rayleigh-Plesset-type equations considering liquid compressibility.

Since the vigour of bubble collapse is similar for SBSL bubbles driven by strong sound fields and laser-induced bubbles with little permanent gas content, the peak pressures and temperatures reached upon collapse will be similar. Their exact values depend on the composition of the bubble content and are not just given by its adiabatic compression but also influenced by water dissociation and a multitude of endothermic chemical reactions. Studies considering these factors predicted temperatures between 10000 and 15000 K in accordance with luminescence spectra (Lin, Storey and Szeri 2002; Toegel and Lohse 2003), somewhat lower than the value of 31530 K obtained through our modelling approach that neglects water dissociation and chemical reactions.

Numerical modelling usually assumes that a boundary between two regions with different EOS representing liquid and gas exists during the entire bubble oscillations. This is justified for bubbles containing only non-condensable gas but questionable for vapour bubbles, where the phase boundary between bubble content and surrounding liquid vanishes during the late collapse phase and reappears only when pressure and temperature drop below the critical point. In reality, the situation is even more complex both for SBSL bubbles and laser-induced bubbles because they contain both gas and vapour, whereby the vapour content of the expanded bubble is in both cases much larger than the gas content (Storey and Szeri 1999; Brenner, Hilgenfeldt and Lohse 2002). Upon first collapse, about 1/6 of the molecules in SBSL bubbles are water molecules (Storey and Szeri 1999), whereas we found a water vapour content of about 2/3 for the laser-induced bubble of figure 9. Hilgenfeldt, Grossmann and Lohse (1999) introduced a time-varying adiabatic exponent to account for the change from isothermal to adiabatic conditions during collapse and the changing composition of the bubble content. This approach worked well for SBSL bubbles but the use of a gas EOS seems to be more questionable for the collapse phase of a bubble containing mainly water vapour, where the phase boundary vanishes during the collapse phase. Nevertheless, it is commonly used for practical reasons. In future, the temperature dependence of surface tension and viscosity should be considered for the time when the phase boundary exists and a more advanced EOS incorporated for the final collapse phase. First attempts have been made by combining the IAPWS-95 EOS for water (Wagner and Pruß 2002) with a finite element model of bubble formation around gold nanoparticles that is linked to the Gilmore model to cover the later phases of bubble dynamics (Dagallier et al. 2017).

The pressure jump at the shock front is often so high that energy dissipation causes a temperature rise beyond the spinodal limit. This extends the region in which vaporization occurs after breakdown and upon rebound in a more effective way than heat conduction does. For the bubble's rebound phase, shock wave induced phase transitions have not yet been captured by time-resolved shadow or Schlieren photographs. However, in figure 8 it is visualized during the expansion of a high-density plasma produced by an energetic laser pulse. After plasma formation, reproducible timing of photographs is easier than for the rebound phase immediately after collapse, and the size of the affected region is large enough to be resolved by optical imaging. When the energy density in the liquid produced by heat dissipation at the shock front exceeds the spinodal limit, the liquid becomes instable and starts to expand, which results in a refractive index change visible on the photographs. The outer boundary of the superheated liquid region is influenced by local fluctuations of the plasma energy density and becomes visible as rugged appearance of the interface between superheated and colder liquid in figure 8 (a). During the rebound of a collapsed spherical bubble, similar processes will occur. For example, in the case shown in figure 14, a shock front exhibiting a maximum pressure jump of ≈ 8 GPa develops, which results in a temperature jump to 436°C (Rice and Walsh 1957) that will produce a phase transition. However, the interface around a spherically rebounding bubble is smooth, and the spatial extent of the superheated region is at or below the optical resolution limit. This makes its visualization much harder than in the vicinity of spatially inhomogeneous plasmas.

Modelling approaches for heat and mass transfer by evaporation and condensation at the bubble wall and heat conduction are already available. However, in the initial phase after laser-induced breakdown and of the bubble's rebound after collapse, the "convective" heat transport by the shock wave becomes more important than heat conduction from the bubble interior because a large amount of energy is dissipated at the shock front within a few micrometre propagation distance. To the best of our knowledge, convective transport by acoustic radiation and energy dissipation at the shock front has not yet been considered in any bubble model. Its inclusion remains a challenge for the future.

### 5.4. *Pressure decay during shock wave propagation*

The first studies on acoustic emission during spherical bubble collapse by Hickling and Plesset (1964) and Akulichev (1968, 1971), and the simulations by Fujikawa and Akamatsu (1980) were based on assumptions about the bubble's gas content upon collapse. Hickling and Plesset presented a full decay curve only for a case where the effect of the geometrical decay of the spherical pressure transient was stronger than the steepening of its front by the nonlinearity of sound propagation but did not continue the calculations for a second case with larger collapse pressure, in which a shock front would have formed. Such calculations were first presented by Akulichev (1968), who applied Landau & Lifschitz's method for determining the shock front position that is also adopted in the present paper. Ebeling (1978) extended the modelling to the shock wave emission upon laser-induced bubble generation and used different values for the equilibrium radius $R_n$ during bubble generation and rebound that were obtained by fitting $R(t)$ curves to the experimentally observed bubble dynamics. Vogel et al. (1996) followed the same fitting approach for simulating the shock wave emission during laser-induced bubble generation, and Akhatov et al. (2001) and Koch et al. (2016) used it for simulating the bubble pressure during its collapse. However, Akhatov et al. (2001) presented only a few $p(r)$ curves and undertook no systematic study of the pressure decay, and Koch et al. (2016) presented $p(r)$ curves only for the conditions simulated previously by Hickling and Plesset (1964). Thus, the present paper presents the first evidence-based simulation of shock wave formation and pressure decay after a vigorous collapse of spherical laser-induced cavitation bubbles. Our results differ strongly from earlier results for lower collapse pressure, which yielded a pressure decay resembling that of acoustic transients with $p_{peak} \propto r^{-1}$ (Hickling and Plesset 1964, Fujikawa and Akamatsu 1980, Koch et al. 2016). We observe a rapid formation of shock waves as previously reported by Akulichev (1968) and Ebeling (1978), together with a pressure decay as fast as $p_{peak} \propto r^{-1.75}$.

A similar decay rate, with a maximum slope in the log-log plot of -1.79, has been obtained in previous simulations for the breakdown shock wave produced by a 10 mJ ns laser pulse that was emitted from high-density plasma with 8.8 GPa initial pressure (Vogel, Busch and Parlitz 1996). Lai et al. (2021) investigated shock wave emission after breakdown with 12-ns, 1064 nm pulses of 22 – 80 mJ pulse energy and obtained slopes around -2.0. It is interesting to note that the experimentally observed pressure decay measured by Vogel, Busch and Parlitz (1996) in the direction perpendicular to the laser beam axis was initially weaker and then faster (with a slope up to -2.4) than the decay predicted by the simulations for spherical bubble dynamics. This discrepancy is likely caused by the elongated plasma shape in the experiments. Here, the shock wave emission in the near field had a cylindrical component, and the pressure decay in the direction perpendicular to the cylinder axis was initially slower than for spherical shock waves (Schoeffmann, Schmidt Kloiber and Reichel 1987). However, the shock wave propagation in the far field exhibits radial symmetry even when the breakdown region is elongated (Tagawa et al. 2016). Therefore, the pressure decay must be faster in the transition zone between near and far field to make up for the slower near-field decay. A similar phenomenon was already observed in figure 18 for the velocity of bubble expansion around elongated plasmas, which in the direction perpendicular to the long plasma axis is first faster and then slower than in the case of spherical symmetry. Future experiments with laser-induced bubbles exhibiting minimum deviations from spherical shape will enable a more precise comparison with numerical simulations than possible to date.

### 5.5. *Energy partitioning*

In this paper, a more detailed energy balance than in previous studies (Vogel et al. 1999; Tinguely et al. 2013; Linz et al. 2020) could be established based on the combination of experimental input ($R_0$, $T_{osc}$) and modelling. The present approach cannot only quantify the total amount of shock wave energy emitted after breakdown and collapse but also distinguish between fractions of the collapse shock wave originating from energy stored in the compressed content of the collapsed bubble and from the compressed liquid surrounding it. This way we find that energy partitioning after optical breakdown and bubble collapse is very different, although in both cases strong shock waves are emitted (Table 2). During laser-induced bubble formation with moderate plasma energy density, 59.2% of the absorbed energy were transformed into mechanical energy, and from this fraction 55.4% went into bubble energy, and 38.9% into shock wave emission. By contrast, during the rebound after the first bubble collapse, less than 2% were transformed into bubble energy, and 96.4% of the energy stored in the compressed bubble content and liquid are radiated away acoustically. During breakdown, the laser energy is stored inside the plasma, and a compression wave affects the surrounding liquid only after the plasma has started to expand (figure 12). Under these conditions, a relatively large fraction of the plasma energy can be converted into bubble energy. By contrast, during collapse, both the bubble content and the surrounding liquid are compressed (figure 14), with the energy content of the compressed liquid being much larger than that of the bubble (Table 2). Therefore, most energy is radiated away acoustically upon rebound and only a small fraction originating from the internal energy of the compressed bubble content can contribute to bubble formation. For the bubble investigated in detail in the present paper, the ratio of the energy contributions from compressed liquid and bubble content was $E_{SWL}/E_{SWB} = 18.45$.

These findings agree with the picture on acoustic emission after bubble collapse that were obtained with a full models based on solutions of the Navier-Stokes equations (Fuster, Dopazo and Hauke 2011). A quantitative comparison between their results and the present results is difficult because the collapse pressures investigated differ. The ratio $E_{SWL}/E_{SWB}$ will increase with decreasing gas content of the collapsing bubble, when less internal energy is stored in the bubble itself and more in the surrounding liquid. This goes along with an increasing collapse pressure, stronger acoustic emission from the liquid surrounding the bubble, and a smaller rebound bubble. In our case, the collapse pressure was 13.5 GPa, whereas the largest collapse pressure investigated by Fuster, Dopazo and Hauke (2011) was only 1.5 GPa according to the Gilmore model, and 2.6 GPa according to their full model. Therefore, $E_{SWL}/E_{SWB}$ is particularly high in the present paper.

With increasing plasma energy density, the fraction $E_v$ required for vaporization of the plasma volume decreases and an ever-larger percentage of the absorbed energy is converted into mechanical energy (Vogel et al. 1999). Furthermore, an ever-larger part of the mechanical energy appears as shock wave energy (Lai et al. 2021). For a 6-ns, 10-mJ pulse with $\varepsilon = 40$ kJ/cm$^3$, the shock wave energy was found to be more than two times larger than the bubble energy (Vogel et al. 1999), different from the case of figure 9 in the present paper, where for $\varepsilon = 8.7$ kJ/cm$^3$ the shock wave energy is smaller than the bubble energy. This finding warrants future detailed investigations of the dependence of energy partitioning on plasma energy density.

In the case of spherical bubble dynamics investigated here, 91.5% of the initial internal energy of the bubble is carried away by shock waves emitted after optical breakdown and bubble collapse, and only 4.7% are lost by viscous damping. The picture changes for non-spherical dynamics near solid or free boundaries, where due to jet formation during collapse shock wave emission is reduced and a large fraction of the bubble energy converted into the kinetic energy of a vortex flow that is finally dissipated through viscous damping (Vogel and Lauterborn 1988, Tian et al. 2020).

Figure 20 shows that for very small bubble sizes, the $R_{max1}/R_{max2}$ ratio increases strongly with decreasing bubble size. The increase is most likely related to the increasing role of surface tension and viscosity with decreasing bubble size as indicated by the $1/R$ proportionality of the last two terms of Eq. (3.3). This conclusion is supported by a theoretical analysis of mechanical driving forces involved in the oscillation of laser-induced nanobubbles around nanoparticles (Lombard, Biben and Merabia 2015). The increase of the pressure arising from surface tension enhances the vigour of the collapse while, at the same time, an ever larger part of the deposited energy is dissipated by viscous damping. The tools presented in this paper will enable a detailed investigation of changes in bubble dynamics and energy partitioning for $R_{max} \to 0$.

## 6. Conclusions

We established a hybrid experimental/simulation approach for providing a rapid and comprehensive characterization of laser-induced bubble oscillations and shock wave emission. The experimental part consists of a photographic characterization of the size of the laser-induced plasma and a single-shot probe beam scattering method for recording the bubble oscillation times with high temporal resolution. The excellent time resolution, large dynamic range and high sensitivity of the method enables to cover the entire bubble lifetime from early large-amplitude nonlinear oscillations to late oscillations of the residual gas bubble with sub-nanometre amplitude.

Simulations are performed based on the Gilmore model with a van der Waals hard core that has been extended by a description of laser-induced bubble formation, automated determination of the shock front location for pressure transients with very large amplitude, and an energy balance encompassing the entire bubble life time. The results of the experiments and calculations complement each other and yield a rich and detailed picture of the events during spherical laser-induced cavitation bubble oscillations.

Like SBSL, laser-induced bubble dynamics is a microlaboratory for high-pressure/density/temperature hydrodynamics, plasma physics, and chemistry that enables to study a large number of nonlinear phenomena and extreme states in a tabletop environment. Laser-induced bubbles offer the option to study both spherical and aspherical bubble dynamics under controlled conditions and are of great practical importance in biophotonics and biomedicine as well as in laser ablation in liquids. Their investigation will continue to provide fruitful insights if modelling and experimental tools are further advanced. However, the challenges for experimental coverage with high spatial and temporal resolution are extremely high because the collapse times show much larger shot-to shot fluctuations than the oscillation times in SBSL. Modelling is also very demanding because of the simultaneous occurrence of a multitude of nonlinear phenomena and the additional challenges posed by aspherical dynamics. Volume of Fluid methods are a versatile tool providing new insights. However, for a better understanding of spherical bubble dynamics in the context of biomedical applications, it is important to study its dependence on laser parameters and properties of the breakdown medium. The combination of relative simplicity and large information content of our spherical bubble model combined with readily available experimental data as starting conditions makes our hybrid approach a useful tool for evidence-based investigations of parameter dependencies. Here we demonstrated the large potential of this approach on one example, where the bubble dynamics could be traced through more than 100 oscillations. The next step is a detailed investigation of the changes in bubble dynamics and energy partitioning for $R_{\max} \to 0$ using the tools presented in this paper. This will be the subject of a future publication.

**Funding**

This work was supported by U.S. Air Force Office of Scientific Research (X.-X.L, A.V., grant numbers FA9550-15-1-0326, FA9550-18-1-0521).

**Declaration of interest**

The authors report no conflict of interest.

**Author ORCID**

X.-X. Liang, https://orcid.org/0000-0002-8325-1627;
N. Linz, https://orcid.org/0000-0002-7843-5966;
S. Freidank, https://orcid.org/0000-0002-5453-269X;
G. Paltauf, https://orcid.org/0000-0002-5431-430X;
A. Vogel, https://orcid.org/0000-0002-4371-9037

**Liang et al. / Table 1**

| Instant | Time (ns) [Oscillation period $T_{osci}$ (ns)] | Bubble radius $R$ (μm) | Equilibrium radius $R_n$ (μm) | Vapour bubble radius $R_v$ (μm) | Vapour mass $m_v$ ($10^{-18}$ kg) |
|---|---|---|---|---|---|
| After Plasma formation | $0.53 \times 10^{-3}$ | 1.33 | 13.718 | 14.56 | 9835 |
| $R_{max1}$ | 3244 | 35.88 | | 10.25 | 3430 |
| 1st collapse | 6488.0 [6488.0] | - | 3.615 | 3.21 | 106 |
| $R_{max2}$ | 7268.8 | 9.10 | | 2.60 | 56 |
| 2nd collapse | 8049.6 [1561.6] | - | 2.415 | - | - |
| 3rd collapse | 9014.4 [964.8] | - | 2.415 | - | - |

TABLE 1. Characteristic breakdown and bubble parameters corresponding to the probe beam signal of figure 9 that were used in the simulations yielding the results presented in figures 10 – 17. The initial bubble radius $R_0$ matches up with the plasma size at a laser pulse energy $E_L = 155$ nJ and is taken from figure 5(c). The times for bubble generation and collapse are read from the signal, and were used to determine the oscillation periods $T_{osci}$ and the times corresponding to $R_{maxi}$, assuming that maximum expansion is reached at the middle of each oscillation period. The equilibrium radii $R_n$ are determined by fitting the predicted $R(t)$ curve to the measured oscillation periods. Specifically, $R_{nbd}$ is used for fitting $T_{osc1}$, $R_{nc1}$ for fitting $T_{osc2}$, and $R_{nc2}$ for $T_{osc3}$. Afterwards, $R_n$ is kept constant for the rest of the calculation. The vapour bubble radius after plasma formation corresponds to the vaporized liquid volume with radius $R_0$ and is given by Eq. (3.52). At $R_{maxi}$, it represents the amount of vapour contained in the expanded bubble at the vapour pressure under ambient condition, $p_v = 2.33$ kPa, and is determined using Eq. (3.53). At first collapse, it is calculated as $R_{vc1} = (R_{nc1}^3 - R_{nc2}^3)^{1/3}$, assuming that the bubble content at the second collapse consists only of non-condensable gas. The respective values for the vapour mass $m_v$ were calculated with $\rho_v = 0.761$ kg/m$^3$.

**Liang et al. / Table 2**

| Bubble expansion | | | Collapse | | | Rebound | | | Afterbounces | | | Entire bubble life | | |
|---|---|---|---|---|---|---|---|---|---|---|---|---|---|---|
| Partition | Energy [nJ] | Fraction [%] | Partition | Energy [nJ] | Fraction [%] | Partition | Energy [nJ] | Fraction [%] | Partition | Energy [nJ] | Fraction [%] | Partition | Energy [nJ] | Fraction [%] |
| $E_{abs}$ | 62.49 | 100 | $E_B^{max1}$ | 30.75 | 100 | $E_{compr}^{total}$ | | 100 | $E_B^{max2}$ | 0.549 | 100 | $E_{abs}$ | 62.49 | 100 |
| $E_{cond}^{exp}$ | 16.61 | 26.59 | $E_{cond}^{coll1}$ | 9.96 | 32.38 | $E_{cond}^{reb}$ | 0.14 | 0.70 | $E_{cond}^{coll2}$ | 0.159 | 29.03 | $E_{SW} + E_{acoust}$ | 33.87 | 54.2 |
| $(\Delta E_v^{exp})$ | (16.59) | (26.56) | $(\Delta E_v^{coll1})$ | (8.61) | (28.01) | $(\Delta E_v^{reb})$ | (0.13) | (0.65) | $(\Delta E_v^{coll2})$ | (0.145) | (26.39) | $E_{cond}$ | 26.87 | 43.0 |
| $(\Delta U_{int,cond}^{exp})$ | (0.02) | (0.03) | $(\Delta U_{int,cond}^{coll1})$ | (1.35) | (4.37) | $(\Delta U_{int,cond}^{reb})$ | (0.01) | (0.05) | $(\Delta U_{int,cond}^{coll2})$ | (0.014) | (2.64) | $W_{visc}$ | 1.72 | 2.76 |
| $E_{SW}^{bd}$ | 14.38 | 23.02 | $W_{visc}$ | 0.81 | 2.65 | $E_{SW}^{reb}$ | 19.24 | 96.30 | $E_{acoust}$ | 0.253 | 46.08 | $U_{int}^{res}$ | 0.02 | 0.04 |
| $W_{visc}$ | 0.74 | 1.19 | | | | $(E_{SWL})$ | (18.24) | (91.32) | $W_{visc}$ | 0.114 | 20.85 | | | |
| | | | | | | $(E_{SWB})$ | (1.0) | (4.98) | $U_{int}^{res}$ | 0.022 | 4.04 | | | |
| | | | | | | $W_{visc}$ | 0.05 | 0.26 | | | | | | |
| **$R_{max1}$** | | | **$R_{min1}$** | | | **$R_{max2}$** | | | | | | | | |
| $E_v^{max1}$ | 8.89 | 14.22 | $E_v^{coll1}$ | 0.27 | 0.89 | $E_v^{max2}$ | 0.14 | 0.72 | | | | | | |
| $U_{int}^{max1}$ | 1.35 | 2.15 | $U_{int}^{min1}$ | 1.46 | 4.75 | $U_{int}^{max2}$ | 0.02 | 0.08 | | | | | | |
| $E_{pot}^{max1}$ | 20.51 | 32.83 | $E_{compr}^{coll1}$ | 18.24 | 59.33 | $E_{pot}^{max2}$ | 0.39 | 1.94 | | | | | | |
| $(W_{stat})$ | (19.34) | (30.95) | | | | $(W_{stat})$ | (0.32) | (1.57) | | | | | | |
| $(W_{surf})$ | (1.17) | (1.88) | | | | $(W_{surf})$ | (0.07) | (0.37) | | | | | | |
| $E_B^{max1} = U_{int}^{max1} + E_{pot}^{max1} + E_v^{max1}$ | | | $E_{compr}^{total} = E_{compr}^{coll1} + U_{int}^{min1} + E_v^{coll1}$ | | | $E_B^{max2} = U_{int}^{max2} + E_{pot}^{max2} + E_v^{max2}$ | | | | | | | | |

TABLE 2 Energy TABLE 2 Energy balance for different time periods and instants during laser-induced bubble oscillations. Laser and bubble parameters are the same as in figures 10-17. The table lists all energy values contained in the flow diagram of figure 4. Starting point is the absorbed laser energy, which at the end of the laser pulse partitions into $E_{abs} = E_v + \Delta U_{int}$, with $E_v$ = 25.49 nJ (40.8 %), and $\Delta U_{int}$ = 37.0 nJ (59.2%). The first column shows how $E_{abs}$ set as 100% splits into various parts, whereby the fractions dissipated during expansion are listed in the upper part of the column, and the parts remaining at $R = R_{max1}$ are given in the lower part. The energy of the expanded bubble, $E_B^{max1}$, is the starting point for partitioning in the collapse phase, the energy $E_{compr}^{total}$ of the compressed bubble content and liquid at $R = R_{min1}$ is the reference point for the rebound phase, and $E_B^{max2}$ for the afterbounces. Terms in brackets denote subfractions of the energy parts listed above them.

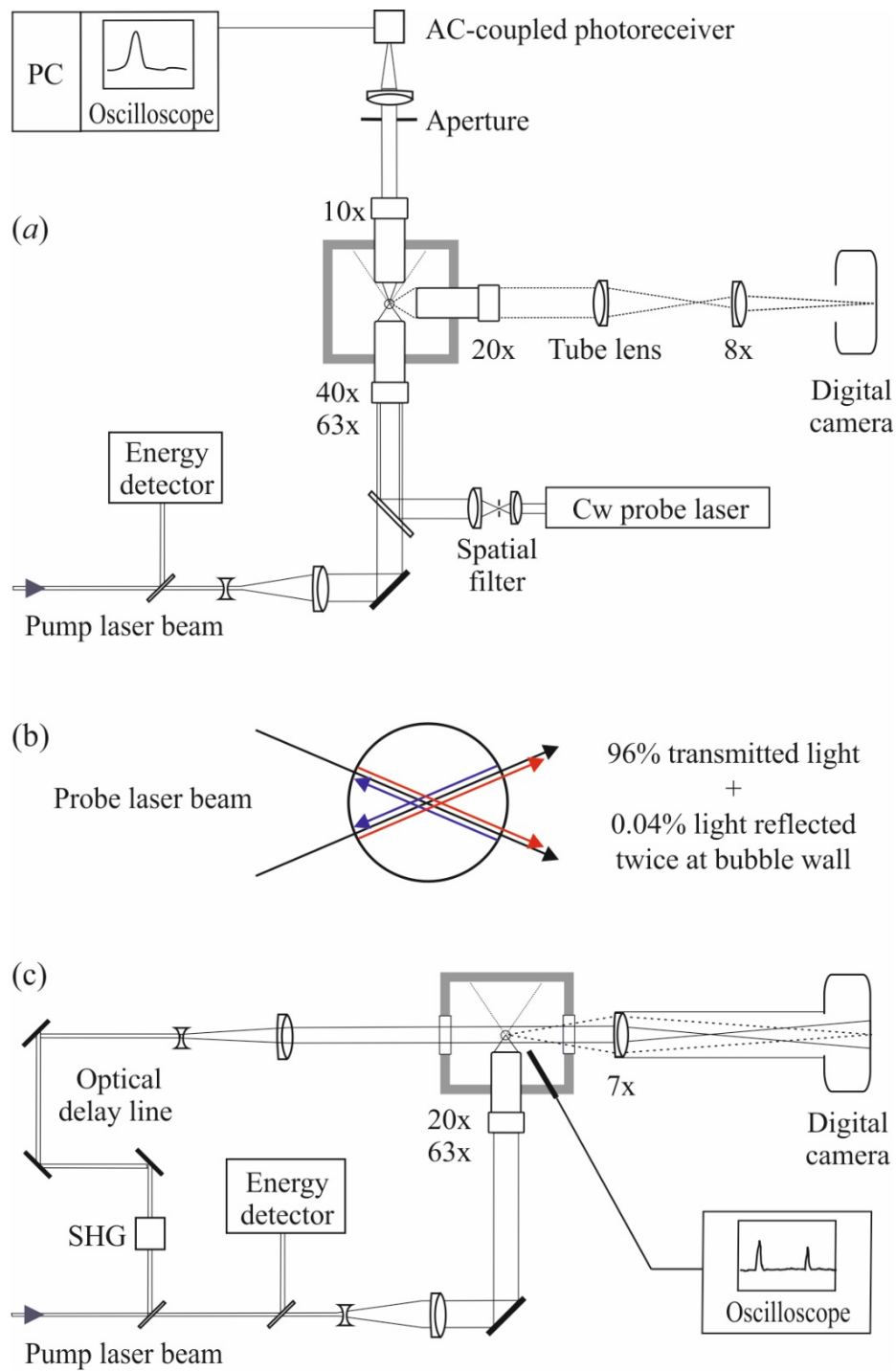

FIGURE 1. Experimental arrangements for the investigation of the behaviour of spherical laser-induced bubbles in water. (a) Setup with confocal adjustment of three microscope objectives enabling to generate highly spherical bubbles, record their oscillations with a cw probe laser beam, and take high-resolution images of plasma luminescence. (b) Illustration of the directly transmitted and multiply reflected parts of the probe laser beam that interfere behind large bubbles. Light scattering and interference is detected by the AC-coupled photoreceiver in (a) and recorded using a digital oscilloscope. (c) Setup for time-resolved photography of the breakdown dynamics at different $NA$s. At early times up to $t = 120$ ns, an optically delayed frequency-doubled portion of the pump laser pulse is used for illumination, and at later times a 20-ns plasma flash. Hydrophone signals of breakdown and collapse shock waves are recorded to monitor the bubble oscillation time for each shot.

**Liang et al. / Figure 2**

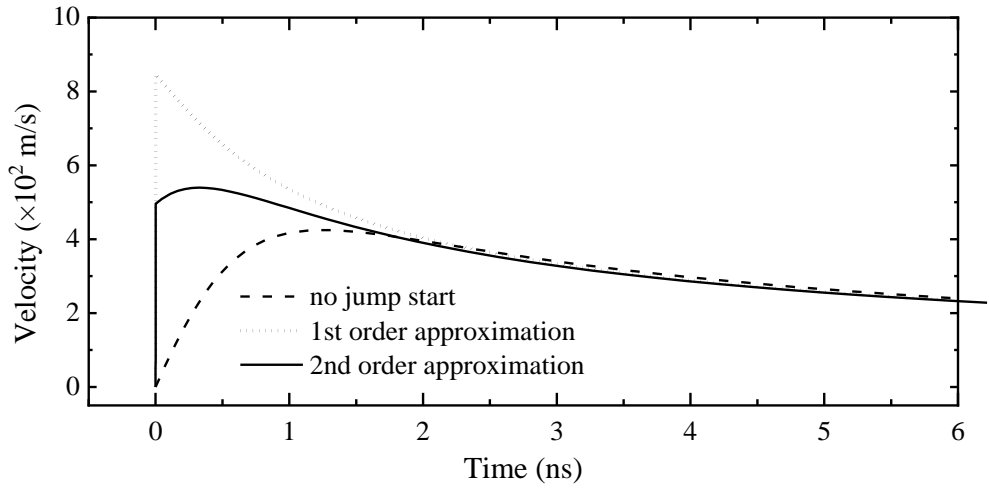

FIGURE 2. Time evolution of bubble wall velocity during the early expansion phase for three modelling approaches: 1. Particle velocity behind the shock wave front is not considered (no jump-start); 2. The first order approximation of $\dot{u}_p$ in Eq. (3.23) is used to consider the evolution of particle velocity during the laser pulse; 3. The second order approximation in Eq. (3.30) is used. For all simulations, the input parameters are $R_0 = 1.33$ μm and $R_{nbd} = 13.88$ μm, which correspond to the signal in figure 9 that will later be later analysed in detail.

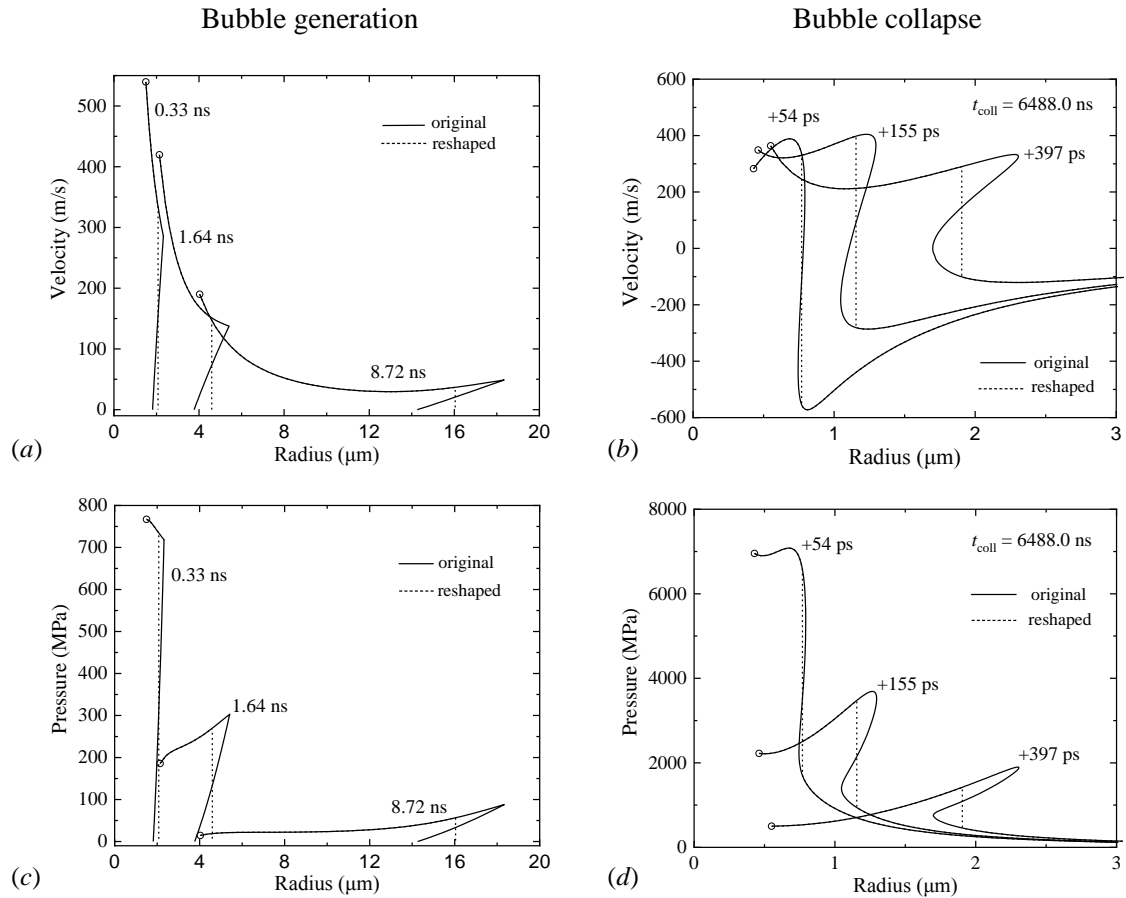

FIGURE 3. Determination of the shock front position illustrated on velocity distributions $u(r)$ and pressure distributions $p(r)$ for acoustic transients emitted after optical breakdown [(*a*) and (*c*)] and during rebound after the first bubble collapse [(*b*) and (*d*)]. For this illustration, the same parameters were used as in figures 12 to 14 below. The vertical lines in the $u(r)$ plots cut off the same area from the ambiguous part of the curve as that added below the curve. The location of the front is then transferred to the $p(r)$ plots and defines the shock front position.

**Liang et al. / Figure 4**

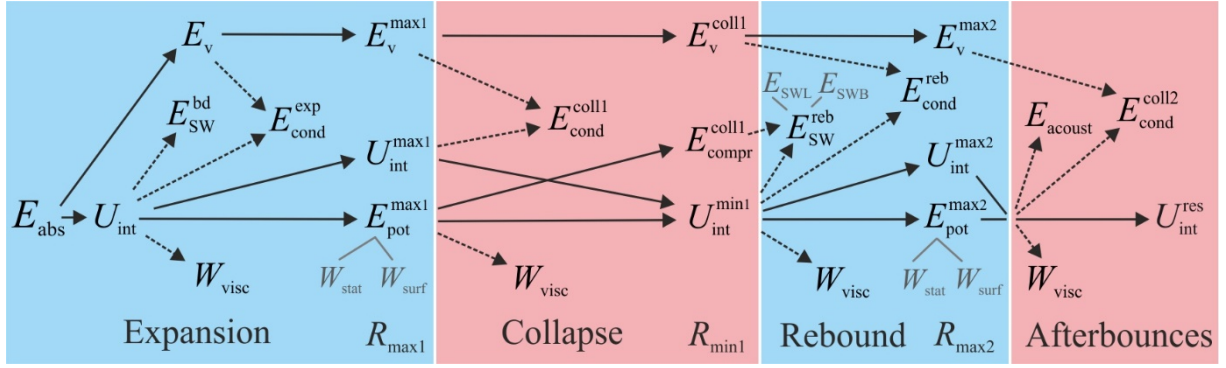

FIGURE 4. Energy partitioning pathways for laser-induced cavitation bubbles separated into four phases: bubble expansion, first collapse, first rebound, and afterbounces. The respective phases are indicated by the superscripts exp, coll, reb, and res in the symbols denoting the energy fractions. Solid arrows indicate the conversion of energy fractions that enter the next bubble oscillation phase, which include the vaporization energy $E_v$, the internal energy $U_{int}$ of the bubble content, and the potential energy $E_{pot}$ of the expanded or collapsed bubble. The potential energy is, at each stage, given by the work done against (or by) hydrostatic pressure, $W_{stat}$, plus the work done against (or by) surface tension, $W_{surf}$. The dashed arrows represent energy dissipation in each phase via viscous damping, $W_{visc}$, vapour condensation, $E_{cond}$, and shock wave emission, $E_{SW}$. The rebound shock wave emission is driven partly by the internal energy of the collapsed bubble, and partly by the energy stored in the liquid compressed during the collapse phase, $E_{compr}^{coll1}$. Correspondingly, the rebound shock wave energy is composed of two fractions: $E_{SWB}$ arising from the bubble rebound, and $E_{SWL}$ arising from the re-expansion of the compressed liquid. A complete energy balance can be established only at particular times, when the kinetic energy is zero, i.e. at $R_{max1}$, $R_{min1}$, and $R_{max2}$. The energy of the shock wave emitted after optical breakdown is obtained by subtracting the balance established for $R = R_{max1}$ from $E_{abs}$, and the energy of the shock wave emitted during the rebound is evaluated by comparing the balance for $R = R_{max2}$ with the total energy of the compressed bubble and liquid upon collapse.

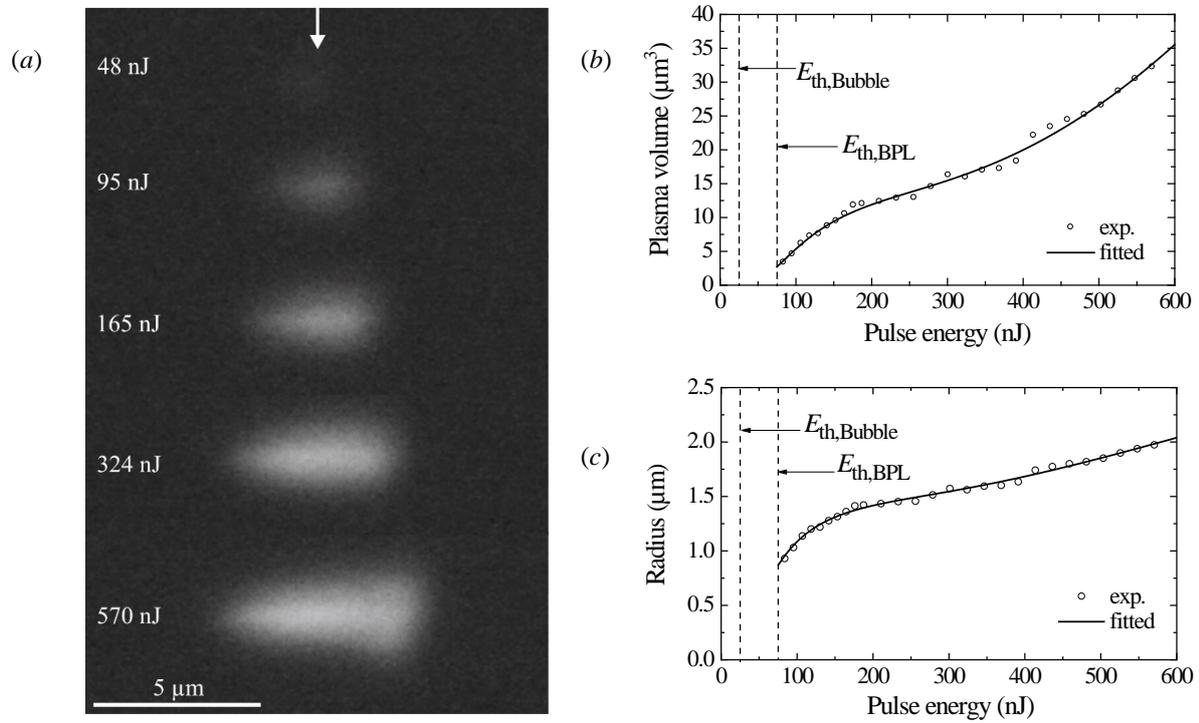

FIGURE 5. Determination of the plasma size produced with 350-fs, 1040-nm pulses of different energies focused at *NA* = 0.8. (a) Photographs of plasma luminescence taken with the setup of figure 1(*a*) and integrated over 70 laser pulses at ISO 3200. (b) Plasma volume determined from photographs as a function of laser pulse energy. (c) Energy dependence of the radius of a sphere having the same volume as the plasma in (b).

**Liang et al. / Figure 6**

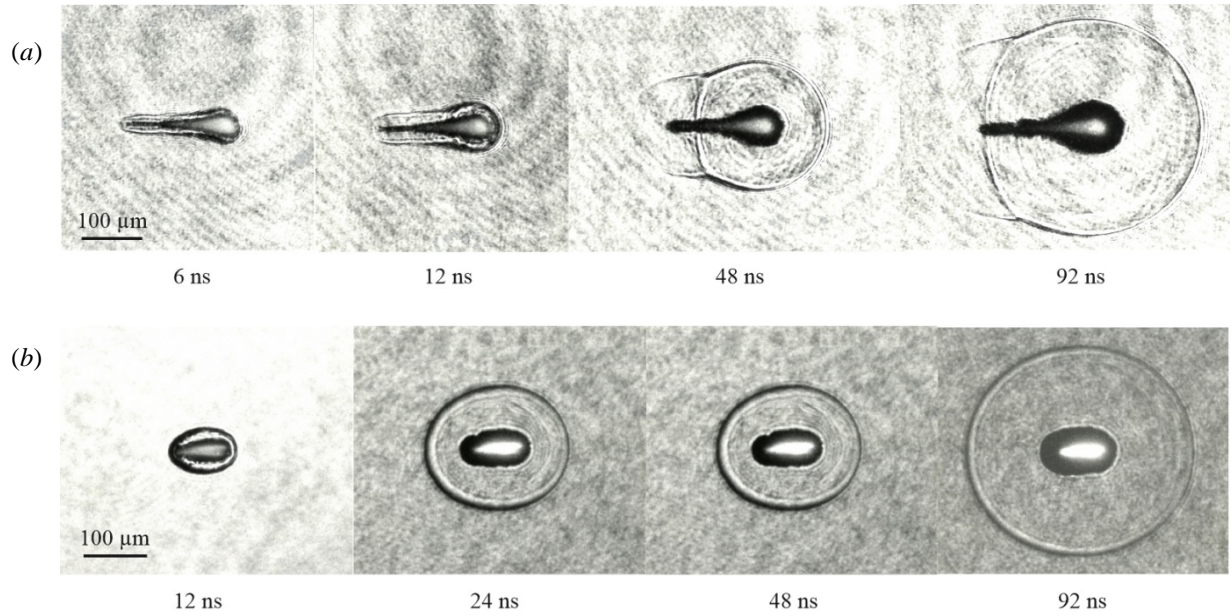

FIGURE 6. Influence of plasma shape on the initial phase of shock wave emission and cavitation bubble expansion after laser-induced breakdown. (*a*) Plasmas were produced with 240-µJ, 30-ps, 1040-nm pulses focused at *NA* = 0.16. (*b*) Plasmas were generated with 250-µJ, 30-ps, 1040-nm pulses focused at *NA* = 0.25. Laser light was incident from the right; photographs were taken with the setup of figure 1(*c*). For both laser pulse durations, the plasmas are elongated. This results in aspherical shock wave emission in the near field that transforms into approximately spherical propagation after little more than 100 ns. The bubble is initially also aspherical and assumes a spherical shape later than the shock wave.

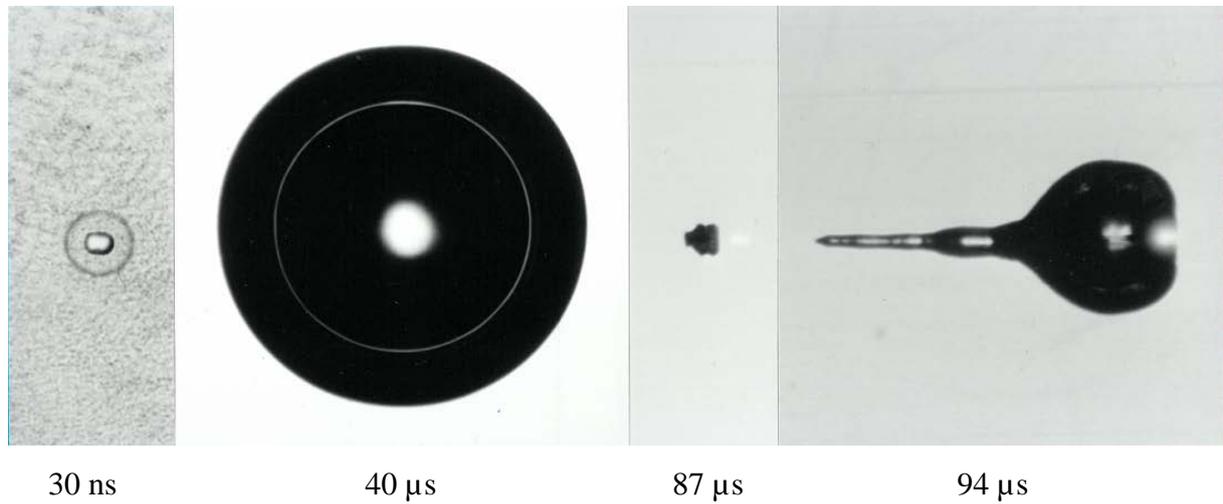

| 30 ns | 40 µs | 87 µs | 94 µs |

FIGURE 7. Spherical expansion and aspherical collapse of a bubble produced by a 1064-nm, 6-ns pulse of 182 µJ energy that was focused through a microscope objective with 2.2 mm working distance at $NA = 0.9$. The oscillation time was 86.6 µs, the maximum bubble size $R_{max} = 473$ µm (Venugopalan et al. 2002), and the dimensionless standoff distance from the front lens of the objective was $\gamma = 4.65$.

**Liang et al. / Figure 8**

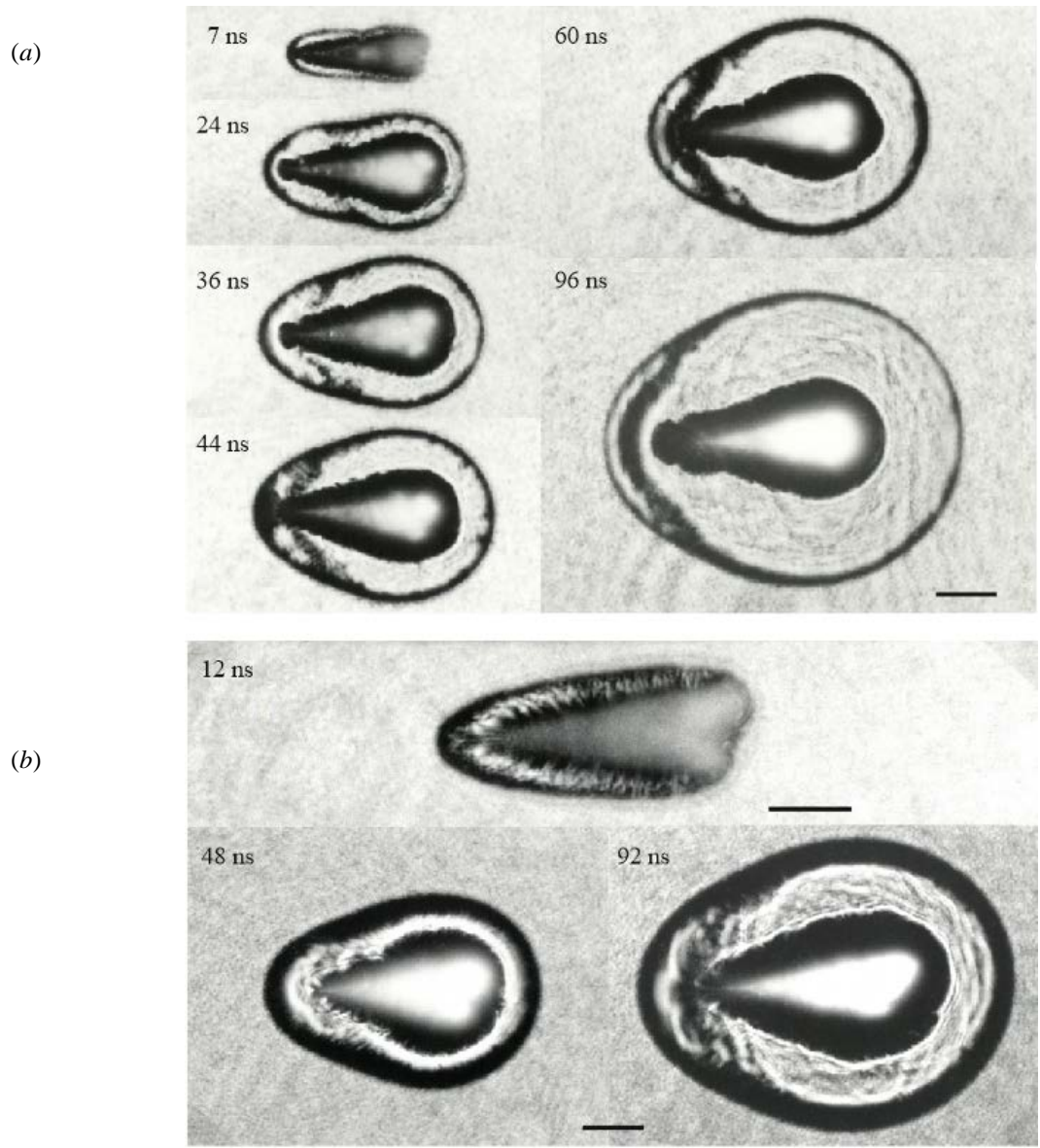

FIGURE 8. Initial phase of shock wave emission and bubble expansion after plasma formation produced by 1064-nm, 6-ns laser pulses with pulse energies of (*a*) 10 mJ and (*b*) 20 mJ that were focused at *NA* = 0.25. The laser light was incident from the right. Scale bars represent 100 µm. The self-luminescent plasma appears on all images because photographs were taken with open shutter in a darkened room. Bright-field illumination was done with a 6 ns laser pulse from a collimated laser beam as shown in figure 1(*c*).

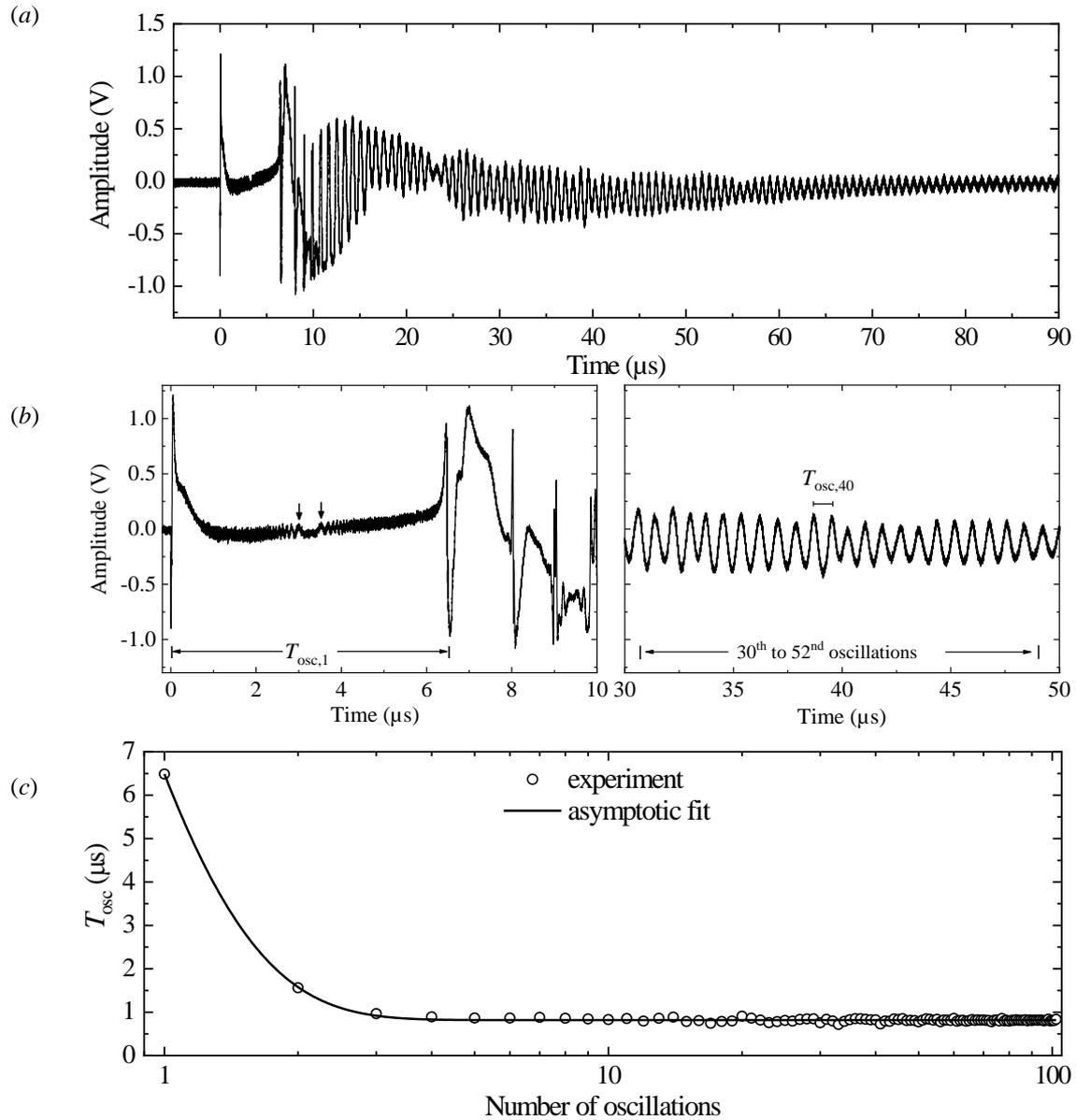

FIGURE 9. Confocal probe beam forward scattering signal from a bubble with 35.8 µm maximum radius produced by a 265-fs, 755-nm laser pulse of 155 nJ energy focused at $NA = 0.9$ The dimensionless stand-off distance from the microscope objective's front lens is $\gamma = 70$. (a) presents the entire signal portraying the transition from nonlinear cavitation bubble oscillations to linear oscillations of the residual gas bubble. (b) shows enlarged views of the first four oscillations and the part from $30^{th}$ to $52^{nd}$ oscillation. The signal undulations during the first oscillation are interference fringes reflecting the radius-time evolution. The arrows mark the time interval around $R_{max}$. Later, each undulation represents one period of the small-amplitude oscillations of the residual bubble. In (c), the oscillation time, $T_{osc}$, is plotted as a function of the oscillation number, $i$. The experimental data are fitted with an asymptotic regression model curve given by $T_{osc} = a - b \times c^i$, with fitting coefficients $a = 0.817$, $b = -42.267$ and $c = 0.134$. The mean oscillation time from $50^{th}$ to $102^{nd}$ oscillation is $813.3 \pm 6.7$ ns.

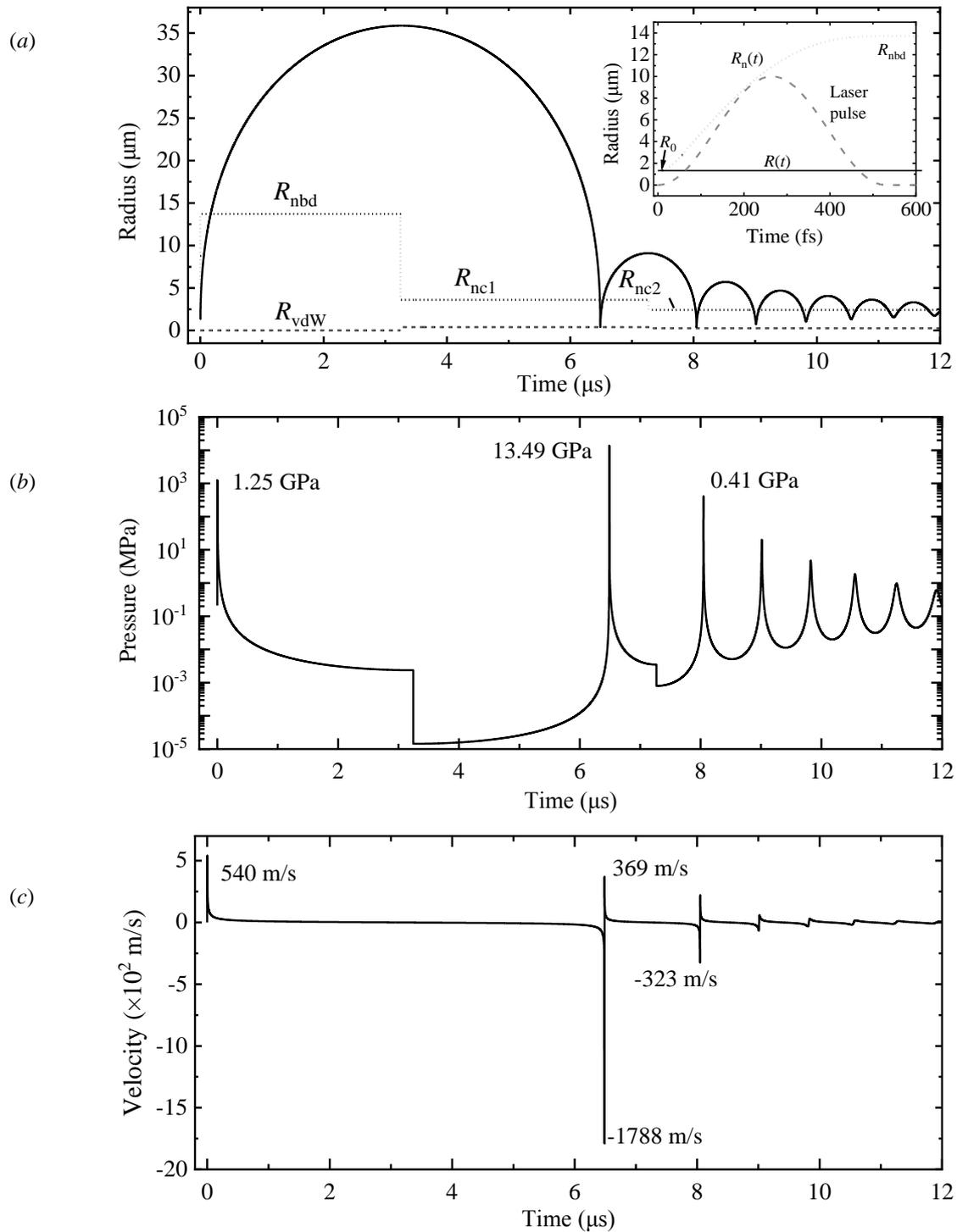

FIGURE 10. Time evolution of (*a*) bubble radius, (*b*) internal pressure, and (*c*) bubble wall velocity corresponding to the signal of figure 9. The insert in (a) shows the laser pulse shape assumed in the calculations and the increase of $R_n$ during the pulse from $R_0$ to $R_{nbd}$. The reduction of equilibrium bubble pressure at $R = R_{max}$ shown in (a) goes along with a drop of internal bubble pressure that represents the net amount of vapour condensation during the first bubble oscillation. Peak pressures upon breakdown and collapse and peak velocities are indicated in the figure.

**Liang et al. / Figure 11**

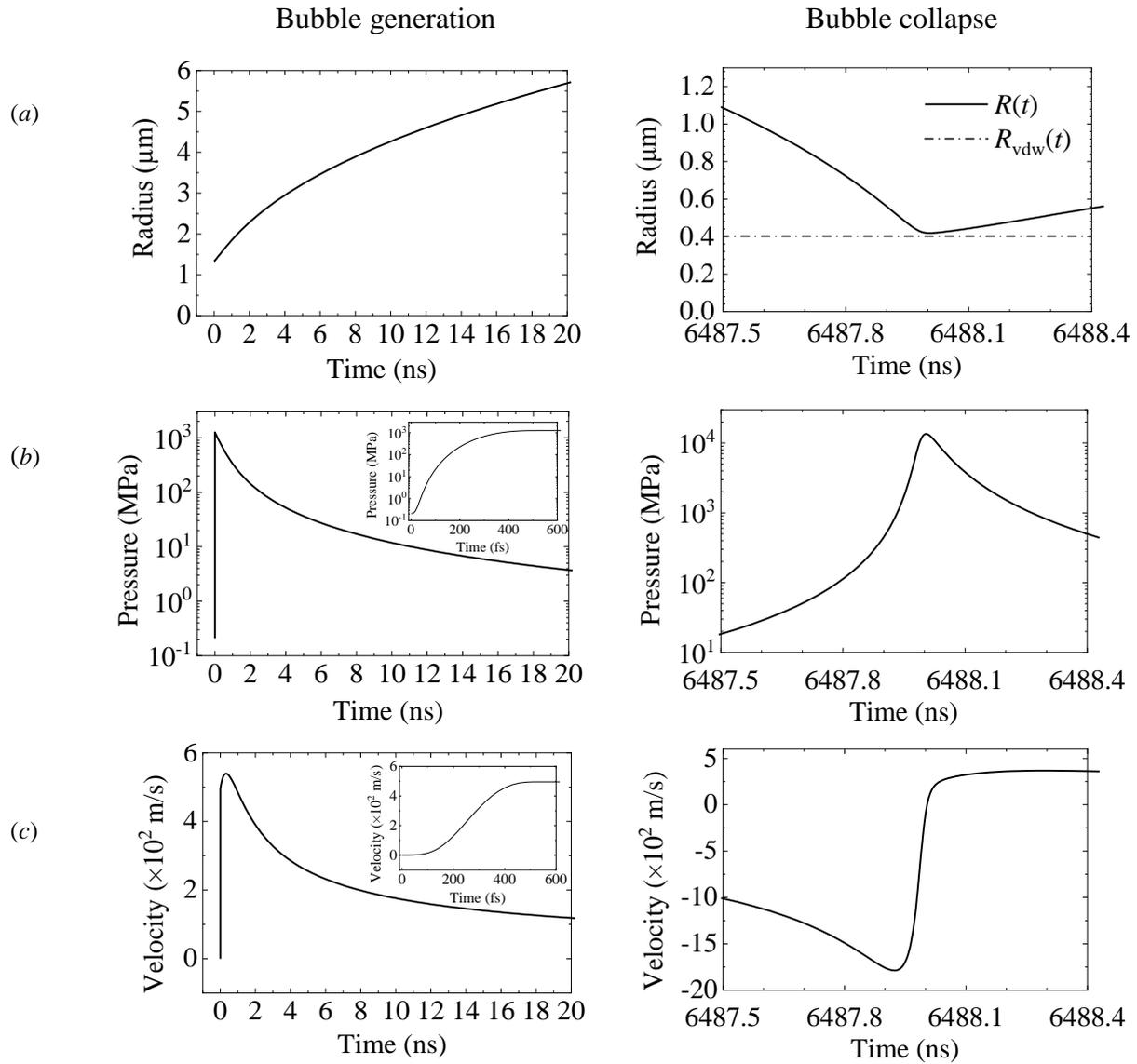

FIGURE 11. Enlarged views of the time evolution of (*a*) bubble radius, (*b*) internal pressure, and (*c*) bubble wall velocity after breakdown and around the first bubble collapse for the same parameters as in figure 10. The displayed time interval is 20 ns for the bubble growth and 1 ns for the collapse-rebound phase. The dashed line in the *R(t)* plot for the collapse phase in (*a*) represents the van der Waals hardcore. Inserts in (*b*) and (*c*) show the time evolution of *P(t)* and *U(t)* during the laser pulse.

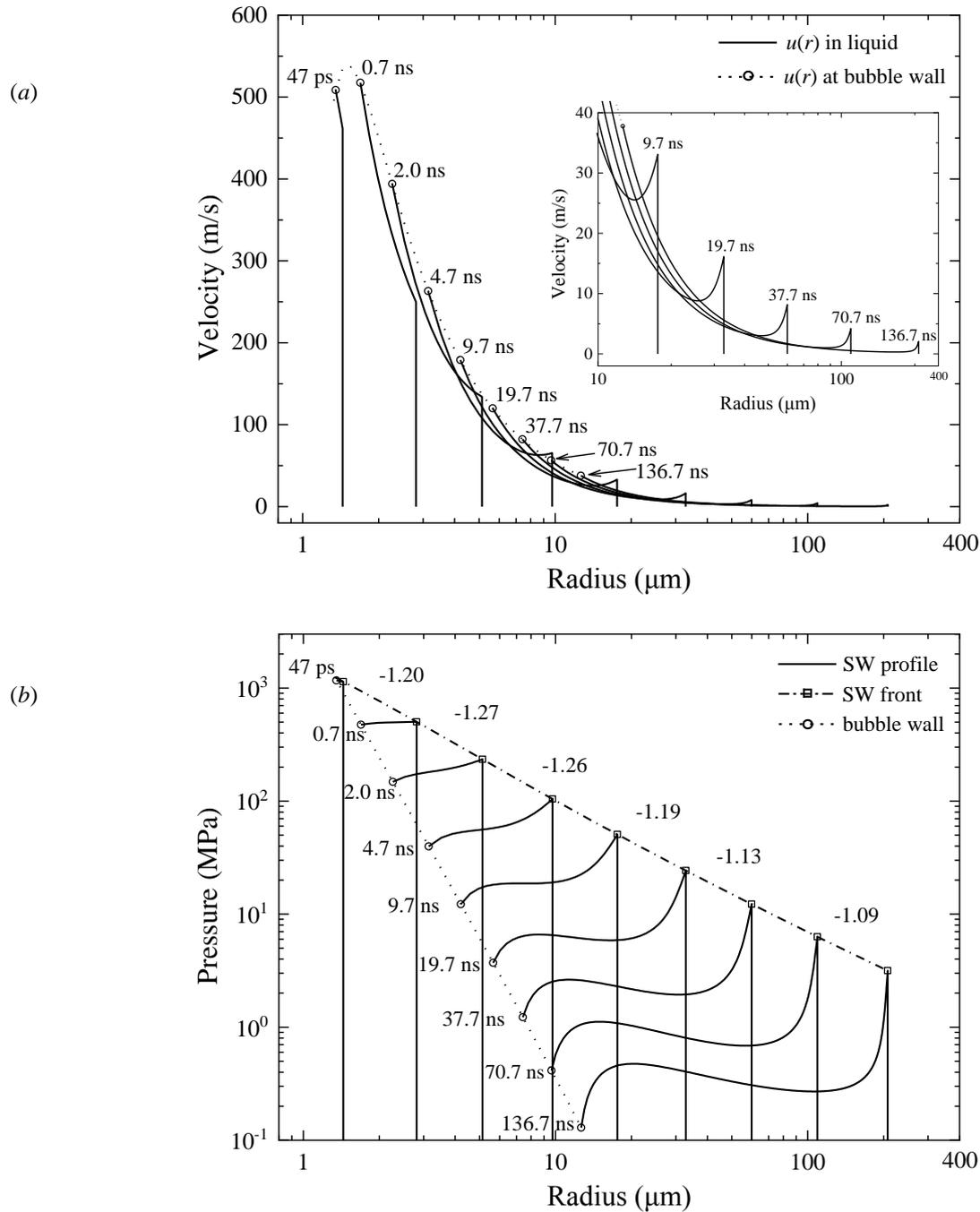

FIGURE 12. Shock wave emission after breakdown for the parameters of figure 10, with velocity distributions in the liquid, $u(r)$, at different times in (*a*), and the corresponding pressure distributions, $p(r)$, presented in (*b*). The circles indicate the respective velocity and pressure values at the bubble wall and its position. The insert in (*a*) shows an enlarged view of the shock wave propagation when it has detached from the outward going radial flow in the bubble's vicinity. The dash-dotted line in (*b*) represents a decay curve of the shock wave's peak pressure, $p_{\text{peak}}(r)$ that was derived from 144 $p(r)$ profiles. The slopes of the $p_{\text{peak}}(r)$ curve are indicated for various propagation distances. The pressure decay is faster than for acoustic waves for which the attenuation would be proportional to $r^{-1}$.

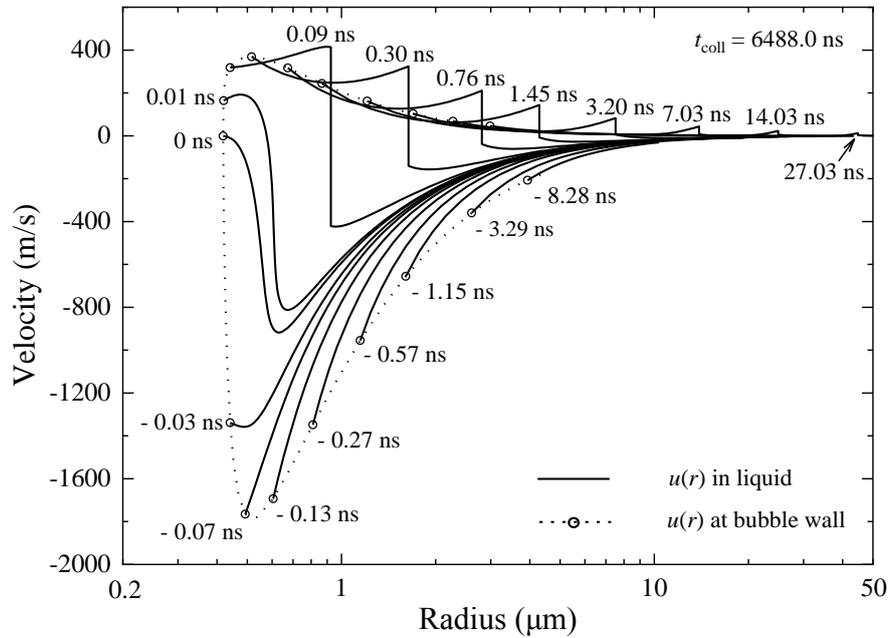

FIGURE 13. Evolution of the velocity distribution in the liquid during the late stage of bubble collapse and during the bubble's rebound for the parameters of figure 10. The time evolution of the $u(r)$ curves is shown with the circles representing the respective pressures at the bubble wall and its position. The times given for the individual $u(r)$ curves refer to the instant at which the bubble reaches its minimum radius, which is set as $t = 0$. On a time scale starting with bubble generation, it corresponds to $t_{coll} = 6488.0$ ns. Upon rebound, the flow around the expanding bubble collides with the still incoming flow from outer regions, and a shock front develops within about 50 ps and 750 nm propagation distance that continues to exist even in the far field.

**Liang et al. / Figure 14**

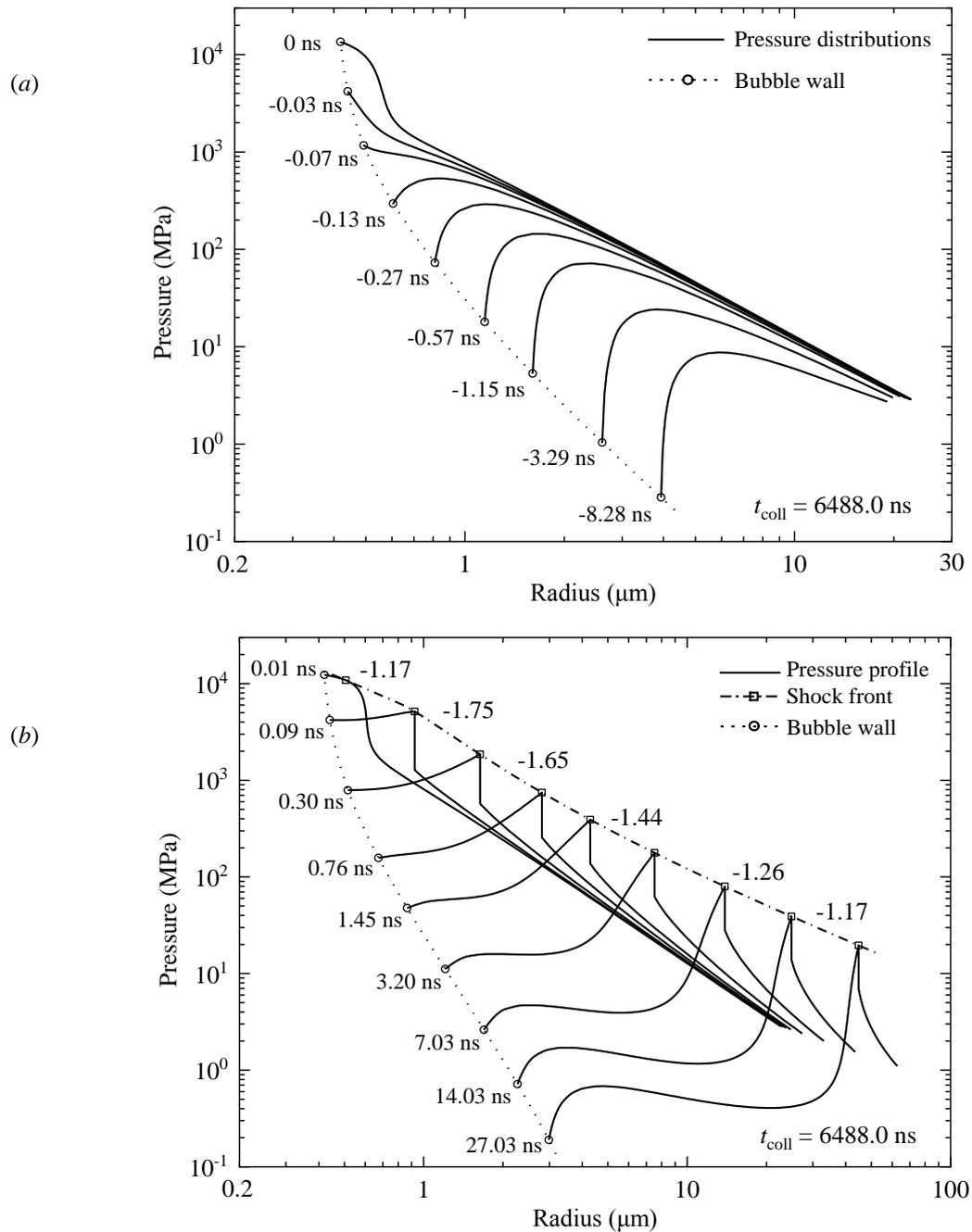

FIGURE 14. Evolution of the pressure distribution in the liquid during the late stage of bubble collapse (*a*) and during the bubble's rebound (*b*) for the parameters of figure 10. The time evolution of the $p(r)$ curves is shown with the circles representing the respective pressures at the bubble wall and its position. The times given for the individual $p(r)$ curves refer to the instant at which the bubble reaches its minimum radius, which is set as $t = 0$. The curve for bubble wall position in (a) was determined from 28 shock wave profiles, and the respective curve in (b) as well as the $p_{peak}(r)$ curve were derived from 53 $p(r)$ profiles. After the shock front has formed, the amplitude of the outgoing pressure wave drops initially very rapidly and later more slowly. However, even in the far field, the shock front continues to exist and the pressure decay is faster than for acoustic waves.

**Liang et al. / Figure 15**

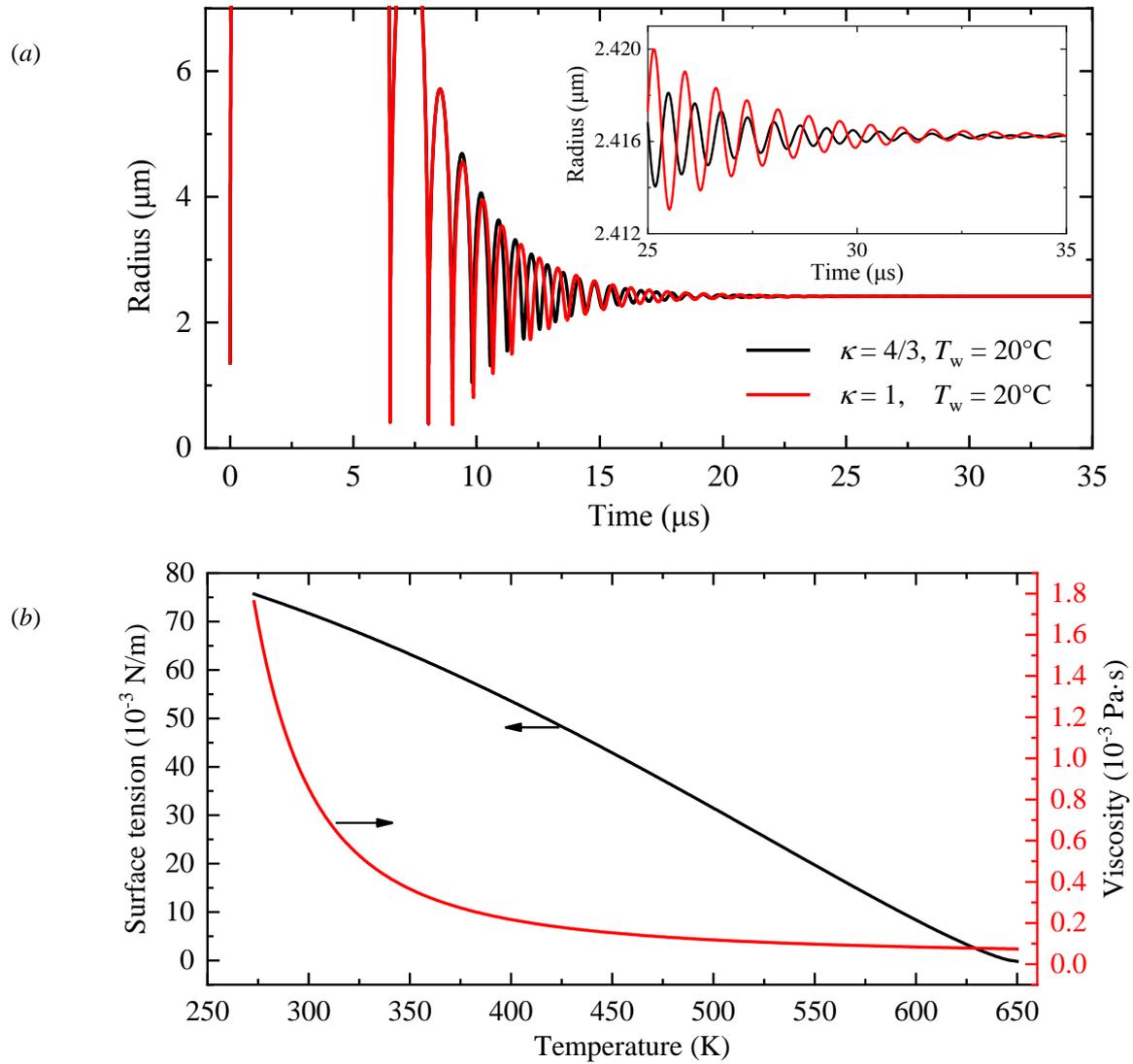

FIGURE 15. (*a*) Time evolution of the bubble radius at late times for the parameters of figure 10, calculated with values for surface tension $\sigma$ and viscosity $\mu$ at room temperature, $T_W = 20\ °C$. The insert shows an expanded view of the radius scale around the equilibrium gas bubble radius in the time interval between 25 µs and 35 µs. Simulations were performed for adiabatic conditions during the entire bubble life time, with $\kappa = 4/3$, and for initially adiabatic conditions followed by isothermal conditions after the maximum of the third oscillation, with $\kappa = 1$. (b) Temperature dependence of surface tension and viscosity of water at ambient pressure.

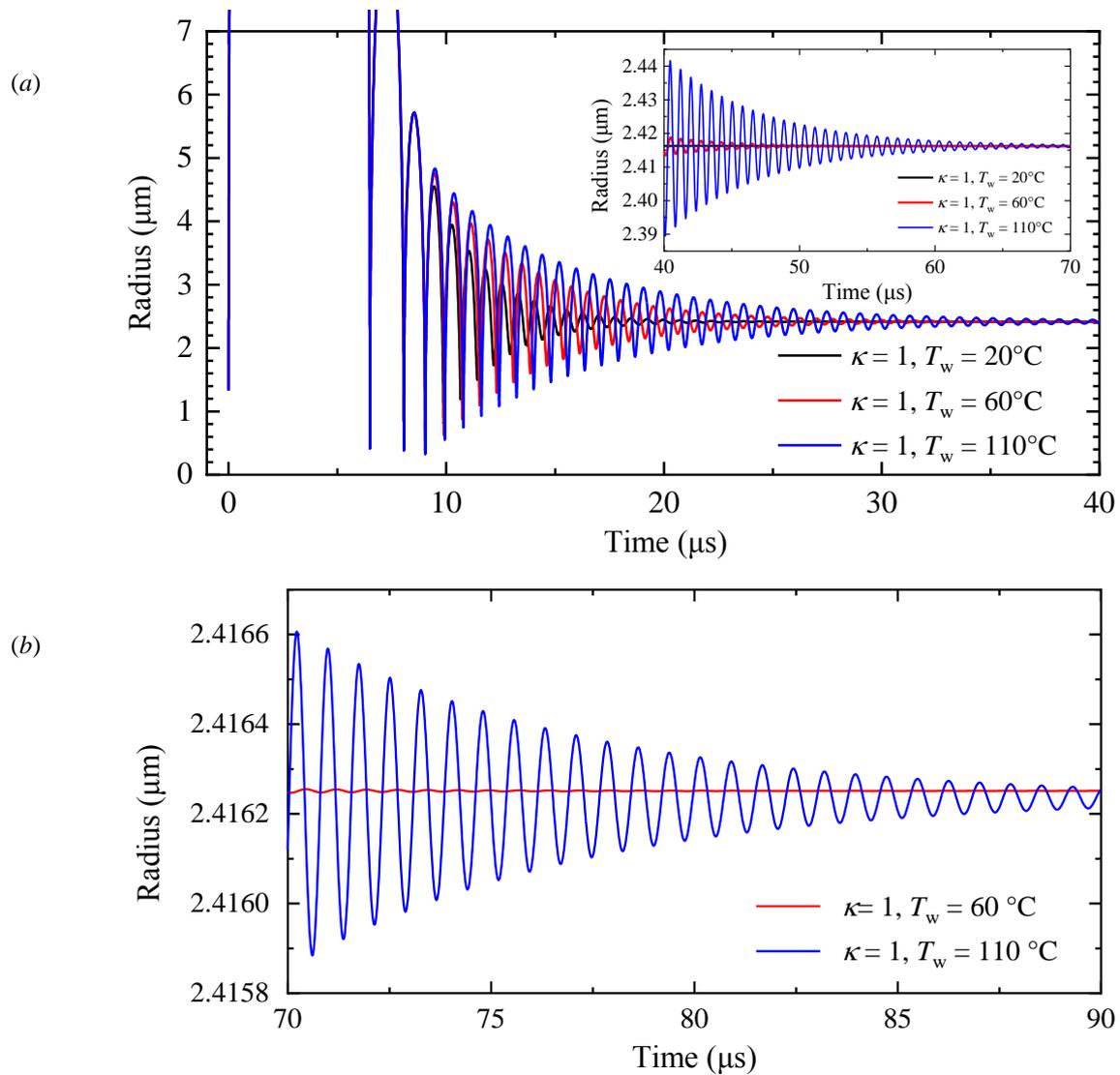

FIGURE 16. Radius-time curves at late times for the parameters of figure 10, calculated with values for surface tension $\sigma$ and viscosity $\mu$ at different water temperatures, with room temperature as reference. Simulations were performed for initially adiabatic conditions followed by isothermal conditions after the maximum of the third oscillation, with $\kappa = 1$. The curves in (a) show the time evolution up to 40 µs, and the insert shows an expanded view of the radius scale around the equilibrium gas bubble radius for the time interval between 40 µs and 70 µs. The curves in (b) show a further expanded radius scale for the time interval between 70 µs and 90 µs, i.e. up to the end of the experimentally observed bubble oscillations.

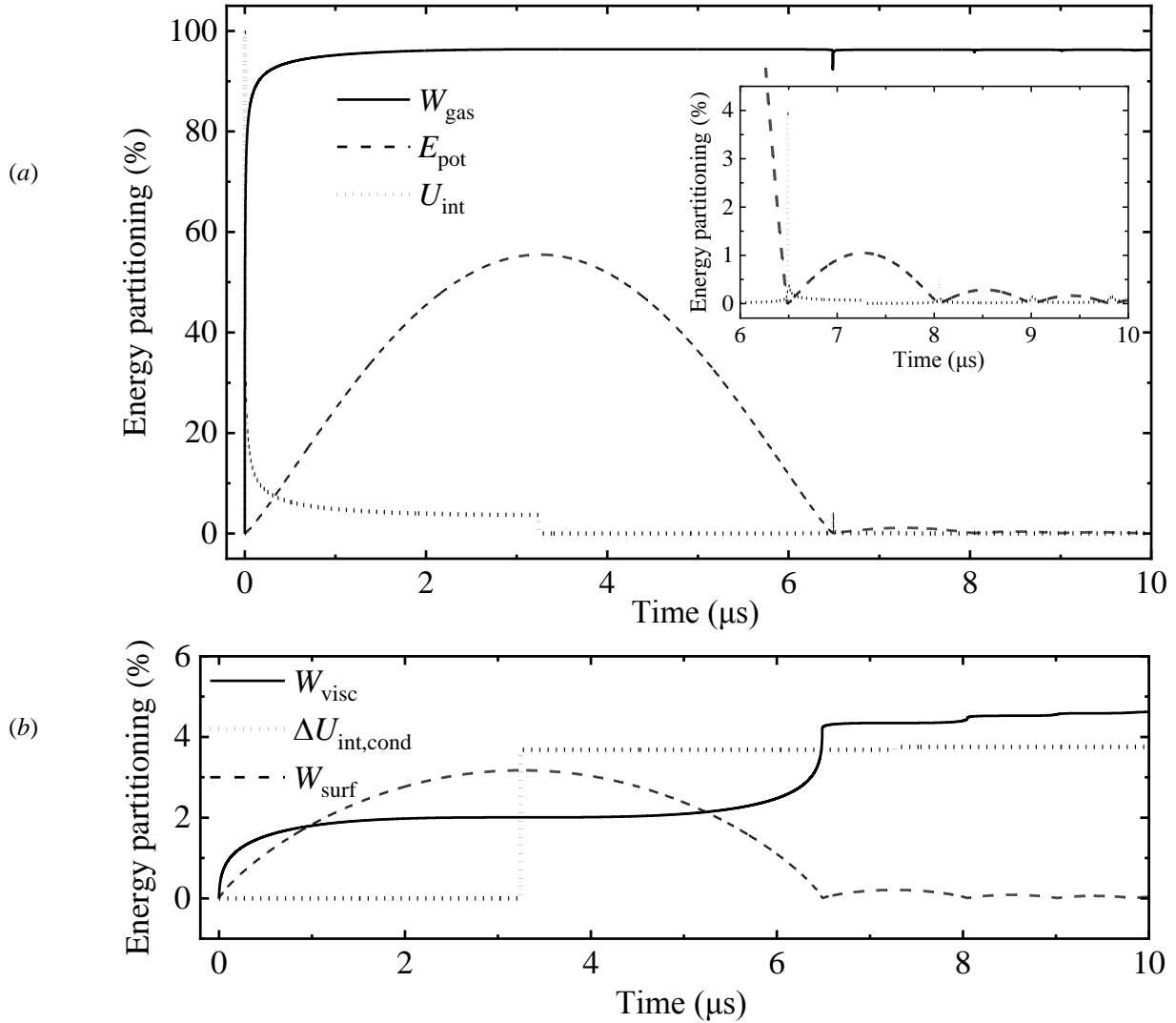

FIGURE 17. Time evolution of the energy partitioning of internal bubble energy $U_{int}$ for the same parameters as in figure 10 and table 2. Part (a) shows the increase of work $W_{gas}$ done on the liquid by the expanding gas together with the corresponding decrease of internal energy, followed by a reversal upon collapse, where the inrushing liquid does work while compressing the bubble content and increasing its internal energy (see insert). Also shown is the evolution of the bubble energy during the first five oscillations. The shock wave energy is included in $W_{gas}$ but can be explicitly evaluated only when $E_{kin} = 0$ at $R_{max1}$ and $R_{max2}$. The corresponding values are listed in table 2. Part (b) presents on an expanded scale the evolution of the energy fractions needed to overcome viscous damping, $W_{visc}$, and surface tension, $W_{visc}$ [part of $E_{pot}$ in (a)], as well as the internal energy lost via condensation, $E_{cond}$. The change of $E_{cond}$ reflects the changes of the bubble's equilibrium radius during the bubble oscillations from $R_{nbd}$ through $R_{nc1}$ to $R_{nc2}$. While the amount of the changes reflects the actual energy dissipation by condensation, the jumps at $R_{max1}$ and $R_{max2}$ are non-physical and due to the numerical procedure.

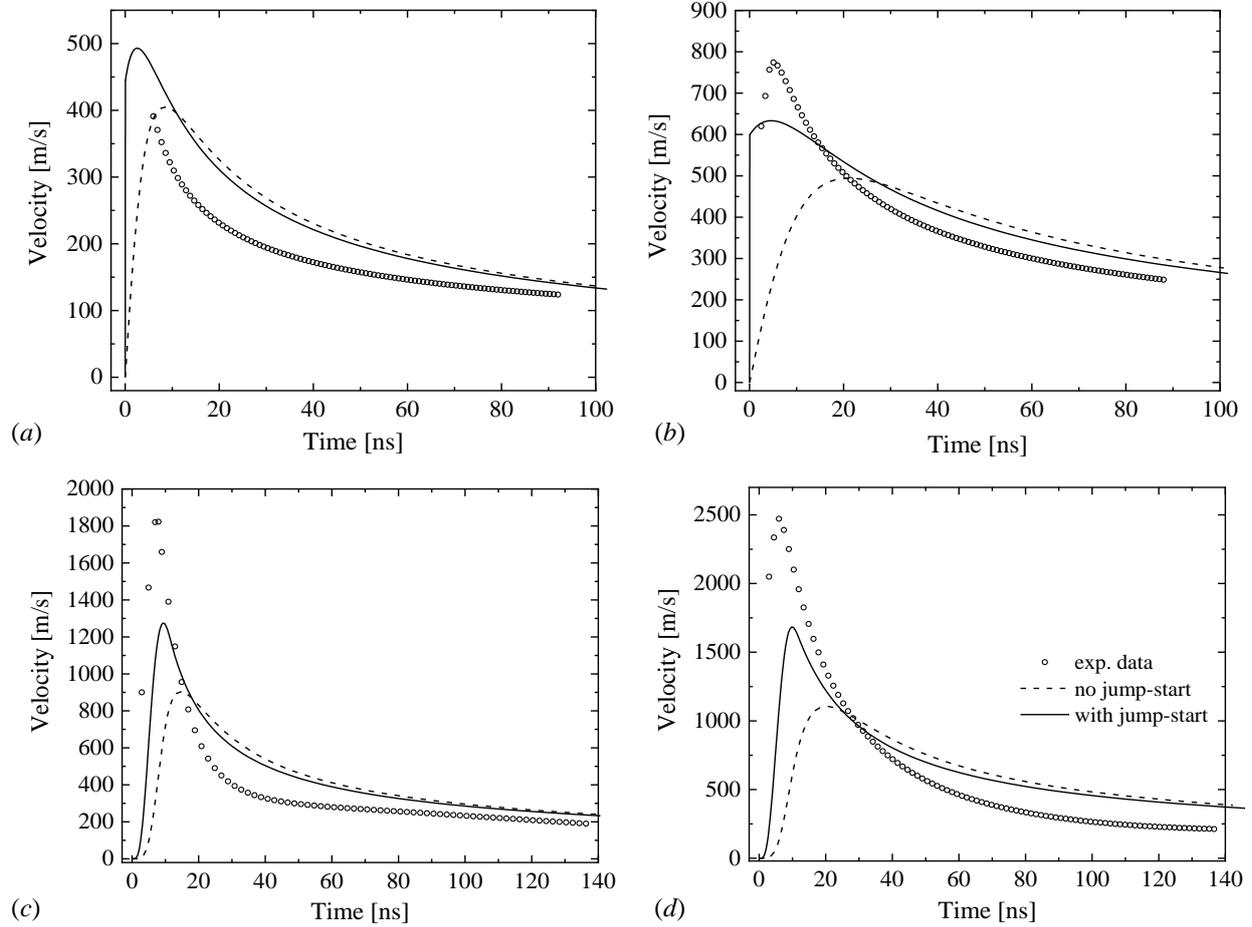

FIGURE 18. Comparison of simulated $U(t)$ curves with the experimental data from Vogel et al. (1996) for laser-induced bubbles generated at 1064 nm wavelength. Pulse durations and energies were 30 ps and 50 µJ in (*a*), 30 ps and 1 mJ in (*b*), 6 ns and 1 mJ in (*c*), and 6 ns and 10 mJ in (*d*). Simulations were performed with and without consideration of the contribution of particle velocity behind the shock front to the bubble wall velocity after optical breakdown. Simulation parameters providing an optimum fit to experimentally determined $R_{max1}$ values with and without 'jump-start' condition are $R_0 = 8.5$ µm, $R_{nbd} = 86.1$ µm, $R_{nbd} = 87.2$ µm in (*a*), $R_0 = 26$ µm, $R_{nbd} = 294$ µm, $R_{nbd} = 298.3$ µm in (*b*), $R_0 = 19$ µm, $R_{nbd} = 291$ µm, $R_{nbd} = 297$ µm in (*c*), and $R_0 = 37$ µm, $R_{nbd} = 660$ µm and $R_{nbd} = 671$ µm in (*d*).

**Liang et al. / Figure 19**

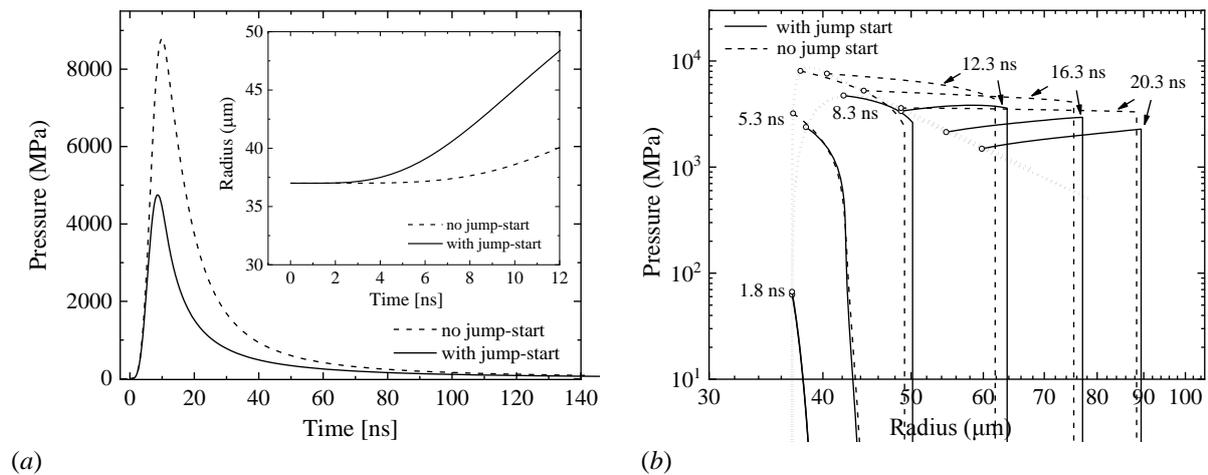

FIGURE 19. (*a*) Simulated *P*(*t*) curves and *R*(*t*) curves (insert) for the initial phase of laser-induced bubble expansion for the same parameters as in figure 18 (*d*). Solid lines show the results with consideration of the contribution of particle velocity behind the shock front to the bubble wall velocity. leading to a jump start of bubble wall velocity, and dashed lines show results without its consideration. (*b*) Pressure distributions in the liquid at different time instants showing shock front formation and the initial phase of shock wave emission. The circles indicate the pressure values at the bubble wall and its position corresponding to the respective *p*(*r*) curves, and the dotted lines show the *P*(*R*) trajectories of the bubble wall. The rapid start of the bubble motion with jump-start of the bubble wall velocity seen in the insert of (a) results in a lower maximum bubble pressure (4749 MPa) than without jump-start (8803 MPa). However, the shock front has formed after about 8 ns in both cases, as seen in (*b*).

**Liang et al. / Figure 20**

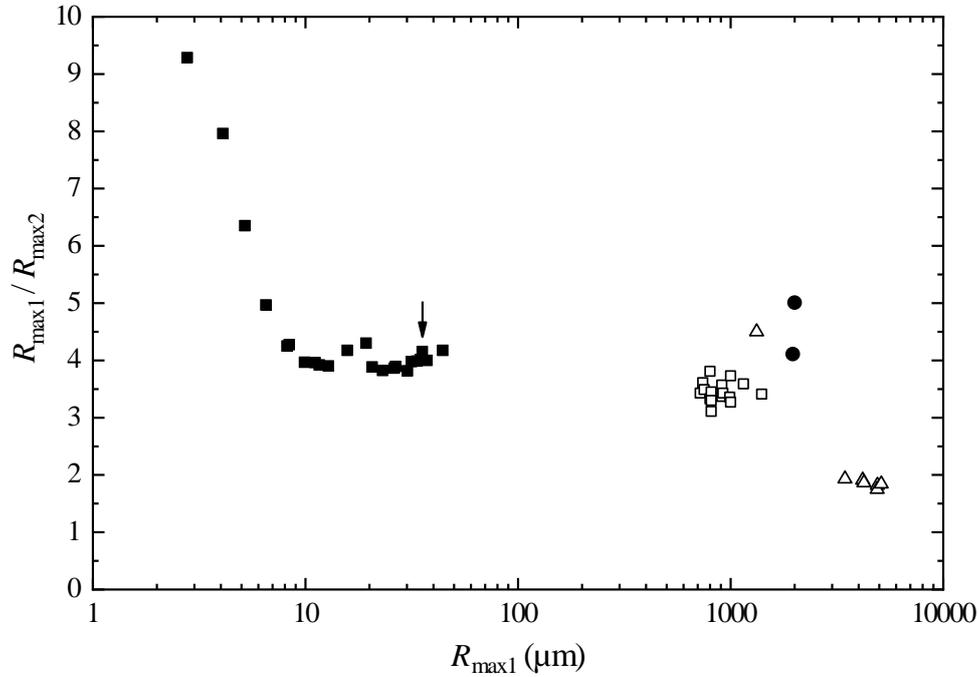

FIGURE 20. Ratio $R_{max1}/R_{max2}$ under ambient conditions as a function of maximum bubble radius. ■ Present study; △ Vogel & Lauterborn (1988); □ Akhatov et al. (2001); ● Sinibaldi et al. (2019). The ratio is a measure of energy dissipation during the first collapse and, since most energy is carried away by acoustic radiation, for the vigour of the collapse that determines the amplitude of the collapse pressure. Data from the present study are compared with data from previous studies on laser-induced cavitation. An arrow marks the data point corresponding to the signal of figure 9. The collapse of the highly spherical laser-induced bubbles investigated in this paper is more vigorous than that of larger, millimetre-sized bubbles, where usually smaller focusing angles were used for plasma generation and buoyancy leads to deviations from spherical shape upon collapse. The largest $R_{max1}/R_{max2}$ value for millimetre-sized bubbles was observed by Sinibaldi et al. (2019) for tight focusing at NA = 0.6. The $R_{max1}/R_{max2}$ ratio exhibits a pronounced increase for $R_{max1} < 10$ μm, where the pressure exerted by surface tension becomes a significant contribution to the outer pressure driving the collapse, and viscous damping increases.